\documentclass[11pt]{article}

\usepackage{amsmath,amsfonts,amssymb,amsthm,epsfig,epstopdf,titling,url,array}
\usepackage{graphicx, psfrag}
\usepackage{dynkin-diagrams}
\usepackage{appendix}
\usepackage{cite}
\usepackage{longtable}
\usepackage{multirow}
\usepackage{float}
\usepackage{hyperref}
\usepackage[utf8]{inputenc}
\usepackage[english]{babel}
\usepackage{color}
\usepackage{booktabs,caption}
\usepackage[flushleft]{threeparttable}
\usepackage{lscape}
\definecolor{Mygrey}{gray}{0.8}
\definecolor{Mywhite}{gray}{1.0}

\newcommand{\be}{\begin{equation}}
\newcommand{\ee}{\end{equation}}
\newcommand{\bea}{\begin{eqnarray}}
\newcommand{\eea}{\end{eqnarray}}
\usepackage{colortbl}

\begin{document}

\begin{center}
\textbf{\Large \bf Classifying three-character RCFTs with Wronskian index equalling 3 or 4}
\end{center}

\vskip .6cm
\medskip

\vspace*{4.0ex}

\baselineskip=18pt

\centerline{\large \rm   Chethan N. Gowdigere$^{1\,ab}$, Sachin Kala$^{2\,ab}$, Jagannath Santara$^{3\,c}$}

\vspace*{4.0ex}

\centerline{\large \it  $^a$National Institute of Science Education and Research Bhubaneshwar,}

\centerline{\large \it  P.O. Jatni, Khurdha, 752050, Odisha, INDIA}

\vspace*{1.0ex}

\centerline{\large \it  $^b$Homi Bhabha National Institute, Training School Complex, }

\centerline{\large \it  Anushakti Nagar, Mumbai 400094, INDIA}

\vspace*{1.0ex}

\centerline{\large \it $^c$Department of Physics,}

\centerline{\large \it  Indian Institute of Technology Madras,}

\centerline{\large \it Chennai 600036, INDIA}

\vspace*{4.0ex}
\centerline{E-mail: $^1$chethan.gowdigere@niser.ac.in, $^2$sachin.kala@niser.ac.in, $^3$jagannath.santara@physics.iitm.ac.in}

\vspace*{5.0ex}
\centerline{\bf Abstract} \bigskip

In the Mathur-Mukhi-Sen (MMS) classification scheme for rational conformal field theories (RCFTs), a RCFT is identified by a pair of non-negative integers $\mathbf{[n, \ell]}$, with $\mathbf{n}$ being the number of characters and $\mathbf{\ell}$ the Wronskian index.  The modular linear differential equation (MLDE) that the characters of a RCFT solve are labelled similarly.  All RCFTs with a given $\mathbf{[n, \ell]}$ solve the modular linear differential equation (MLDE) labelled by the same $\mathbf{[n, \ell]}$. With the goal of classifying $\mathbf{[3,3]}$ and $\mathbf{[3,4]}$ CFTs, we set-up and solve those MLDEs, each of which is a three-parameter non-rigid MLDE, for character-like solutions. In the former case,  we obtain four infinite families  and a discrete set of $15$ solutions, all in the range $0 < c \leq 32$. Amongst these $\mathbf{[3,3]}$ character-like solutions, we find pairs of them that form coset-bilinear relations with meromorphic CFTs/characters of central charges $16, 24, 32, 40, 48, 56,  64$. There are six families of coset-bilinear relations where both the RCFTs of the pair are drawn from the infinite families of solutions. There are an additional  $23$ coset-bilinear relations between character-like solutions of the discrete set.  The coset-bilinear relations should help in identifying the $\mathbf{[3,3]}$ CFTs.  In the  $\mathbf{[3,4]}$ case, we obtain nine character-like solutions each of which is a  $\mathbf{[2,2]}$ character-like solution adjoined with a constant character.

\vfill \eject

\baselineskip=18pt

\tableofcontents

\section{Introduction \label{1s}}

Two dimensional conformal field theory (CFT) has proved to be  very useful in multiple areas of both physics and mathematics. A partial list of references are \cite{Belavin:1984vu, Witten:1983ar, DiFrancesco:1997nk, Moore:1989vd, Fuchs:2009iz, Gaberdiel:1999mc}. Given it’s ubiquitousness and usefulness, one would desire  a classification of $2d$ CFTs. But perhaps this task is too hard. A more tractable subclass of 2d CFTs are the rational conformal field theories (RCFT). The definition is in terms of torus partition functions. The torus partition function of a RCFT is a finite sum\footnote{This is also the definition that can be found in the textbook \cite{DiFrancesco:1997nk}, see section 10.8.7.} of products of a finite number of building blocks known as the characters. The number of characters of a RCFT is  important information about the RCFT. This is the finiteness aspect of the definition of a RCFT. Another aspect is rationality, which is in the name itself. The central charge $c$ and conformal dimensions of primary fields, the $h$s, of a RCFT are all rational numbers. One does not need to include rationality in the definition. Defining it by the finiteness condition suffices. It was shown in \cite{Anderson:1987ge} that finiteness implies rationality. The classification problem for RCFT is also deemed to be quite hard. One can read the following statement in section 10.8.7 of \cite{DiFrancesco:1997nk} : ``The classification of all RCFTs is a formidable task and it will probably remain an open problem for a while.''

There exists a classification scheme for RCFTs that was discovered by Mathur, Mukhi and Sen (MMS) in \cite{Mathur:1988na, Mathur:1988gt}. In this scheme, one identifies a RCFT by a pair of  integers, $\mathbf{[n, \ell]}$. $\mathbf{n}$ is the number of characters and $\mathbf{\ell}$ is the Wronskian index\footnote{The reader may be disappointed not find this term anywhere in the original papers. It was introduced into the subject by one of the founders only in \cite{Mukhi:2019xjy}.}  The Wronskian index is described now. The $\mathbf{n}$ characters of a RCFT solve a $\mathbf{n}$-th order ordinary linear differential equation on the moduli space of the torus, a modular linear differential equation (MLDE) \cite{Eguchi:1987qd}. The Wronskian of these characters is modular of weight $\mathbf{n} ( \mathbf{n} - 1)$. Modularity restricts the zeroes and poles of a modular object according to the weight. The Wronskian index of a RCFT is some information concerning the zeroes of the Wronskian of the characters of that RCFT. Another essential element of the MMS classification scheme is the MLDE that the characters solve. The MLDE, remarkably, turns out to be also labelled by the same kind of pair of integers, $\mathbf{[n, \ell]}$. To be more precise, given a $\mathbf{[n, \ell]}$, one can construct a more-or-less unique MLDE. The MLDE will come with some parameters, making it not completely unique. The parameter space of a MLDE is an important object of consideration in this subject.  The main statement of the MMS classification scheme is the following : Every $\mathbf{[n, \ell]}$ RCFT solves a $\mathbf{[n, \ell]}$ MLDE; different $\mathbf{[n, \ell]}$ RCFTs solve the $\mathbf{[n, \ell]}$ MLDE  but for (typically) different values of the parameters. The MMS classification scheme thus urges one to break up the classification program of RCFTs from one big problem to many small classification problems viz. the classification of $\mathbf{[n, \ell]}$ RCFTs.

The working practitioners of this subject, in the first stage $\mathbf{(I)}$ set up the $\mathbf{[n, \ell]}$ MLDE. Then in the second stage $\mathbf{(II)}$ they scan the parameter space of the MLDE and obtain admissible character solutions. An admissible character solution is a set of solutions that have all the desired features of CFT characters, mainly integrality, i.e. the $q$-series expansion coefficients are all non-negative. It is important to recognise that at this stage, we do not yet have a CFT. Then in the last stage $\mathbf{(III)}$, there is an extra layer of study and analysis, that asks and settles the question of which admissible character solutions of the MLDE are actually CFT characters and which are not\footnote{All three situations can arise. Some  admissible character solutions can correspond to no CFTs while others can  correspond to a unique CFT and yet others can correspond to multiple CFTs.}. Only at the end of this stage, one would have.a classification of $\mathbf{[n, \ell]}$ RCFTs.

For low $\mathbf{n}$ and low $\mathbf{\ell}$s, it happens that there is typically no separate stage $(\mathbf{III})$. Either every admissible character solution corresponds to a CFT or not much analysis is needed in identifying which of the admissible character solutions corresponds to CFTs. But the situation changes for higher $\mathbf{n}$ and/or higher $\mathbf{\ell}$. In this paper we study the $\mathbf{[3,3]}$ and the $\mathbf{[3,4]}$ MLDEs and get their admissible character solutions.  We can complete the stage $(\mathbf{III})$ of the analysis only for the latter case because a quick glance settles the issue while for the former  it is going to be involved and is postponed for future work.  

Our paper is a chapter in the MMS classification program. To arrive at the context of our work, we start our discussion with a survey of completed chapters of the program. We begin with $\mathbf{n} = 1$, i.e. one-character RCFTs a.k.a meromorphic CFTs. It turns out (see \eqref{windex} below) that the Wronskian index of these are always even and non-zero.  The smallest possible value of the Wronskian index is $2$. The $\mathbf{[1,2]}$ MLDE has a unique admissible character solution, $j^{\frac13}$; this is  the character of a single RCFT viz. $E_{8,1}$. The next possible value of the Wronskian index is $4$. The $\mathbf{[1,4]}$ MLDE has a unique admissible character solution, $j^{\frac23}$. And it is known that this can be the character of either of two CFTs viz. $E_{8,1} \otimes E_{8,1}$ or the one-character extension of $D_{16,1}$. Going on, the $\mathbf{[1,6]}$ MLDE has an infinite number of admissible character solutions viz. $j  + {\cal N}, \quad {\cal N} \geq -744$. In \cite{Schellekens:1992db}, it was shown that only $29$ these admissible character solutions correspond to ($71$) CFTs. For  $\mathbf{[1,\ell \geq 8]}$, the MLDE admits an infinite number of admissible character solutions, but the stage ($\mathbf{III}$) of the analysis has not been completed  and there is no complete classification of RCFTs yet; work in this  direction include \cite{Kervaire, King, Das:2022slz}. 

Next we survey two-character RCFTs.  The original paper  \cite{Mathur:1988na} that announced the program was  for a vanishing Wronskian index. The $\mathbf{[2,0]}$ MLDE has one parameter \footnote{\label{rigidnr}The parameter is what is known as a rigid parameter. A rigid parameter is a MLDE parameter which is fixed by the indices (central charge and conformal dimensions) of the characters. On the other hand a non-rigid parameter is a MLDE parameter which is not-fixed by the indices; it is free to take arbitrary values independent of the indices. Another way to discuss rigid and non-rigid parameters is via the indicial equation (the MLDE is usually solved by a Frobenius-like method and the zeroth order equation is the indicial equation). Rigid parameters occur in the indicial equation while non-rigid parameters do not.}. This parameter space was scanned and $9$ admissible character solutions were obtained. In this case, it turned out that stage $\mathbf{(III)}$ was quite simple and it was quickly recognised that $8$ of them corresponds to CFTs (one CFT for one admissible character), One of the admissible characters was shown to correspond to no CFT \cite{Mathur:1988gt}. The $\mathbf{[2,2]}$ MLDE has one rigid parameter and $9$ admissible character solutions were obtained \cite{Naculich:1988xv, Hampapura:2015cea}.  The analysis of stage $\mathbf{(III)}$  involves the meromorphic theory of cosets \cite{Gaberdiel:2016zke} and it was shown that $7$ of the admissible characters each correspond to multiple CFTs for a total of $21$ CFTs. The next Wronskian indices to consider are $4, 6, \ldots$\footnote{It was shown in \cite{Naculich:1988xv} that for two characters, the only relevant Wronskian indices to consider are the even ones. See also appendix of \cite{movpole}.} The meromorphic theory of cosets has turned out to be crucial tool in stage $\mathbf{(III)}$ of identifying genuine CFT amongst admissible character solutions; as we will review below, it played a crucial role for the $\mathbf{[3,0]}$ case.  The $\mathbf{[2,4]}$ MLDE was studied in \cite{Chandra:2018pjq, Tener:2016lcn}. As the Wronskian indices increase, typically the number of parameters in the MLDE increase. Roughly speaking, when the number of parameters exceeds the number of indices (the number of characters), we get non-rigid parameters.  MLDEs with more parameters are harder to solve. For two characters, one starts getting non-rigid MLDEs\footnote{Non-rigid MLDEs and non-rigid parameters display some phenomenon which are absent in  MLDEs with only rigid parameters. One could have two different admissible character solutions that exist at two points in the parameter space which differ only in the values taken by the non-rigid parameters. And those two solutions would have the same indices and hence the same central charges and conformal dimensions. And if they correspond to CFTs, we would have a situation where we would have two completely different CFTs but with the same central charge and conformal dimensions. Thus solving non-rigid MLDEs would lead us to such interesting situations.}  for $\mathbf{\ell} \geq 6$. Initially, admissible character solutions to the non-rigid $\mathbf{[2,6]}$ MLDE were obtained in \cite{Chandra:2018ezv}, not by directly solving the MLDE, but in an indirect manner via the  theory of quasi-characters \cite{Chandra:2018pjq}.  A very recent paper \cite{movpole} gives a general theory of non-rigid MLDEs and their solutions. Other studies of two-character RCFTs and MLDEs, including some in the math literature can be found in \cite{GradyLam, Kaidi:2020ecu, Bae:2017kcl, Bae:2020xzl, Bantay:2010uy, Harvey:2018rdc,  Harvey:2019qzs, kaneko5, kaneko1, kaneko2, kaneko3, Gannon:2013jua, arike1, mason1}.

The theory of quasi-characters was introduced to the RCFT literature in \cite{Chandra:2018pjq}. The mathematical basis of this theory has applications in pure mathematics \cite{kaneko5}. Quasi-characters are solutions to the MLDE which are more general than admissible character solutions; quasi-character solutions are allowed to have negative integral Fourier coefficients. $\mathbf{[2,0]}$, $\mathbf{[2,2]}$  and $\mathbf{[2,4]}$ quasi-characters were constructed in \cite{Chandra:2018pjq}. According to quasi-character theory, admissible character solutions to $\mathbf{[2, \ell]}$ MLDE for $\mathbf{\ell} \geq 6$ can be constructed by taking sums of quasi-characters in appropriate ways. In particular all admissible character solutions to $\mathbf{[2, 6]}$ MLDE were obtained as a sum of two $\mathbf{[2,0]}$ quasi-characters \cite{Chandra:2018ezv}. The general theory is the following\footnote{Although there are cogent arguments why this should be so, there is no proof for the general case. The recent work \cite{movpole} proves, starting from the MLDE, that every Frobenius solution of the $\mathbf{[2, 6]}$ MLDE (and $\mathbf{[2, 8]}$ MLDE) can be constructed using two Frobenius solutions of the $\mathbf{[2, 0]}$ MLDE (resp. $\mathbf{[2, 2]}$ MLDE).} : (i) admissible character solutions to $\mathbf{[2,6r]}$ MLDE can be obtained as sums of $r+1$ $\mathbf{[2,0]}$ quasi-characters, (ii) admissible character solutions to $\mathbf{[2,6r+2]}$ MLDE can be obtained as sums of $r+1$ $\mathbf{[2,2]}$ quasi-characters (iii) admissible character solutions to $\mathbf{[2,6r+4]}$ MLDE can be obtained as sums of $r+1$ $\mathbf{[2,4]}$ quasi-characters. Precise details have not yet been worked out and work in this direction is expected soon \cite{explorations}. Thus one can think of (quasi-character) solutions to the $\mathbf{[2,0]}$, $\mathbf{[2,2]}$ and the $\mathbf{[2,4]}$ MLDEs as building blocks of admissible character solutions to $\mathbf{[2, \ell]}$ MLDEs for all $\mathbf{\ell} \geq 6$.

Now we survey the case of three-characters. Even in one of  original papers of the MMS program \cite{Mathur:1988gt}, the $\mathbf{[3,0]}$ MLDE was set-up and some solutions were studied. It should be noted that this is a two-parameter MLDE (both parameters are rigid) and is technically harder. But all it’s admissible character solutions were obtained decades later in \cite{Kaidi:2021ent, Das:2021uvd, Bae:2021mej}.  (see also \cite{franc1,kaneko4}). The first of these is a study based on modular data and $SL(2, \mathbb{Z})$ representation theory while the latter two directly solve the two-parameter MLDE. The stage $\mathbf{(III)}$ of the analysis that identifies genuine CFT amongst the large number ($303$, excluding an infinite class) of admissible character solutions was done in \cite{Das:2022uoe}. This again involved the application of the theory of meromorphic cosets \cite{Gaberdiel:2016zke}. The final classification of $\mathbf{[3,0]}$ RCFT was presented in  \cite{Das:2022uoe}. The $\mathbf{[3,2]}$ MLDE is also a two-parameter MLDE (both parameters rigid) and a classification of all it’s admissible character solutions and that of CFT was given in \cite{Das:2021uvd}.  

This brings us to the setting of this paper. Here, we study the next two Wronskian indices : the $\mathbf{[3,3]}$ and  $\mathbf{[3,4}]$ MLDEs and their admissible character solutions. Both of them are three-parameter MLDEs : in a certain presentation of the coefficient functions, two of the parameters are rigid and one is non-rigid. This is the first time, according to the best of our knowledge,  a three-parameter MLDE has been solved. We follow the methodology that was set up even in the early paper \cite{Mathur:1988gt} and was implemented in \cite{Das:2021uvd}: (i) we use the first three orders of the Frobenius solution to write the three parameters in terms of objects of the identity character viz. the central charge, and the first two Fourier coefficients $\text{m}_1$ and $\text{m}_2$, (ii) the Frobenius solution at the fourth order then becomes a Diophantine equation for four Diophantine variables, (iii) the indicial equation now becomes a  polynomial equation whose coefficients are made of $\text{m}_1$, $\text{m}_2$ etc and one of it’s solutions is readily known resulting into a quadratic equation whose discriminant needs to be perfect square which provides us with a second Diophantine equation, (iv) one then solves the two Diophantine equations simultaneously and for every solution, (v) compute higher orders of the Frobenius solution and checks for integrality. We thus obtain a classification of $\mathbf{[3,3]}$ and $\mathbf{[3,4]}$ admissible character solutions for central charge $c < 96$. 

The stage $\mathbf{(III)}$ of the analysis that identifies genuine CFT from the admissible character solutions for $\mathbf{[3,3]}$ has not been done here. It is an involved analysis which will be reported later. We do some preliminary work in this direction. Given that the theory of meromorphic cosets \cite{Gaberdiel:2016zke} played a crucial role in the stage $\mathbf{(III)}$ analysis in earlier chapters of the MMS program (\cite{Gaberdiel:2016zke} for $\mathbf{[2,2]}$ and \cite{Das:2022uoe} for $\mathbf{[3,0]}$), we identify pairs of $\mathbf{[3,3]}$ admissible character solutions which form coset-bilinear relations \ref{34ss}. We find coset-bilinear relations involving meromorphic CFTs with central charges up to $64$. The stage $\mathbf{(III)}$ analysis for $\mathbf{[3,4]}$ is quite straight-forward and we are able to give a classification of all $\mathbf{[3,4]}$ CFTs. It turns out that, for $\mathbf{[3,2]}$ in \cite{Das:2021uvd} and for $\mathbf{[3,4]}$ here in this paper, each of the CFTs are a two-character CFT adjoined with a constant character. Thus these CFTs have modular data that are reducible representations. It is known even from \cite{Naculich:1988xv} (see \cite{Kaidi:2021ent} for a complete analysis) that three-character CFTs with modular data that are irreducible representations of $SL(2,\mathbb{Z})$ have Wronskian indices that are multiples of three. Thus true three-character CFTs (in the sense that their modular data are irreducible representations) can only come from the $\mathbf{[3,3]}$ admissible characters. 

It also needs to be remarked that there are no known $\mathbf{[3,3]}$ CFTs. This is very different from the $\mathbf{[3,0]}$ case, where an infinite number of $\mathbf{[3,0]}$ CFTs were known to exist even before the $\mathbf{[3,0]}$ MLDE study : the $B_{r,1}, r \geq 2$ and $D_{r,1}, r \geq 5$. One can find the MMS identification labels $\mathbf{[n, \ell]}$ for many infinite classes of CFTs such as Virasoro minimal models, super-Virasoro minimal models, WZW CFTs (with the associated finite Lie-algebra being simple) in \cite{Das:2020wsi}. Perusing the tables of that paper reveals that there is not a single $\mathbf{[3,3]}$ CFT.  This makes the study of the $\mathbf{[3,3]}$ MLDE that much more interesting. 

One last motivation for this present paper is in connection to the theory of quasi-characters, which for $\mathbf{n = 3}$ is not as well-developed (compared to $\mathbf{n = 2}$), see \cite{Mukhi:2020gnj}. For $\mathbf{n} = 2$, the $\mathbf{[2,0]}$, $\mathbf{[2,2]}$ and $\mathbf{[2,4]}$ quasi-characters are the building blocks for $\mathbf{[2, 6r]}$, $\mathbf{[2, 6r+2]}$ and $\mathbf{[2, 6r+4]}$ admissible characters. Similarly the expectation for $\mathbf{n} = 3$ is the following : the $\mathbf{[3,0]}$ and $\mathbf{[3,3]}$ quasi-characters are  building blocks for $\mathbf{[3, 6r]}$ and $\mathbf{[3, 6r+3]}$ admissible characters. We have not carried out the study of quasi-character solutions to the  $\mathbf{[3,3]}$ MLDE here but we hope that the study of admissible character solutions of this paper will serve as a stepping stone.

This paper is organised as follows. In section \ref{2s}, we quickly recall the various elements of the MLDE approach to RCFT that is at the heart of the MMS classification program. Section \ref{3s} is all about the study of the $\mathbf{[3,3]}$ MLDE. In \ref{31ss} we give the analysis that results in a solution-generating procedure.  In \ref{32ss}, the admissible character solutions of the $\mathbf{[3,3]}$ MLDE are tabulated. There is a discrete set of $15$ solutions and there are four infinite classes. The appendix \ref{app1} should be considered as adjoining \ref{32ss} and contains all relevant details of the admissible character solutions.  Section \ref{33ss} provides a discussion of various aspects of the solutions. In particular, the Kaidi-Lin-Parra-Martinez theory \cite{Kaidi:2021ent} is reviewed and we elaborate on the relation of that theory to the solutions of the present paper. The appendix \ref{app2} adjoins \ref{332ss}.  Section \ref{34ss} is about coset-bilinear relations between $\mathbf{[3,3]}$ admissible characters. We find coset-relations involving meromorphic theories with central charges $16, 24, 32, 40, 48, 56, 64$. These results are meant to prepare the ground for the future study of identifying genuine $\mathbf{[3,3]}$ RCFTs from the admissible character solutions of this paper.  Section \ref{4s} studies the $\mathbf{[3,4]}$ MLDE and classifies all $\mathbf{[3,4]}$ CFT; details are in the accompanying appendix \ref{app3}. In the last section \ref{5s}, we conclude and discuss future directions.

\section{MLDE Approach to RCFTs \label{2s}}
The torus partition function of a  RCFT, 
\be \label{1n}
Z(\tau,\bar{\tau})=\sum_{i,j=0}^{\mathbf{n}-1} M_{ij}\chi_i(\tau)\chi_j(\bar{\tau})
\ee
is constructed out of  building blocks, the $\mathbf{n}$ functions, $\chi_i(\tau)$, on the moduli space of the torus, termed as characters. We will use the term characters generically to denote the building blocks of the partition function.  In a discussion of RCFTs with reference to chiral algebras, these characters are characters in the representation-theoretic sense, and hence the name. 
The characters have a $q$-series expansion\footnote{Here, we assume $c > 0$ and $h_i > 0$. This makes the identity character the ``most singular’’ character and non-identity characters with larger conformal dimensions are ``less singular’’ than the ones with smaller conformal dimensions. This also indicates a unitary RCFT. But non-unitary RCFTs are not excluded in this set-up. We discuss the characters of a non-unitary CFT in terms of it’s indices. The most-singular character would not then correspond to the identity character. },  around the infinite cusp $\tau = i \infty~ (q = 0)$ :
\bea \label{2n}
\chi_0(\tau) &=& q^{\alpha_0} \left( 1 + \sum_{k=1}^\infty \text{m}_k q^{k}\right) \nonumber \\
\chi_i(\tau) &=& q^{\alpha_i}\,\sum_{k=0}^\infty a_k^{(i)} q^{k},\qquad i = 1,\cdots,\mathbf{n}-1
\eea
where $q=e^{2\pi i\tau}$. Here $\alpha_i$ are the indices, corresponding to $h_i-\frac{c}{24}$ where $h_i$ are the conformal dimensions of the primaries and $c$ is the central charge. The $q$-series expansion is also a Fourier expansion and the coefficients are referred to as Fourier coefficients. The Fourier coefficients of the identity character ($\text{m}_k, k\geq 1$) are non-negative integers. The torus partition function is required to be modular invariant : 
\be \label{3n}
Z(\gamma \tau,\gamma\bar{\tau}) = Z(\tau,\bar{\tau}),\quad \gamma=
\begin{pmatrix}
a~ & ~b~\\ c~ & ~d~
\end{pmatrix}
\in SL(2,\mathbb{Z})
\ee
which further requires the characters to transform like vector-valued modular forms:
\be \label{4n}
\chi_i(\gamma\tau)= \sum_kV_{ik}(\gamma)\chi_k(\tau)
\ee
where $V_{ik}$ are matrices in the $\mathbf{n}$-dimensional representation of $SL(2, \mathbb{Z})$.

Let us denote the upper half plane $\mathbb{H}$ as,
\begin{align} \label{5n}
\mathbb{H} = \{\tau\in\mathbb{C} \ | \ \text{Im}(\tau)>0\}. 
\end{align}
The {\it Serre-Ramanujan} derivative operator (see, for example, section 2.8 of \cite{kilford} and exercise 5.1.8 of \cite{rammurthy1}) is defined as follows:
\begin{align} \label{6n}
\mathcal{D} := \frac{6}{\pi i}\frac{d}{d \tau} - k E_2(\tau) = 12\,q\frac{d}{d q} - k E_2(q), 
\end{align}
where $E_2$ is the weight-$2$ Eisenstein series. The operator $\mathcal{D}$  is a linear operator which maps weight $m$ modular objects to weight $m+2$ modular objects and satisfies the Leibniz rule. This is the differential operator with which one sets up the modular linear differential equations and defines the associated Wronskians. With this we can write the general $\mathbf{n}^{th}$ order MLDE \cite{Mathur:1988na}:
\begin{align}
\mathcal{D}^{\mathbf{n}} \chi_i + \sum_{r=0}^{\mathbf{n}-1} \phi_r(\tau)\mathcal{D}^r \chi_i = 0, \label{7n}    
\end{align}
For the solutions of the differential equation \eqref{7n}, define the generalised Wronskians,
\be
W_r = \begin{pmatrix}
\chi_1 & \chi_2  & \cdots & \cdots & \cdots &  \chi_{n} \\
\mathcal{D}\chi_1 & \mathcal{D}\chi_2  & \cdots & \cdots & \cdots & \mathcal{D}\chi_{n} \\
\vdots & \ddots & \ddots & \ddots & \ddots & \vdots \\
\mathcal{D}^{r-1}\chi_1 & \mathcal{D}^{r-1}\chi_2  & \cdots & \cdots & \cdots & \mathcal{D}^{r-1}\chi_{n} \\
    \mathcal{D}^{r+1}\chi_1 & \mathcal{D}^{r+1}\chi_2  & \cdots & \cdots & \cdots & \mathcal{D}^{r+1}\chi_{n} \\
    \vdots & \ddots & \ddots & \ddots & \ddots & \vdots \\
 \mathcal{D}^n\chi_1 & \mathcal{D}^n\chi_2  & \cdots & \cdots & \cdots &  \mathcal{D}^n\chi_{n} \\
\end{pmatrix}
\ee
where
\be
\mathcal{D}^k \,\chi = \mathcal{D}_{2k-2}\ldots \mathcal{D}_{2}\,\mathcal{D}_{0}\chi
\ee
The Wronskian, $W$, of the solutions of \eqref{7n} is given by $W\equiv W_n$. The coefficient functions of the MLDE can be expressed in terms of the generalized Wronskians, 
\begin{align}
\phi_r = (-1)^{n-r}\frac{W_r}{W}. \label{phir}    
\end{align}
Now let us analyse the weights of various objects in \eqref{7n}. $\chi_i$s being characters of a RCFT are weight $0$ modular functions. The first term, $\mathcal{D}^{\mathbf{n}} \chi_i$, is of weight $2\mathbf{n}$ as each operation of $\mathcal{D}$ increases the weight by $2$. Similarly  $\mathcal{D}^r \chi_i$ is of weight $2r$ and hence $\phi_r(\tau)$ are modular functions of weight $2(\mathbf{n}-r)$. Note that since $W$ can have zeros in $\mathbb{H}$, so $\phi_r$s are not modular forms but rather are modular functions. Also, note that $W$ is a modular function of weight $\mathbf{n}(\mathbf{n}-1)$.

\subsubsection*{The Valence Formula and the Wronskian Index}
For a modular function $f$ of weight $k$, $\nu_{\tau_0}(f)$ denotes the order of pole/zero of $f$ at $\tau=\tau_0$. Let us now look at the valence formula (see, for example, section 3.1 of \cite{kilford}) which states that,
\begin{align}
\nu_{\iota\infty}(f) + \frac{1}{2}\nu_\iota(f) + \frac{1}{3}\nu_\omega(f) + \sum^{'}_{\tau\neq \iota,\omega; \ \tau\in\mathcal{F}} \nu_\tau(f) = \frac{k}{12}, \label{VF}   
\end{align}
where $\omega$ is one of the cube root of unity, $\mathcal{F}$ is the Fundamental Domain and the $'$ above the summation means that the sum excludes $\tau\in\mathbb{H}$ which have $\text{Re}(\tau)=\frac{1}{2}$ or, which have both $|\tau|=1$ and $\text{Re}(\tau)>0$. Now, at $\tau\rightarrow i \infty$, $\chi_i\sim q^{\alpha_i}$ and hence, $W\sim q^{\sum_{i=0}^{n-1}\alpha_i}$. Hence, $W$ has a $-\sum_{i=0}^{n-1}\alpha_i$ order pole at $i \infty$. The last three terms on the left hand side of \eqref{VF} together can be written as  $\frac{\mathbf{\ell}}{6}$ where $\mathbf{\ell}$ is a non-negative integer. Hence, the valence formula now reads
\begin{align}
\sum_{i=1}^{n}\alpha_i + \frac{\mathbf{\ell}}{6} = \frac{n(n-1)}{12}  . \label{RR}    
\end{align}
whose alternate form in terms of central charge and conformal dimensions is 
\be \label{windex}
\frac{\mathbf{\ell}}{6} = \frac{n(n - 1)}{12} + \frac{n\,c}{24} - \sum_i h_i.
\ee
$\mathbf{\ell}$, which is some information about the zeroes of the Wronskian (at the cusp points and the interior of the fundamental domain), is the Wronskian index of the RCFT. This Wronskian index $\mathbf{\ell}$, which is expressed purely in terms of the RCFT data (as seen in \eqref{windex}), proves to be important for the classification of RCFTs. The pair of integers $\mathbf{[n, \ell]}$ identify a RCFT within the MMS classification program. Given a $\mathbf{[n, \ell]}$, one can construct a MLDE using \eqref{phir}, upto a few constants (the parameters of the MLDE).

\section{$\mathbf{[3,3]}$ Admissible Characters \label{3s}}
In this section, we study the $\mathbf{[3,3]}$ MLDE and it's admissible character solutions. In the first sub-section \ref{31ss}, we set up the $\mathbf{[3,3]}$ MLDE and outline a step-by-step procedure to obtain their admissible character solutions. In the subsequent subsection \ref{32ss} we implement the procedure and obtain $\mathbf{[3,3]}$ admissible characters.

\subsection{The $\mathbf{[3,3]}$ MLDE  and a Solution-generating Procedure \label{31ss}}
A three-character RCFT has three rational numbers associated with it: $c$ the central charge, $h_1$ and $h_2$ (with $h_1 <  h_2$) the conformal dimensions of the non-identity characters. The three characters are denoted by $\chi_0, \chi_1, \chi_2$ where $\chi_0$ is the identity character, $\chi_1$ and $\chi_2$ are the non-identity characters. The characters of every $\mathbf{[3,3]}$ RCFT are solutions to the $\mathbf{[3,3]}$ MLDE : 
\bea \label{16p}
\mathcal{D}^3 \chi_i  + \mu_{2,-1}~\frac{E_4^2}{E_6} ~ \mathcal{D}^2\chi_i  + \mu_{1,0} ~ E_4 ~ \mathcal{D} \chi_i + \mu_{3,-1}~\frac{E_4^3}{E_6} ~ \chi_i + \mu ~ \frac{\Delta}{E_6} ~ \chi_i =  0, 
\eea
with $i = 0,1,2$. As it stands, there are four parameters in the MLDE  viz. $\mu_{2,-1}$, $\mu_{1,0}$, $\mu_{3,-1}$, $\mu$. The characters have a $q$-expansion,
\be \label{17p}
\chi_i(q)= q^{\alpha_i}\sum\limits_{n=0}^{\infty} f^{(i)}_n q^n
\ee
where the leading behaviour are the indices/characteristic exponents of the CFT :
\be \label{18p}
\alpha_0 = - \frac{c}{24}, \quad \alpha_1 = h_1 - \frac{c}{24}, \quad \alpha_2 = h_2 - \frac{c}{24}.
\ee
We will employ the Frobenius-like method to solve the linear differential equation in \eqref{16p}. In this method, one determines the ratios $\frac{f^{(i)}_n}{f^{(i)}_0}$. For the identity character one imposes $f^{(0)}_0 = 1$ and we will write $f^{(0)}_n = \text{m}_n$ for $n > 0$ and further impose that the $\text{m}_n$’s are all non-negative integers. For the non-identity characters, the $f^{(i)}_0$'s  can be any non-zero positive integer. In our search for character-like solutions we will require that the ratios $\frac{f^{(i)}_n}{f^{(i)}_0}$ for $i = 1, 2$ need not be non-negative integers but only non-negative rational numbers such that upon multiplying by a suitable positive integer all of them become non-negative integers. This step depends on the order $n$ up to which one solves the differential equation. We will denote this positive integer by $D_1$ for the $\chi_1$ character and by $D_2$ for the $\chi_2$ character.  $f^{(1)}_0$ and $f^{(2)}_0$ cannot be determined per se; they can be any multiple of $D_1$ and $D_2$ respectively. 

At the leading order in the $q$-expansion, the differential equation \eqref{16p} gives the following cubic equation
\begin{equation} \label{19p}
    1728 \, \alpha ^3 + 144 \, \alpha ^2 \, \left(\mu _{2,-1}-6\right) + 12 \,\alpha \, \left(\mu _{1,0} - 2 \mu _{2,-1} + 8 \right)+\mu _{3,-1}=0
\end{equation}
which is the indicial equation whose solutions are the indices \eqref{18p}. Now,  from the valence formula, we have the following constraint for the indices $ \sum_{r=0}^{n-1}\alpha_r=\frac{n(n-1)}{12}-\frac{l}{6}$ which for $l = 3$ gives $\sum_{r=0}^{2}\alpha_r=0$. Imposing this constraint in \eqref{19p} sets  $\mu_{2,-1} =6$ and the indicial equation is 
\begin{equation} \label{20p}
    1728 \alpha ^3+12 \alpha  \left(\mu _{\text{1,0}}-4\right)+\mu _{\text{3,-1}}=0.
\end{equation}
More importantly, the number of parameters in the MLDE \eqref{16p} has reduced to three viz. $\mu_{1,0}$, $\mu_{3,-1}$, $\mu$. Note that only two of them are present in the indicial equation \eqref{20p}. This means that these two parameters are completely determined by the indices. Such parameters are termed ``rigid'' parameters. The other kind of parameters, the ones which are completely independent of the indices are termed ``non-rigid'' parameters. Here, for us in \eqref{18p}, we have that $\mu_{1,0}$ and $\mu_{3,-1}$ are rigid parameters and $\mu$ is a non-rigid parameter. Typically this kind of a clean separation of parameters into rigid and non-rigid ones may not be possible. Here, in \eqref{18p}, we could achieve this separation because of the way the co-efficient of the no-derivative term was written : the numerator of this coefficient which needs to be a weight-$12$ modular form was written as $\mu_{3,-1}\, E_4^3 + \mu\, \Delta$ instead of say $a\,E_4^3 + b\,E_6^2$. In the latter representation, both $a$ and $b$ would appear in the indicial equation and neither can be called a non-rigid parameter. In any case, we are left to have to solve the following three-parameter MLDE to obtain $\mathbf{[3,3]}$ CFTs :
\bea \label{21p}
\mathcal{D}^3 \chi_i  + 6~\frac{E_4^2}{E_6} ~ \mathcal{D}^2\chi_i  + \mu_{1,0} ~ E_4 ~ \mathcal{D} \chi_i + \mu_{3,-1}~\frac{E_4^3}{E_6} ~ \chi_i + \mu ~ \frac{\Delta}{E_6} ~ \chi_i =  0. 
\eea

At the next to leading order in the $q$-expansion, the differential equation \eqref{21p} gives the following equation :
\begin{equation}\label{22p}
    \frac{f_1^{(i)}}{f_0^{(i)}}=\frac{870912 \alpha_i ^3-870912 \alpha_i ^2+\alpha_i  \left(3168 \mu _{\text{1,0}}+111744\right)-720 \mu _{\text{3,-1}}-\mu}{1728 \alpha_i ^3+5184 \alpha_i ^2 + \alpha_i  \left(12 \mu _{\text{1,0}}+5136\right)+12 \mu _{\text{1,0}}+\mu _{\text{3,-1}}+1680}.
\end{equation}

At the next to the next to the leading order in the $q$-expansion, the differential equation \eqref{21p} gives the following equation :

\begin{footnotesize}
\begin{align} \label{23p}
    &\frac{f_2^{(i)}}{f_0^{(i)}}=\Big(808150597632 \alpha_i ^6+808150597632 \alpha_i ^5+\alpha_i ^4 \left(8671297536 \mu _{\text{1,0}}-697071992832\right)+\alpha_i ^3 (-1700352 \mu+13598171136 \mu _{\text{1,0}}\notag\\&-1535169024 \mu _{\text{3,-1}}-759347675136)+\alpha_i ^2 \left(29538432 \mu _{\text{1,0}}^2-746496 \mu+8484258816 \mu _{\text{1,0}}-1613924352 \mu _{\text{3,-1}}-33119705088\right)\notag\\&+\alpha_i  (-6048 \mu \mu _{\text{1,0}}-5088096 \mu _{\text{1,0}} \mu _{\text{3,-1}}+29538432 \mu _{\text{1,0}}^2-971136 \mu+3557385216 \mu _{\text{1,0}}-1698827904 \mu _{\text{3,-1}}+29155977216)\notag\\&-2880 \mu \mu _{\text{1,0}}+1464 \mu \mu _{\text{3,-1}}+\mu^2-4432320 \mu _{\text{1,0}} \mu _{\text{3,-1}}+339120 \mu _{\text{3,-1}}^2-381646080 \mu _{\text{3,-1}}-71424 \mu \Big)\notag\\&\times\Big( 2985984 \alpha_i ^6+26873856 \alpha_i ^5+\alpha_i ^4 \left(41472 \mu _{\text{1,0}}+98371584\right)+\alpha_i ^3 \left(248832 \mu _{\text{1,0}}+3456 \mu _{\text{3,-1}}+187121664\right)\notag\\&+\alpha_i ^2 \left(144 \mu _{\text{1,0}}^2+558720 \mu _{\text{1,0}}+15552 \mu _{\text{3,-1}}+194837760\right)+\alpha_i  \left(24 \mu _{\text{1,0}} \mu _{\text{3,-1}}+432 \mu _{\text{1,0}}^2+556416 \mu _{\text{1,0}}+25824 \mu _{\text{3,-1}}+105262848\right)\notag\\&+36 \mu _{\text{1,0}} \mu _{\text{3,-1}}+288 \mu _{\text{1,0}}^2+\mu _{\text{3,-1}}^2+205056 \mu _{\text{1,0}}+15408 \mu _{\text{3,-1}}+23063040\Big)^{-1}
\end{align}
\end{footnotesize}

We now evaluate the equations \eqref{20p}, \eqref{22p} and \eqref{23p} for the identity character. This means we replace $\alpha_i$ in these three equations by $\alpha_0$ and we replace the left hand sides of the latter two equations by $\text{m}_1$ and $\text{m}_2$ respectively. Then we interpret these three equations as the ones to be solved simultaneously for the three parameters in terms of $\alpha_0$, $\text{m}_1$ and $\text{m}_2$. The solution is as follows : 

\begin{footnotesize}
\begin{align}
    &\mu_{1,0}=\frac{1}{338328 \alpha_0+\text{m}_1^2+240 \text{m}_1-2 \text{m}_2}\Big(-32441472 \alpha _0^3+\alpha_0^2 \big(-432 \text{m}_1^2-207360 \text{m}_1+864 \text{m}_2+6531840\big)\notag\\&+\alpha_0 \big(-432 \text{m}_1^2-62208 \text{m}_1+1728 \text{m}_2-264096\big)-140 \text{m}_1^2-5952 \text{m}_1+1144 \text{m}_2\Big)
\label{24p}
\end{align}
\end{footnotesize}

\begin{footnotesize}
\begin{align}
    &\mu _{\text{3,-1}} = \frac{1}{338328 \alpha _0+\text{m}_1^2+240 \text{m}_1-2 \text{m}_2}\Big(-195333120 \alpha _0^4+\alpha _0^3 \big(3456 \text{m}_1^2+2073600 \text{m}_1-6912 \text{m}_2-78382080\big)\notag\\&+\alpha _0^2 \big(5184 \text{m}_1^2+746496 \text{m}_1-20736 \text{m}_2+19408896\big)+\alpha _0 \big(1728 \text{m}_1^2+82944 \text{m}_1-13824 \text{m}_2\big)\Big)
\label{25p}
\end{align}
\end{footnotesize}

\begin{footnotesize}
\begin{align}
    &\mu = \frac{1}{338328 \alpha _0+\text{m}_1^2+240 \text{m}_1-2 \text{m}_2}\Big(332519178240 \alpha _0^4+\alpha _0^3 \big(-2985984 \text{m}_1^2-3305484288 \text{m}_1+5971968 \text{m}_2-217525948416\big)\notag\\&+\alpha _0^2 \big(-4727808 \text{m}_1^2-2775845376 \text{m}_1+22146048 \text{m}_2+22995062784\big)+\alpha _0 \big(-2073600 \text{m}_1^2-10368 \text{m}_2 \text{m}_1-616978944 \text{m}_1\notag\\&+13353984 \text{m}_2\big)-331776 \text{m}_1^2-10368 \text{m}_1 \text{m}_2\Big)
\label{26p}
\end{align}
\end{footnotesize}

We have thus determined the three parameters of the MLDE in terms of the objects associated with the identity character viz. the index $\alpha_0$, the first Fourier coefficient $\text{m}_1$ and second Fourier coefficient $\text{m}_2$. We now consider the differential equation \eqref{21p} at the next order in the $q$-expansion, which when evaluated for the identity character, is an equation for $\text{m}_3$. After using \eqref{24p}, \eqref{25p} and \eqref{26p}, we obtain the following polynomial equation in the variables $\alpha_0$, $\text{m}_1$, $\text{m}_2$ and $\text{m}_3$. 

\begin{tiny}
\begin{align} \label{27p}
    &\alpha _0^4
    +\frac{\alpha _0^3}{427622160}\Big(-457 \text{m}_1^3-289940 \text{m}_1^2+1371 \text{m}_2 \text{m}_1-23696162 \text{m}_1+579880 \text{m}_2-1371 \text{m}_3+527592240 \Big)\notag\\&
    +\frac{\alpha _0^2 }{10262931840}\Big(-19437 \text{m}_1^3+10 \text{m}_2 \text{m}_1^2-5995448 \text{m}_1^2+81795 \text{m}_2 \text{m}_1+30 \text{m}_3 \text{m}_1-40 \text{m}_2^2-121345812 \text{m}_1+22936256 \text{m}_2-128763 \text{m}_3-2759184576\Big)\notag\\&
    +\frac{\alpha _0}{123155182080} \Big(-126156 \text{m}_1^3+432 \text{m}_2 \text{m}_1^2-3 \text{m}_3 \text{m}_1^2+\text{m}_2^2 \text{m}_1-17597216 \text{m}_1^2+901800 \text{m}_2 \text{m}_1-864 \text{m}_3 \text{m}_1-288 \text{m}_2^2+3 \text{m}_2 \text{m}_3-811350960 \text{m}_1\notag\\&+149967232 \text{m}_2-1516860 \text{m}_3\Big)\notag\\&
    +\frac{1}{123155182080}\Big(-24528 \text{m}_1^3+280 \text{m}_2 \text{m}_1^2-4 \text{m}_3 \text{m}_1^2+\text{m}_2^2 \text{m}_1-1360512 \text{m}_1^2+265380 \text{m}_2 \text{m}_1-1056 \text{m}_3 \text{m}_1-16 \text{m}_2^2+5 \text{m}_2 \text{m}_3\Big)=0
\end{align}
\end{tiny}
We observe that if we define a new variable 
\be \label{28p}
N = -  427622160 \, \alpha_0   
\ee
the polynomial equation \eqref{27p} becomes the following polynomial equation with indeterminates $N$, m$_1$, m$_2$ and m$_3$ : 
\begin{equation} \label{29p}
    \text{N}^4+p_1(\text{m}_1,\text{m}_2,\text{m}_3)\text{N}^3+p_2(\text{m}_1,\text{m}_2,\text{m}_3)\text{N}^2+p_3(\text{m}_1,\text{m}_2,\text{m}_3)\text{N}+p_4(\text{m}_1,\text{m}_2,\text{m}_3)=0
\end{equation}
where the $p_i(\text{m}_1,\text{m}_2,\text{m}_3)$'s are the following polynomials of the first three Fourier coefficients :
\begin{footnotesize}
\begin{align}
    p_1(\text{m}_1,\text{m}_2,\text{m}_3)=&\quad 457 \text{m}_1^3+289940 \text{m}_1^2-1371 \text{m}_2 \text{m}_1+23696162 \text{m}_1-579880 \text{m}_2+1371 \text{m}_3-527592240\nonumber\\
    p_2(\text{m}_1,\text{m}_2,\text{m}_3)=&-346320496830 \text{m}_1^3+178175900 \text{m}_2 \text{m}_1^2-106824434330320 \text{m}_1^2+1457389774050 \text{m}_2 \text{m}_1\notag\\&+534527700 \text{m}_3 \text{m}_1-712703600 \text{m}_2^2-2162089926433080 \text{m}_1+408668805543040 \text{m}_2\notag\\&-2294246341170 \text{m}_3-49162019509491840 \nonumber\\
    p_3(\text{m}_1,\text{m}_2,\text{m}_3)=&\quad 80100610931024527200 \text{m}_1^3-274291067584598400 \text{m}_2 \text{m}_1^2+1904799080448600 \text{m}_3 \text{m}_1^2\notag\\&-634933026816200 \text{m}_2^2 \text{m}_1+11173053618418463699200 \text{m}_1^2-572582603582849160000 \text{m}_2 \text{m}_1\notag\\&+548582135169196800 \text{m}_3 \text{m}_1+182860711723065600 \text{m}_2^2-1904799080448600 \text{m}_2 \text{m}_3\notag\\&+515153520843029613552000 \text{m}_1-95219148537007286758400 \text{m}_2+963104511056421132000 \text{m}_3 \nonumber\\
    p_4(\text{m}_1,\text{m}_2,\text{m}_3)=&-6659632413477502969579776000 \text{m}_1^3+76023201067094782757760000 \text{m}_2 \text{m}_1^2\notag\\&-1086045729529925467968000 \text{m}_3 \text{m}_1^2+271511432382481366992000 \text{m}_2^2 \text{m}_1\notag\\&-369394561893554489569019904000 \text{m}_1^2+72053703925662905172336960000 \text{m}_2 \text{m}_1\notag\\&-286716072595900323543552000 \text{m}_3 \text{m}_1-4344182918119701871872000 \text{m}_2^2\notag\\&+1357557161912406834960000 \text{m}_2 \text{m}_3 \label{30p}
\end{align}
\end{footnotesize}

Now let us consider \eqref{29p} as a polynomial equation to determine $N$, which at this stage is a rational number. Given that the coefficient of $N^4$ is $1$ (which is the reason we made the definition \eqref{28p}) and that all other coefficients are integers, we can use the integer-root-theorem to conclude that $N$ has to be an integer. This is an important stage of our analysis. $N$ which was just rational is now restricted to be an integer. This means that the central charge, given in terms of $N$ by
\be \label{31p}
    c=\frac{\text{N}}{17817590},
\ee
is a rational number, whose denominator (not necessarily in the $\frac{p}{q}$ form with $p$, $q$ coprime) is always $17817590$. Our problem has become a ``finite'' problem in the following sense.  If one is searching for CFTs in a certain range of central charge, say $0 < c \leq 1$, one only needs to do $17817590$ computations for $ c = \frac{1}{17817590}, \frac{2}{17817590}, \ldots \frac{17817589}{17817590}, \frac{17817590}{17817590}$. $17817590$ computations is certainly a lot of computations, but it is less than an infinite number of computations. 

Every computation involves solving two Diophantine equations. One of the Diophantine equations is \eqref{29p}. The other Diophantine equation arises as follows. Via equations \eqref{24p}, \eqref{25p} and \eqref{26p}, we can think of the MLDE parameters as functions of $N$, $\text{m}_1$ and $\text{m}_2$ and subsequently the indicial equation \eqref{20p} as having coefficients which are polynomials of $N$, $\text{m}_1$ and $\text{m}_2$. The solutions are thus now functions of $N$, $\text{m}_1$ and $\text{m}_2$. In these terms, we know one of the solutions  viz. $-\frac{N}{427622160}$, which is nothing but $\alpha_0$ (see \eqref{28p}). Factoring out this solution, the indicial equation \eqref{20p} reduces to a quadratic equation in $\alpha$ with  coefficients which are polynomials of $N$, $\text{m}_1$ and $\text{m}_2$. The roots of this quadratic equation determine the other two indices $\alpha_1$ and $\alpha_2$ and eventually $h_1$ and $h_2$. For rational roots the discriminant has to be the square of a rational number. But since the discriminant is an integer (on account of all coefficients being integers), it needs to be a perfect square to be able to result in rational roots. If we denote the discriminant by $k^2$ with $k$ chosen to be a positive integer, we have 
\begin{equation}
    -122264458903381444848576\text{N}^4+q_1(\text{m}_1,\text{m}_2)\text{N}^3+q_2(\text{m}_1,\text{m}_2)\text{N}^2+q_3(\text{m}_1,\text{m}_2)\text{N}+q_4(\text{m}_1,\text{m}_2)=k^2
    \label{32p}
\end{equation}
where the $q_i(\text{m}_1, \text{m}_2)$ are the following functions of $\text{m}_1$ and $\text{m}_2$ :
\begin{footnotesize}
\begin{align}
    q_1(\text{m}_1,\text{m}_2)=&-16457464055075904 \big(241784696300 \text{m}_1^2+138404188305600 \text{m}_1-483569392600 \text{m}_2-5063679255196800\big) \nonumber \\
    q_2(\text{m}_1,\text{m}_2)=&\quad 5224693733104078274079228422400 \big(\text{m}_1^4+800 \text{m}_1^3-4 \text{m}_2 \text{m}_1^2+565344 \text{m}_1^2-1600 \text{m}_2 \text{m}_1+4 \text{m}_2^2\notag\\&+60120576 \text{m}_1-1764096 \text{m}_2+1688933376\big)\nonumber\\
    q_3(\text{m}_1,\text{m}_2)=&-2978926425984572608494442225493440512000 \big(\text{m}_1^4+384 \text{m}_1^3-6 \text{m}_2 \text{m}_1^2+151080 \text{m}_1^2-1248 \text{m}_2 \text{m}_1\notag\\&+8 \text{m}_2^2+6311808 \text{m}_1-909696 \text{m}_2\big)\nonumber\\
    q_4(\text{m}_1,\text{m}_2)=&\quad 424618317586867688507075910820237365857648640000 \big(\text{m}_1^4+288 \text{m}_1^3-10 \text{m}_2 \text{m}_1^2+11520 \text{m}_1^2-2016 \text{m}_2 \text{m}_1+16 \text{m}_2^2\big).
\end{align}
\end{footnotesize}

The equation \eqref{32p} is a polynomial equation with indeterminates $N$, $\text{m}_1$, $\text{m}_2$ and $k$, all non-negative integers and hence a Diophantine equation. We thus have two Diophantine equations, \eqref{29p} and \eqref{32p}, involving five Diophantine variables viz. $N$, $\text{m}_1$, $\text{m}_2$, $\text{m}_3$ and $k$. $\text{m}_3$ and $k$ are each present in only one of the Diophantine equations while the other three are simultaneously present in both. 

The procedure to obtain character like solutions to the $\mathbf{[3,3]}$ MLDE can now be stated. 

\textbf{Step I}: We solve the two Diophantine equations \eqref{29p} and \eqref{32p} simultaneously.  We fix a value of $N$ and and then obtain solutions to \eqref{29p}. Each such solution is a set of non-negative integers for $N$, $\text{m}_1$, $\text{m}_2$ and $\text{m}_3$. We take the $\text{m}_1$, $\text{m}_2$ and $N$ values of each such solution and see if we can find non-negative integer values for $k$ that satisfy \eqref{32p}. At the end of this procedure, we will have a set of five non-negative integers, for $N$, $\text{m}_1$, $\text{m}_2$, $\text{m}_3$ and $k$ which simultaneously solve \eqref{29p} and \eqref{32p}.

\textbf{Step II}: At this stage, we have the three indices: $\alpha_0$, $\alpha_1$ and $\alpha_2$ ($N$ gives $\alpha_0$ while $k$ the discriminant, together with $\text{m}_1$, $\text{m}_2$ and $\text{m}_3$ gives the other two indices), which further means that we have the $c$, $h_1$ and $h_2$ of the putative CFTs. But we have imposed positivity on the Fourier coefficients of the character only upto order $q^3$ and only for the identity character.  Hence, now in this next stage of the procedure, we impose positivity. That is we determine the Fourier coefficients of all the characters upto some high order (for us $q^{2000}$), which are after all functions of the index of the character and the parameters in the MLDE which in turn are functions of the known $\alpha_0$, $\text{m}_1$ and $\text{m}_2$ (\eqref{24p}, \eqref{25p}, \eqref{26p}) and check if they are positive. We reject solutions to the Diophantine equations which do not survive the positivity constraints. We keep all the other solutions and in the process also compute $D_1$ and $D_2$, the positive integers that when multiplied to $\chi_1$ and $\chi_2$, makes all their Fourier coefficients to be positive integers as well. These are often referred to as  apparent degeneracies. 

\subsection{Solutions to the $\mathbf{[3,3]}$ MLDE\label{32ss}}

We implement the just delineated procedure for  $c < 96$. To search for solutions within a central charge range of $1$, we will need to perform around $17$ million computations and for a central charge range of $96$ that is around $1.7$ billion computations. But we undertake a limited search. The prime factorisation of the denominator in \eqref{31p} is as follows : $17817590 = 2\times 5\times 7\times 254537$. We look for solutions with central charges given by (a) $c=\frac{n}{2}$ with $2\nmid n$ (these are $96$ computations), (b) $c = \frac{n}{5}$ with $5\nmid n$ (these constitute  $384$ computations), (c) $c = \frac{n}{7}$ with $7\nmid n$ (these constitute $576$ computations) and (d) $c = n$ with $n \in \mathbf{Z}_{\geq 0}$ (these are $96$ computations). We thus perform a total of $1152$ computations. 

In a previous paper on the subject, for the $\mathbf{[3,0]}$ MLDE, the authors had obtained the analog of the formula \eqref{31p} which was $c = \frac{N}{70}$. Then, in a search for solutions with $0 < c \leq 96$, one only needs to perform $6720$ computations and all of them were done. Note the prime factorisation of the denominator $70 = 2 \times 5 \times 7$. It was found that all the admissible character solutions  had a central charge, which after having been written in the $\frac{p}{q}$ form with $p$, $q$ coprime,  whose  denominator was either $1$ or $2$ or $5$ or $7$. The denominator always is only one of the prime factors never two or more. This pattern to the central charges of admissible character solutions of the $\mathbf{[3,0]}$ MLDE is corroborated by studies of modular data for three-character RCFTs or modular tensor categories (MTCs). Our own limited search (limiting to $1152$ cases as opposed to a $1.7$ billion cases) here for solutions of $\mathbf{[3,3]}$ MLDE, is motivated similarly.

\begin{table}[!hbtp] 
\begin{center}
\resizebox{\textwidth}{!}{
\renewcommand{\arraystretch}{1.4}
\begin{threeparttable}
\caption{Discrete set of solutions to the Diophantine equations \eqref{29p} and \eqref{32p} with $0 < c \leq 96$}\label{t1}
\begin{tabular}{c||ccccc||ccc||ccc||}
\hline \hline
\rowcolor{Mywhite}
\# & $N$ &  $\text{m}_1$ & $\text{m}_2$ & $\text{m}_3$ &  $k$ & $\mu_{1,0}$ & $\mu_{3,-1}$ & $\mu$ & $c$ & $h_1$ & $h_2$   \\  
\hline \hline
\rowcolor{Mygrey}
S-1. & 89087950& 27& 106& 433 & 10342630691459386216552857600& $-\frac{155}{16}$& $-\frac{595}{32}$&  $-68310$& 5 & $\frac{1}{16}$ & $\frac{9}{16}$  \\ \rowcolor{Mywhite}
S-2. & 106905540& 26& 79& 326 & 12678063428240537942871244800& $-\frac{47}{4}$& $-\frac{81}{4}$&  $-86832$& 6 & $\frac{1}{8}$ & $\frac{5}{8}$  \\ \rowcolor{Mygrey}
S-3. & 122177760& 78& 784& 5271 & 13670984061093056577243033600& $-\frac{116}{7}$& $-\frac{10368}{343}$&  $-\frac{41015808}{343}$& $\frac{48}{7}$ & $\frac{1}{7}$ & $\frac{5}{7}$  \\ \rowcolor{Mywhite}
S-4. & 124723130& 25& 53& 246 & 15013496165021689669189632000& $-\frac{227}{16}$& $-\frac{665}{32}$&  $-111186$& $7$ & $\frac{3}{16}$ & $\frac{11}{16}$  \\ \rowcolor{Mygrey}
S-5. & 142540720& 24& 28& 192 & 17348928901802841395508019200& $-17$& $-20$&  $-141696$& $8$ & $\frac{1}{4}$ & $\frac{3}{4}$  \\ \rowcolor{Mywhite}
S-6. & 142540720& 134& 1920& 15904 & 12999967382473207851576960000& $-\frac{524}{25}$& $-\frac{896}{25}$&  $-\frac{3884544}{25}$& $8$ & $\frac{1}{5}$ & $\frac{4}{5}$  \\ \rowcolor{Mygrey}
S-7. & 162903680& 136& 2417& 24520 & 15375588244086679187252889600& $-\frac{164}{7}$& $-\frac{10240}{343}$&  $-\frac{65622528}{343}$& $\frac{64}{7}$ & $\frac{2}{7}$ & $\frac{6}{7}$  \\ \rowcolor{Mywhite}
S-8. & 264718480& 188& 17260& 442300 & 22829833773044248767796224000& $-44$& $\frac{18304}{343}$&  $-\frac{188103168}{343}$& $ \frac{104}{7}$ & $\frac{5}{7}$ & $\frac{8}{7}$  \\ \rowcolor{Mygrey}
S-9. & 285081440& 232& 31076& 946432 & 22829833773044248767796224000& $-\frac{1244}{25}$& $\frac{2048}{25}$&  $-\frac{16436736}{25}$& $ 16$ & $\frac{4}{5}$ & $\frac{6}{5}$  \\ \rowcolor{Mywhite}
S-10. & 305444400& 156& 28926& 1053508 & 33196534784141788026353433600& $-\frac{404}{7}$& $\frac{34560}{343}$&  $-\frac{264964608}{343}$& $ \frac{120}{7}$ & $\frac{6}{7}$ & $\frac{9}{7}$  \\ \rowcolor{Mygrey}
S-11. & 407259200& 40& 60440& 5474720 & 80410293873630197603223552000& $-\frac{740}{7}$& $\frac{81920}{343}$&  $-\frac{517501440}{343}$& $ \frac{160}{7}$ & $\frac{8}{7}$ & $\frac{12}{7}$  \\ \rowcolor{Mywhite}
S-12. & 427622160& 30& 87786& 9614200 & 98877947399322534305678592000& $-\frac{2924}{25}$& $\frac{6912}{25}$&  $-\frac{42259968}{25}$& $ 24$ & $\frac{6}{5}$ & $\frac{9}{5}$  \\ \rowcolor{Mygrey}
S-13. & 447985120& 14& 66512& 8878186 & 90093424404042464671113523200& $-\frac{884}{7}$& $\frac{119680}{343}$&  $-\frac{655907328}{343}$& $ \frac{176}{7}$ & $\frac{9}{7}$ & $\frac{13}{7}$  \\ \rowcolor{Mywhite}
S-14. & 549799920& 3& 52254& 20440112 & 75231152349388320481013664000& $-\frac{1268}{7}$& $\frac{279936}{343}$&  $-\frac{1127913984}{343}$& $ \frac{216}{7}$ & $\frac{12}{7}$ & $\frac{15}{7}$  \\ \rowcolor{Mygrey}
S-15. & 570162880& 3& 62500& 31015600 & 74986027395393703935850560000& $-\frac{4844}{25}$& $\frac{23296}{25}$&  $-\frac{18095616}{5}$ & $ 32$ & $\frac{9}{5}$ & $\frac{11}{5}$  \\
 \hline \hline
\end{tabular} 
\end{threeparttable}
}
\end{center}
\end{table}

\begin{table}[!hbtp]
\begin{center}
\resizebox{\textwidth}{!}{
\renewcommand{\arraystretch}{1.4}
\begin{threeparttable}
\caption{Infinite families of solutions to the Diophantine equations \eqref{29p} and \eqref{32p} with $0 < c \leq 96$}\label{t2}
\begin{tabular}{c||ccccc||ccc||ccc||}
\hline \hline
\rowcolor{Mywhite}
\# & $N$ &  $\text{m}_1$ & $\text{m}_2$ & $\text{m}_3$ &  $k$ & $\mu_{1,0}$ & $\mu_{3,-1}$ & $\mu$ & $c$ & $h_1$ & $h_2$   \\
\hline \hline
\rowcolor{Mygrey}
S-16. & $213811080$& $22 + n$& $26 + 44 n$& $1652 + 718 n$ & $-162906859429488820195200 (-163452 + 196 n + n^2)$& $-32$& $0$&  $-331776$& $12$ & $\frac{1}{2}$ & $1$  \\ \rowcolor{Mywhite}
S-17. & $285081440$& $497+n$& $69912 + 160 n$& $2120356 + 5348 n$ & $0$& $-44$& $128$&  $-718848$& $16$ & $1$ & $1$  \\ \rowcolor{Mygrey}
S-18. & $356351800$& $13+n$& $354 + 548 n$& $32322 + 31114 n$ & $-162906859429488820195200 (-279359 - 830 n + n^2)$& $-80$& $160$&  $-1105920$& $20$ & $1$ & $\frac32$  \\ \rowcolor{Mywhite}
S-19. & $498892520$& $4$& $2 + 3 n$& $61816 + 668n$ & $977441156576932921171200 (65624 + n)$& $-152$& $560$&  $-2543616$& $28$ & $\frac{3}{2}$ & $2$  \\ \hline \hline
\end{tabular} 
\end{threeparttable}
}
\end{center}
\end{table}

For each of  the $1152$ cases that we listed above, we implemented the solution-generating procedure that we have described in section \ref{31ss}. We first give an account of the (simultaneous) solutions to the Diophantine equations \eqref{29p} and \eqref{32p}. We collect the Diophantine solutions in tables \ref{t1} and \ref{t2}. We found a discrete set of $15$ solutions which are given in table \ref{t1}; we further found four infinite families of solutions which are given in table \ref{t2}.  We will number the discrete set of solutions S-1 - S-15 and the four infinite families by S-16 - S-19. Note that in each  infinite family, the variable $n \geq 0$. For each solution, we first give the five Diophantine variables $N$, $\text{m}_1$, $\text{m}_2$, $\text{m}_3$ and $k$.  We also give the values of the MLDE parameters or in other words the point in the space of MLDE parameters at which this solution exists. The values for the MLDE parameters are obtained from the Diophantine variables from the formulae in \eqref{24p}, \eqref{25p}, \eqref{26p}. The last detail we include in the tables is the central charge $c$ and the conformal dimensions of the characters $h_1$ and $h_2$; the $c$ follows from $N$ and the conformal dimensions require further the value of $k$, $\text{m}_1$ and $\text{m}_2$. The $19$ solutions presented here are not all the solutions to the Diophantine equations \eqref{29p} and \eqref{32p}. They are those solutions which on computation of higher Fourier coefficients (up to $q^{2000}$) result in admissible characters. There are other solutions to the Diophantine equations which on computation of higher Fourier coefficients result not in admissible characters but in quasi-characters (in the sense of \cite{Chandra:2018pjq}) which we do not report here.

\begin{table}[!hbtp]
\begin{center}
\resizebox{0.8\textwidth}{!}{
\renewcommand{\arraystretch}{1.2}
\begin{threeparttable}
\caption{Details of the individual characters of the discrete set of $\mathbf{[3,3]}$ admissible character solutions}\label{t3}
\begin{tabular}{c||ccc||ccc||cc||}
\hline \hline
\rowcolor{Mywhite}
\# & $c$ &  $h_1$ & $h_2$ & $\text{m}_1$ &  $\text{m}_2$ & $\text{m}_3$ & $D_1$ & $D_2$    \\
\hline \hline
\rowcolor{Mygrey}
S-1. & 5 & $\frac{1}{16}$ & $\frac{9}{16}$ & 27 & 106 & 433  & $1$ & $25$    \\ \rowcolor{Mywhite}
S-2. & 6 & $\frac{1}{8}$ & $\frac{5}{8}$ & 26 & 79 & 326  & $1$ & $13$    \\ \rowcolor{Mygrey}
S-3. & $\frac{48}{7}$ & $\frac{1}{7}$ & $\frac{5}{7}$ & 78 & 784 & 5271  & $1$ & $55$    \\ \rowcolor{Mywhite}
S-4. & $7$ & $\frac{3}{16}$ & $\frac{11}{16}$ & 25 & 53 & 246  & $1$ & $27$    \\ \rowcolor{Mygrey}
S-5. & $8$ & $\frac{1}{4}$ & $\frac{3}{4}$ & 24 & 28 & 192  & $1$ & $7$    \\ \rowcolor{Mywhite}
S-6. & $8$ & $\frac{1}{5}$ & $\frac{4}{5}$ & 134 & 1920 & 15904  & $1$ & $57$    \\ \rowcolor{Mygrey}
S-7. & $\frac{64}{7}$ & $\frac{2}{7}$ & $\frac{6}{7}$ & 136 & 2417 & 24520 & $3$ & $117$    \\ \rowcolor{Mywhite}
S-8. & $\frac{104}{7}$ & $\frac{5}{7}$ & $\frac{8}{7}$ & 188 & 17260 & 442300  & $44$ & $725$    \\ \rowcolor{Mygrey}
S-9. & $16$ & $\frac{4}{5}$ & $\frac{6}{5}$ & 232 & 31076 & 946432  & $9$ & $154$    \\ \rowcolor{Mywhite}
S-10. & $\frac{120}{7}$ & $\frac{6}{7}$ & $\frac{9}{7}$ & 156 & 28926 & 1053508  & $78$ & $2108$    \\ \rowcolor{Mygrey}
S-11. & $\frac{160}{7}$ & $\frac{8}{7}$ & $\frac{12}{7}$ & 40 & 60440 & 5474720   & $285$ & $27170$    \\ \rowcolor{Mywhite}
S-12. & $24$ & $\frac{6}{5}$ & $\frac{9}{5}$ & 30 & 87786 & 9614200 & $11$ & $1102$    \\ \rowcolor{Mygrey}
S-13. & $\frac{176}{7}$ & $\frac{9}{7}$ & $\frac{13}{7}$ & 14 & 66512 & 8878186   & $782$ & $50922$    \\ \rowcolor{Mywhite}
S-14. & $\frac{216}{7}$ & $\frac{12}{7}$ & $\frac{15}{7}$ & 3 & 52254 &  20440112 & $11495$ & $260623$    \\ \rowcolor{Mygrey}
S-15. & $32$ & $\frac{9}{5}$ & $\frac{11}{5}$ & 3 & 62500 & 31015600    & $19$ & $434$    \\ \rowcolor{Mywhite}
 \hline \hline
\end{tabular}
\end{threeparttable}
}
\end{center}
\end{table}

\begin{table}[!hbtp]
\begin{center}
\resizebox{0.8\textwidth}{!}{
\renewcommand{\arraystretch}{1.3}
\begin{threeparttable}
\caption{Details of the individual characters of the infinite families of $\mathbf{[3,3]}$ admissible character solutions}\label{t4}
\begin{tabular}{c||ccc||ccc||cc||}
\hline \hline
\rowcolor{Mywhite}
\# & $c$ &  $h_1$ & $h_2$ & $\text{m}_1$ &  $\text{m}_2$ & $\text{m}_3$ & $D_1$ & $D_2$    \\
\hline \hline
\rowcolor{Mygrey}
S-16. & $12$ & $\frac{1}{2}$ & $1$ & $22+n$ & $26+44 n$ & $1652 + 718 n$  & $1$ & $1$    \\
\rowcolor{Mywhite}
S-17. & $16$ & $1$ & $1$ & $497+n$ & $69912 + 160 n$ & $2120356 + 5348 n$  & $1$ & $1$    \\
\rowcolor{Mygrey}
S-18. & $20$ & $1$ & $\frac32$ & $13+n$ & $354 + 548 n$ & $32322 + 31114 n$  & $1$ & $5$    \\
\rowcolor{Mywhite}
S-19. & $28$ & $\frac{3}{2}$ & $2$ & $4$ & $2 + 3 n$ & $61816 + 668 n$  & $1$ & $3$    \\
\hline \hline
\end{tabular}
\end{threeparttable}
}
\end{center}
\end{table}

After having described the Diophantine aspects of the admissible character solutions, we give further details of the solutions in the tables \ref{t3} and \ref{t4}. These tables collect details about the individual characters. We first give the triple of central charge and conformal dimensions of the two non-identity characters. Then we give the first four Fourier coefficients of the identity character viz. $\text{m}_1$, $\text{m}_2$, $\text{m}_3$ (with $\text{m}_0 =1$). After that we give the leading Fourier coefficient of the two non-identity characters $D_1$ and $D_2$, the apparent degeneracies. It is important to note that the leading Fourier coefficient of the non-identity characters could be any positive integral multiple of what we have given because even then the MLDE would be solved. We have given the minimal values for $D_1$ and $D_2$. We have maintained a consistent numbering of the solutions between tables \ref{t1}, \ref{t2} and tables \ref{t3}, \ref{t4}.

Finally, we give further details of the admissible character solutions in appendix \ref{app1}. There we collect all possible details in one place : (i) the triple of central charge and conformal dimensions of the non-identity characters, (ii) the triple of indices, (iii) the solution to the Diophantine equations, (iv) the minimal values for $D_1$ and $D_2$ pertaining to the non-identity characters, (v) the point in the space of parameters of the MLDE where the solution exists, (vi) the $q$-series expansions of the three characters upto order $q^6$ and (vii) the Kaidi-Lin-Parra-Martinez exponent triplet class (see \ref{33ss} below) that the solution belongs to. Our numbering of the solutions in appendix \ref{app1} are consistent with the tables \ref{t1} - \ref{t4}.

\subsection{\label{33ss}Notes and Comments on $\mathbf{[3,3]}$ Admissible Characters including the Relation with Kaidi-Lin-Parra-Martinez Theory}

Here first we we make notes about the admissible character solutions and provide some comments, in the first sub-section \ref{331ss}. Then we introduce the elements of Kaidi-Lin-Parra-Martinez (KLP) theory \cite{Kaidi:2021ent}, the parts relevant for three characters and Wronskian index equalling $3$, in the second sub-section \ref{332ss} and discuss the connection to  the solutions we have obtained here in tables \ref{t1}, \ref{t2}, \ref{t3}, \ref{t4} and appendix \ref{app1},

\subsubsection{Notes on $\mathbf{[3,3]}$ Admissible Characters \label{331ss}}

\textbf{(A)} From table \ref{t1}, we see that the fifteen solutions in the  discrete set exist at different points in the space of MLDE parameters ($\mu_{1,0}, \mu_{3,-1}, \mu$); there is one solution at each point. And for the infinite families of solutions, we note that (table \ref{t2}), there is one infinite family of solutions at a single point in the space of MLDE parameters. Thus there are four different points in the space of MLDE parameters each of which host an infinite family of admissible character solutions. 
\bigskip

\noindent\textbf{(B)} We note that there are two admissible character solutions with $c = 8$ but at two different points in the space of MLDE parameters (solution S-5 and S-6).
We also note that there are an infinite number of $c = 16$ admissible character solutions that exist at two different points in the space of MLDE parameters. One of them (solution S-9) exists at one of the points and the other point hosts an infinite number of solutions (S-16).

\bigskip

\noindent\textbf{(C)} All the $\mathbf{[3,3]}$ admissible character solutions that we have exist within the range of central charge given by $0 < c \leq 32$. This is an observation that is being made on the solutions that were obtained. At the moment, we do not have an explanation for this. Does the validity of this range continue to be true even after more computations are done (beyond $c = 96$)? Is there a way to derive this independently? We do not have answers to these questions yet. We recall such ranges\footnote{There is an understanding of ranges of central charges of two-character theories coming from the fact that the  $\mathbf{[2,2]}$ CFTs are cosets of $c = 24$ meromorphic theories by $\mathbf{[2,0]}$ CFTs. This explains  the range $16 < c < 24$ for $\mathbf{[2,2]}$ CFTs given that the range of $\mathbf{[2,0]}$ CFTs is $0 < c < 8$. But such an explanation is not there for $\mathbf{[3,3]}$ CFTs because there is no obvious way in which they are the coset of anything. In fact, the simplest way in which $\mathbf{[3,3]}$ CFTs can feature in a coset pair is with themselves. In section \ref{34ss}, we study coset-bilinear relations between pairs of $\mathbf{[3,3]}$ admissible characters.} of central charges for other classes of CFTs: $\mathbf{[2,0]}$ CFTs were limited to $0 < c < 8$, $\mathbf{[2,2]}$ CFTs were limited to $16 < c < 24$. We also recall that this finite range of central charges for $\mathbf{[3,3]}$ admissible characters stands in contrast to $\mathbf{[3,0]}$ admissible characters whose central charges are unbounded (the $B_{r,1}$ and $D_{r,1}$ CFTs provide $\mathbf{[3,0]}$ admissible characters, for all $r$, and their central charges are $r+\frac12$ and $r$ respectively).

\bigskip

\noindent
\textbf{(D)} From tables \ref{t3} and \ref{t4}, we observe that the central charge of a $\mathbf{[3,3]}$ admissible character solution is a rational number, when expressed in the $\frac{a}{b}$ form with $a$ and $b$ being coprime, is such that $b$ is either $1$ or $7$.  In our search, we had looked for the cases $b = 2, 5$ as well but there are no such solutions, for $c < 96$.

\bigskip

\noindent
\textbf{(E)} The admissible character solutions are putative characters of a three-character RCFT.  It is not clear which of them correspond to CFTs and which do not. One clue to answering such questions is the $\text{m}_1$ of the admissible character solution. The chiral algebra of the CFT would contain an affine Lie-algebra as a sub-algebra and the finite Lie-algebra (which need not be simple and can include multiple simple factors) on which the affine Lie-algebra would be based would have a dimension given by $\text{m}_1$. We find that, for the discrete set of solutions (S1-S15) from table \ref{t3},  $\text{m}_1$ takes values ranging from $3$ to $232$. While it is easy to figure out the finite Lie-algebra for small values of $\text{m}_1$  (for example, when $\text{m}_1 = 3$, the only two possibilities are $A_1$ and $U(1)^3$), a more thorough analysis is needed for other values of $\text{m}_1$. We will postpone this for future work. For the infinite families of solutions (S-16 - S-19), we find two interesting situations. In the first situation, all admissible character solutions of a family have the same $\text{m}_1$ value; this happens for solution S-19. The second situation occurs for solutions S-16, S-17, S-18 where the $\text{m}_1$ value depends on the $n$ that labels that solution in the infinite family and  thus can be arbitrarily large. It would be interesting to see if these admissible characters pertain to CFTs and what their associated finite Lie-algebras would be. 

\bigskip

\noindent
\textbf{(F)} Here we note some interesting aspects of the four infinite families of solutions (table \ref{t4}) which also explain their existence. Let us consider the example of solution S-16, which is an infinite family of solutions labelled by $n$ which is a non-negative integer.  Consider the $q$-series expansions of the identity character $\chi_0$ for the $n=0$ member of the infinite family and the non-identity character with conformal dimension $1$ viz. $\chi_1$. They are similar in that both of them are a series of half-integer powers of $q$. Hence they can be added to result in a similar series of only half-integral powers. One says that the two $q$-series are commensurate. We can produce a new solution from the $n = 0$ member of the family as follows. Add the identity character to a positive integral multiple of $\chi_1$; this sum is also a solution to the the MLDE and declare this to  be the identity character of a new solution. The other two characters of the new solution are simply the same as the other two characters of the $n = 0$ member of the family. We thus get an infinite number of admissible character solutions labelled by the positive integer multiple which is nothing but $n$. We can verify the admissible character nature of the new solution by examining the explicit form of the $q$-series, which is available in the appendix \ref{app1}. All our infinite family admissible character solutions follow this story. 

For the solution in S-17 the identity character of the $n = 0$ member of the family and non-identity character with conformal dimension $1$ have commensurate $q$-series : the powers of $q$ are $\frac13 ~\text{mod} ~1$ and hence can be added with an arbitrary positive integer multiplying the non-identity character (see the explicit $q$-series given in appendix \ref{app1}). For the solution in S-18 the identity character of the $n = 0$ member of the family and non-identity character with conformal dimension $1$ have commensurate $q$-series : the powers of $q$ are $\frac16 ~\text{mod} ~1$ and hence can be added with an arbitrary positive integer multiplying the non-identity character (see the explicit $q$-series given in appendix \ref{app1}). In the last case, for the solution in S-19 the identity character of the $n = 0$ member of the family and non-identity character with conformal dimension $2$ have commensurate $q$-series : the powers of $q$ are $\frac56 ~\text{mod} ~1$ and hence can be added with an arbitrary positive integer multiplying the non-identity character (see the explicit $q$-series given in appendix \ref{app1}).

\bigskip

\noindent
\textbf{(G)} Solution S-17 is not the typical solution. It consists of an identity character and the two non-identity characters are the same character repeated twice. We do not expect this admissible character solution to correspond to a three-character RCFT. We include this curiosity here for completeness. 

\subsubsection{Kaidi-Lin-Parra-Martinez Theory of $\mathbf{[3,3]}$ Admissible Characters \label{332ss}}

In the paper \cite{Kaidi:2021ent}, the authors Kaidi, Lin and Parra-Martinez, study admissible characters for CFTs with small number of characters (up to $5$) and with a small Wronskian index (up to the number of characters). They view the characters as constituting a vector valued modular form, for which the modular data is an important part of the definition. In their paper, they classify vector valued modular forms by classifying modular data. We will only consider that part of their paper which is relevant in this section viz. $\mathbf{[3, 3]}$ vector valued modular forms.  We will direct the reader to table $12$ of their paper \cite{Kaidi:2021ent}, where they give their results. Their result  consists of a list of triples of non-negative rational numbers (less than $1$), which we will refer to as ``exponent-triplets.” We will see below how they are related to the set of indices, also referred to as index-set, of characters of a  $\mathbf{[3, 3]}$ admissible character solution (a given  exponent-triplet results in an infinite number of index-sets). We have reproduced the $34$ exponent-triplets of \cite{Kaidi:2021ent} for $\mathbf{[3, 3]}$, here in table \ref{t5}.

\begin{table}[!hbtp] \label{t5}
\begin{center}
\resizebox{\textwidth}{!}{
\renewcommand{\arraystretch}{1.3}
\begin{threeparttable}
\caption{The $34$ Kaidi-Lin-Parra-Martinez exponent-triplets for $\mathbf{[3,3]}$ CFTs, from \cite{Kaidi:2021ent}}\label{t5}
\begin{tabular}{c|c||c|c||c|c||c|c||}
\hline \hline
\rowcolor{Mywhite}
\# & Exponent Triplets &  $\#$ & Exponent Triplets &  $\#$ & Exponent Triplets &  $\#$ & Exponent Triplets\\
\hline \hline
\rowcolor{Mygrey}
   1  & $\left\{0, \frac{1}{3}, \frac{2}{3}\right\}$ &2. & $\left\{0, \frac{1}{4}, \frac{3}{4}\right\}$ &3. & $\left\{0, \frac{1}{5}, \frac{4}{5}\right\}$ &4. & $\left\{0, \frac{2}{5}, \frac{3}{5}\right\}$\\
\rowcolor{Mywhite}
   5. & $\left\{\frac{1}{7}, \frac{2}{7}, \frac{4}{7}\right\}$ &6. & $\left\{\frac{3}{7}, \frac{5}{7}, \frac{6}{7}\right\}$ &7. & $\left\{\frac{1}{8}, \frac{1}{4}, \frac{5}{8}\right\}$ &8. & $\left\{\frac{3}{8}, \frac{3}{4}, \frac{7}{8}\right\}$\\
\rowcolor{Mygrey}
   9. & $\left\{\frac{1}{12}, \frac{1}{ 3}, \frac{7}{12}\right\}$ &10. & $\left\{\frac{5}{12}, \frac{2}{3}, \frac{11}{12}\right\}$ &11. & $\left\{\frac{2}{15}, \frac{1}{3}, \frac{8}{15}\right\}$ &12. & $\left\{\frac{1}{3}, \frac{11}{15}, \frac{14}{ 15}\right\}$\\
\rowcolor{Mywhite}
   13. & $\left\{\frac{7}{15}, \frac{2}{3}, \frac{13}{15}\right\}$ &14. & $\left\{\frac{1}{15}, \frac{4}{15}, \frac{2}{3}\right\}$ &15. & $\left\{\frac{1}{16}, \frac{3}{8}, \frac{9}{16}\right\}$ &16. & $\left\{\frac{1}{8}, \frac{3}{16}, \frac{11}{16}\right\}$\\
\rowcolor{Mygrey}
   17. &  $\left\{\frac{5}{16}, \frac{13}{16}, \frac{7}{8}\right\}$ & 18. & $\left\{\frac{7}{16}, \frac{5}{8}, \frac{15}{16}\right\}$ & 19. & $\left\{\frac{1}{21}, \frac{4}{21}, \frac{16}{21}\right\}$ & 20. & $\left\{\frac{2}{21}, \frac{8}{21}, \frac{11}{21}\right\}$\\
\rowcolor{Mywhite}
   21. & $\left\{\frac{5}{21}, \frac{17}{21}, \frac{20}{21}\right\}$ & 22. & $\left\{\frac{10}{21}, \frac{13}{21}, \frac{19}{21}\right\}$ & 23. & $\left\{\frac{1}{24}, \frac{5}{12}, \frac{13}{24}\right\}$ & 24. & $\left\{\frac{1}{12}, \frac{5}{24}, \frac{17}{24}\right\}$\\
\rowcolor{Mygrey}
   25. & $\left\{\frac{7}{24}, \frac{19}{24}, \frac{11}{12}\right\}$ & 26. & $\left\{\frac{11}{24}, \frac{7}{12}, \frac{23}{24}\right\}$ & 27. & $\left\{\frac{1}{48}, \frac{11}{24}, \frac{25}{48}\right\}$ & 28. & $\left\{\frac{1}{24}, \frac{11}{48}, \frac{35}{48}\right\}$\\
\rowcolor{Mywhite}
   29. & $\left\{\frac{19}{48}, \frac{17}{24}, \frac{43}{48}\right\}$ & 30. & $\left\{\frac{23}{48}, \frac{13}{24}, \frac{47}{48}\right\}$ & 31. & $\left\{\frac{7}{48}, \frac{5}{24}, \frac{31}{48}\right\}$ & 32. & $\left\{\frac{17}{48}, \frac{19}{24}, \frac{41}{48}\right\}$\\
\rowcolor{Mygrey}
   33. & $\left\{\frac{5}{48}, \frac{7}{24}, \frac{29}{48}\right\}$ & 34. & $\left\{\frac{13}{48}, \frac{37}{48}, \frac{23}{24}\right\}$ &&&&\\
\hline \hline
\end{tabular}
\end{threeparttable}
}
\end{center}
\end{table}

The theory of Kaidi, Lin and Parra-Martinez of \cite{Kaidi:2021ent} (henceforth KLP theory) posits that the diagonal elements of the T-matrix of a $\mathbf{[3,3]}$ admissible character can not be anything but the exponentials of the numbers in the exponent-triplets. Let $\{ a_1, a_2, a_3 \}$ denote an exponent-triplet. KLP theory connects this exponent-triplet to  vector-valued modular forms in the following way. 

Let $\{ \alpha_0, \alpha_1, \alpha_2 \}$ be the index-set for a vector valued modular form written following the conventions of this paper viz. $\alpha_0$ is the index of the identity character (and hence negative and this defines the central charge) and $\alpha_1 < \alpha_2$ (which define the conformal dimensions of the non-identity characters $h_1$ and $h_2$).  The T-matrix for such a vector modular form would be $\text{Diag} \{ e^{2\pi i \alpha_0}, e^{2\pi i \alpha_1}, e^{2\pi i \alpha_2}\}$. KLP theory then posits that the (unordered) set of elements  $\{ e^{2\pi i \alpha_0}, e^{2\pi i \alpha_1}, e^{2\pi i \alpha_2} \}$ is identical to the (unordered) set of exponentials of the numbers in the exponent-triplet, $\{ e^{2\pi i a_1}, e^{2\pi i a_2}, e^{2\pi i a_3}  \}$.  One is free to set the T-matrix element associated to the identity character viz. $e^{2\pi i \alpha_0}$ to either of $e^{2\pi i a_1}$ or $e^{2\pi i a_2}$ or $e^{2\pi i a_3}$. Suppose one chooses to set $e^{2\pi i \alpha_0}$ to say $e^{2\pi i a_1}$; this means $\alpha_0 - a_1 $ is any integer and one has an infinite number of choices for $\alpha_0$, even after requiring that it should be negative. Similarly one obtains infinite number of possibilities for $\alpha_1$ and an infinite number of possibilities for $\alpha_2$. These triple of infinite possibilities are somewhat reduced after we impose the requirement that $\alpha_0$ is negative and $\alpha_1 < \alpha_2$ and that $\alpha_0 + \alpha_1 + \alpha_2 = 0$ (which is the requirement that the Wronskian index is $3$), but there are still an infinite number of index-sets that follow from any KLP exponent-triplet. To illustrate the point, consider the exponent-triplet $\{\frac{3}{8}, \frac{3}{4},\frac{7}{8}\}$ (which is S-8 of table \ref{t5}). Each of the following index-sets\footnote{There are more index-sets that are consistent with the given exponent-triplet. In all of the above, the identity character index is associated with $\frac34$. There are other index-sets where the identity character is associated with $\frac38$ and others where it is associated with $\frac78$, etc. Here we just give some of them to illustrate the point that a given exponent-triplet can result in multiple index-sets.} $\alpha_0 = - m - \frac14, \alpha_1 = -\frac{1}{8}, \alpha_2 = m + \frac38 $ where $m$ is a non-negative integer (and hence we have an infinite number of index-sets) has a T-matrix which is consistent with the given exponent-triplet. 

We bring KLP theory to bear on the approach in this paper (in sections \ref{31ss} and \ref{32ss}) in the following way.  For every exponent-triplet of table \ref{t5}, we compute the index-sets that result from it and which have a central charge less than $96$.  Recall that we have searched for $\mathbf{[3,3]}$ admissible characters with $c <96$. In appendix \ref{app2}, for each of the KLP exponent-triplets, we give, not the resulting index-sets, but the equivalent triple of central charges and conformal dimensions $\{ c, h_1, h_2 \}$. There are a finite number of such triples for each exponent-triplet. It seems that the the finite number of triples is either $78$ or $66$. There are a total of $2472$ triples. Thus, KLP theory directs a search for $\mathbf{[3,3]}$ admissible characters with $c < 96$ to these $2472$ computations. 

What we now need to do is run the solution-generating procedure of section \ref{31ss} for each of $2472$ triples. The way we go about it is as follows. Consider an example : the triple $c = 8, h_1 = \frac13, h_2 = \frac23$.  To start step I of the procedure, we need $N$ which can be had from the value of $c$, from \eqref{31p}, to be $N = 142540720$. We run through all of steps I and steps II of section  \ref{31ss}. Note that we have used only the value of $c$ so far and have not used the value of $h_1$ and $h_2$. What this means is that at the end of this procedure, we will obtain all admissible character solutions with $c = 8$. If there is one with $c = 8, h_1 = \frac13, h_2 = \frac23$, we would have found it (it turns out that there is no such admissible character solution). This is what we need to do for each of the $2472$ triples of central charges and conformal dimensions of appendix \ref{app2} that arise from the $34$ KLP exponent triplets. 

We note the following feature of the central charges that appear in each of the $2472$ triples : the central charges, being rational and positive, when written in the $\frac{a}{b}$ form with $a$ and $b$ being coprime, are such that $b$ is either $1$ or $2$ or $5$ or $7$.  Thus to answer the question of which of the $2472$ triples of central charges and conformal dimensions result in admissible character solutions, we need to run the procedure of section \ref{31ss} for these central charges. But, remarkably, in section \ref{32ss}, we had limited our search, from around $1.7$ billion computations to some $1152$ computations, and this limited search is exactly the same as what is required. Thus, the $\mathbf{[3,3]}$ admissible character solutions that arise from the KLP theory are the fifteen solutions given in tables \ref{t1} and \ref{t3} and (the first fifteen solutions of) appendix \ref{app1}.

We note that $13$ of the $34$ KLP exponent-triplets are realised as admissible character solutions.  The rest $21$ exponent-triplets do not seem to have a realisation as an admissible character solution for $c \leq 96$; we suspect that they could be realised as an admissible character with a central charge bigger than $96$. We hope to perform this bigger search in a future work. We also note that there are two  KLP exponent triplets each of which are realized as two different admissible character solutions. This happens for the triplet  $\{\frac{5}{7},\frac{6}{7},\frac{3}{7}\}$ which is realized by the S-3 and S-14 solutions and also for the triplet $\{\frac{2}{3},\frac{7}{15},\frac{13}{15}\}$ which are realized by the S-6 and S-15 solutions. 

Finally we address the four infinite family of solutions (S-16 -S-19) and their relation to KLP theory. Let us compute the T-matrix $\text{Diag} \{ e^{2 \pi i \alpha_0}, e^{2 \pi i \alpha_1}, e^{2 \pi i \alpha_2}\}$ for these solutions. First let us study solution S-18 as an example. Note that  $\alpha_0 - \alpha_1 = -1$ and the T-matrix is $\text{Diag} \{ e^{\frac{\pi i}{3}}, e^{\frac{\pi i}{3}}, e^{\frac{4\pi i}{3}} \}$. Since the T-matrix has equal entries in the first and second diagonal elements\footnote{Another way to understand this equalling of two entries of the T-matrix is the following. When the conformal dimension of a non-identity character is an integer, it's $q$-series is commensurate with that of the identity character, and the corresponding diagonal entries in the T-matrix would be equal. Note that each of the solutions $\#16 - \#19$ have an integral conformal dimension for a non-identity character.}, this results in an S-matrix which is block-diagonal (a $2 \times 2$ block and a $1 \times 1$ block). Thus what we have is a reducible representation of the modular group for the vector valued modular form of solution S-18.  This happens for the solutions S-16 and S-19 also. 
Typical studies of modular data, which includes \cite{Kaidi:2021ent} are for irreducible representations. Hence KLP theory excludes vector valued modular forms such as the ones in solutions S-16 - S-19. There are no KLP exponent-triplets for these solutions. 

\subsection{Coset Bilinear Relations between $\mathbf{[3,3]}$ Admissible Characters\label{34ss}}
In this section, we will show that amongst our admissible character solutions, there are many pairs which  potentially can be the characters of the denominator and coset CFTs of the meromorphic coset construction of Gaberdiel-Hampapura-Mukhi (GHM). 

First let us recall the details of the meromorphic coset construction:
\begin{align} \label{33p}
{\cal C} = \frac{{\cal H}}{{\cal D}}.
\end{align}
Here ${\cal H}$ is a meromorphic CFT, a one-character CFT which has for it's central charge  a multiple of $8$: $c^{{\cal H}} = 8 K$. ${\cal D}$ is a CFT whose affine sub algebra is a sub algebra of the affine sub algebra of  that of ${\cal H}$. ${\cal C}$ is a CFT whose affine sub algebra is such that it’s finite part is the commutant of the finite part of the affine sub algebra of ${\cal C}$ inside the finite part of the affine sub algebra of ${\cal H}$. One of the important details of this construction is that the number of characters of the ${\cal C}$ and the ${\cal D}$ CFTs are the same. Let $p$ denote this common number of characters of the ${\cal C}$ and the ${\cal D}$ CFTs (and in this paper, we only have $p = 3$). Let ${\cal D}$ be a $\mathbf{[p, l]}$ CFT and ${\cal C}$  a $\mathbf{[\tilde{p}, \tilde{l}]}$ CFT.
Let us denote the central charges and conformal dimensions\footnote{Here the conformal dimensions need not be written in an increasing order. They are labelled to follow equation \eqref{34p}.} of the ${\cal D}$ CFT by $\{ c, h_1, \ldots h_{p-1} \}$, similarly of the ${\cal C}$ CFT by $\{ \tilde{c}, \tilde{h}_1, \ldots \tilde{h}_{p-1} \}$. The central charges of the three CFTs are related;  the conformal dimensions of the ${\cal C}$ and the ${\cal D}$ CFTs are such that 
\begin{align} \label{34p}
c + \tilde{c} = c^{{\cal H}} = 8K, \qquad h_i + \tilde{h}_i = n_i,  \qquad n_i\in \mathbf{Z}_{\geq 0}.
\end{align}
Note that the $n_i$'s are at least $1$. We then have the following very pertinent formula that brings all these quantities together : 
\begin{align} \label{GHMmaster}
l + \tilde{l} = p^2+(2 K -1)p-6\sum_{i=1}^{p-1}n_i.
\end{align}
Another important formula in the GHM theory of meromorphic cosets is the coset-bilinear identity, also referred to as the bilinear identity or a coset-relation, which relates the characters of all the CFTs in \eqref{33p} : 
\begin{equation}
    \chi^{\mathcal{H}}_{0}(\tau)=\chi_{0}(\tau).\tilde{\chi}_{0}(\tau)+\sum_{i=1}^{p-1}d_i\chi_{i}(\tau).\tilde{\chi}_{i}(\tau),
\end{equation}
where $ \chi^{\mathcal{H}}_{0}$ is the character of the meromorphic CFT ${\cal H}$, $\chi_{0}$ and $\tilde{\chi}_{0}$ are the identity characters of the ${\cal D}$ and ${\cal C}$ CFTs, $\chi_{i}$'s and 
$\tilde{\chi}_i$'s are the non-identity characters of the ${\cal D}$ and ${\cal C}$ CFTs, $d_i$ are positive integers\footnote{Typically all of these $p-1$ integers are non-zero. But there are examples where some of them can vanish \cite{Das:2022uoe}.} that are sometimes referred  to as the multiplicities of the bilinear relation. In this paper, we are mostly interested in the situation when $p = 3$. The equation \eqref{GHMmaster} now becomes :
\begin{align} \label{37p}
    l+\tilde{l}=6 K + 6 - 6 (n_1 + n_2)
\end{align}
and the coset-relation is 
\begin{equation} \label{38p}
    \chi^{\mathcal{H}}_{0}(\tau)=\chi_0(\tau).\tilde{\chi}_{0}(\tau)+d_1\chi_{1}(\tau).\tilde{\chi}_{1}(\tau)+d_2\chi_{2}(\tau).\tilde{\chi}_{2}(\tau).
\end{equation}
Sometimes it happens that the ${\cal C}$ and the ${\cal D}$ CFTs are identical and we call this a self-coset.  In terms of the coset-relation \eqref{38p}, self-cosets come in two varieties. The first is when  $\chi_0 = \tilde{\chi_0},~\chi_1 = \tilde{\chi_1},~\chi_2 = \tilde{\chi_2}$. In this case, the self-coset relation would be identical to \eqref{38p} and we would have two multiplicities $d_1, d_2$ : 
\begin{equation} \label{39p}
    \chi^{\mathcal{H}}_{0}(\tau)=\chi_0(\tau)^2 + d_1\chi_{1}(\tau)^2 + d_2\chi_{2}(\tau)^2.
\end{equation}
The second variety of self-coset relation has $\chi_0 = \tilde{\chi_0},~\chi_1 = \tilde{\chi_2},~\chi_2 = \tilde{\chi_1}$. In this case the self-coset relation would take the following form and there would be only one multiplicity in the bilinear relation :
\begin{equation} \label{40p}
    \chi^{\mathcal{H}}_{0}(\tau)=\chi_0(\tau)^2 + (d_1 + d_2) \chi_{1}(\tau).\chi_{2}(\tau).
\end{equation}

The study of coset-relations amongst admissible character solutions can be very useful.  In \cite{}, all $\mathbf{[3,0]}$ admissible character solutions were found. Then, in \cite{Das:2022uoe}, all possible coset-bilinear relations amongst these admissible character solutions were computed and tabulated. Note that these papers discuss bilinear-relations \eqref{38p} which follow \eqref{37p} with $l = \tilde{l} = 0$. This exhaustive list of coset-relations for $c^{{\cal H}} = 8, 16, 24, 32, 40$ proved to be very useful. They were the starting point to identify genuine $\mathbf{[3,0]}$ CFTs among  $\mathbf{[3,0]}$ admissible characters and led to a classification of all $\mathbf{[3,0]}$ CFTs \cite{Das:2022uoe}. Furthermore, the coset-relations also led to the discovery of new meromorphic CFTs \cite{Das:2022slz}.

We now undertake the study of coset-bilinear relations for pairs of $\mathbf{[3,3]}$ admissible characters. We set $l = \tilde{l} = 3$ in \eqref{37p} so that we have
\begin{equation} \label{39p}
K = n_1 + n_2.
\end{equation}
We then search for pairs of $\mathbf{[3,3]}$ admissible characters whose central charges add to some multiple ($K$) of $8$ and the conformal dimensions of non-identity characters are such that : $h_1 + \tilde{h}_1 = n_1, ~ h_2 + \tilde{h}_2 = n_2$ and furthermore the $n_1$ and $n_2$ follow \eqref{39p}. For every such pair of $\mathbf{[3,3]}$ admissible characters we then proceed with the computation of the right hand side of \eqref{38p} and ask if there are positive integers $d_1$ and $d_2$ such that the right hand side results in the left hand side, that is a character of the meromorphic CFT. Recall that the character of a meromorphic CFT is a known polynomial of $j$ multiplied either by $j^{\frac13}$ or $j^{\frac23}$. There may be some parameters in the polynomial of $j$ which are also determined in our computation. 

We should mention that when we consider a pair of admissible characters, they are interchangeable.  We may denote one of them by $\{c, h_1, h_2\}$ and the other by $\{\tilde{c}, \tilde{h}_1, \tilde{h}_2\}$ but it is not meant to indicate that  the former is a denominator CFT ${\cal D}$ and the latter is a coset CFT ${\cal C}$. At this stage of the computations and analysis, we have only a bilinear relation between admissible characters. Further analysis (which is beyond the scope of the present paper) is needed to ascertain if these bilinear relations between characters are CFT coset relations and which of the admissible characters forms the denominator CFT and which forms the coset CFT.

We first consider the case when the meromorphic coset has the central charge $c^{{\cal H}} = 8.$ In this case, there is a unique CFT and a unique character : the $E_{8,1}$ CFT with character $j^{\frac13}$. We have $K = 1$. But from \eqref{39p}, we see that we cannot have bilinear identities between two $\mathbf{[3,3]}$ admissible characters and the character of such a  meromorphic CFT. 

\subsubsection{Coset Bilinear Relations with $c^{{\cal H}} = 16$ }

We then consider the case when the meromorphic CFT has the central charge $c^{{\cal H}} = 16.$ In this case, there is a unique character viz. $j^{\frac23}$; there are two CFTs viz. $E_{8,1} \otimes E_{8,1}$ and the one-character extension of the $D_{16,1}$ CFT. But, in this paper, we are working only at the level of characters and  hence these CFTs will not play any role here. We now have $K = 2$ and we see from \eqref{39p} that it is possible to have bilinear identities between two $\mathbf{[3,3]}$ admissible characters and the character of such  meromorphic CFTs. There is a unique solution to $\eqref{39p}$ and hence only one kind of bilinear relation with $n_1 = 1, n_2 = 1$. We will see, that for higher $K$'s,  we can have multiple kinds of bilinear relations, one kind for every solution to \eqref{39p}.

\renewcommand{\arraystretch}{1.5}
\begin{table}[H] 
\centering
\begin{tabular}{| c || c| c  c    c  c|| c| c  c  c   c||c    |}
    \rowcolor{lightgray}\hline
     \# & Soln. & $c$ & $h_1$ & $h_2$ & $\text{m}_1$  & Soln. & $\tilde{c}$ & $\tilde{h}_1$ & $\tilde{h}_2$ &  $\tilde{m}_1$ & $(d_1, d_2)$\\ \hline
     C-1 & S-5 & 8 & $\frac{1}{4}$ & $\frac{3}{4}$ & 24 & S-5 & 8 & $\frac{3}{4}$ & $\frac{1}{4}$ & 24 &   \tiny{$d_1 +  d_2 = 64$}  \\ \rowcolor{lightgray}
     C-2 & S-6 & 8 & $\frac{1}{5}$ & $\frac{4}{5}$ & 24 & S-6 & 8 & $\frac{4}{5}$ & $\frac{1}{5}$ & 24 & \tiny{$d_1 +  d_2 = 4$}  \\
     C-3 & S-3 & $\frac{48}{7}$ & $\frac{1}{7}$ & $\frac{5}{7}$ & 78 & S-7 & $\frac{64}{7}$ & $\frac{6}{7}$  & $\frac{2}{7}$ & 136 & $(1,1)$\\
    \hline
\end{tabular}
\caption{$\mathbf{[3,3]}$ admissible characters in a coset-bilinear relation with $c^{{\cal H}} = 16$ and  $( n_1, n_2 ) = ( 1, 1 )$.} \label{t6}
\end{table}
There are three coset-relations between $\mathbf{[3,3]}$ admissible characters with $n_1 = 1, n_2 = 1$ and they are collected in table \ref{t6}.
The two coset-relations C-1 and C-2 are both self-coset relations of the type in \eqref{40p} with  one multiplicity $d_1 + d_2$ while the coset-relation C-3 is the generic kind of coset relation \eqref{38p} with two multiplicities $d_1, d_2$.

\subsubsection{Coset Bilinear Relations with $c^{{\cal H}} = 24$ }

We now consider the case when the meromorphic CFT has the central charge $c^{{\cal H}} = 24$ or $K = 3$. In this case, there are an infinite number of characters possible viz. $j  + {\cal N}$, where ${\cal N} \geq - 744$ is an integer. Amongst these, only a finite number of them correspond to CFTs; this was worked out in the celebrated paper \cite{Schellekens:1992db} where it was shown that only $16$ of these correspond to ($71$) CFTs. Hence, here at $K = 3$, unlike for $K = 2$, we have that there are many possibilities for the meromorphic theory and we will specify which in our tables below. Also, note that for $K = 3$, we have a unique solution to \eqref{39p} and hence we have only one kind of bilinear relation with $\{n_1,  n_2\} = \{1, 2\}.$ 

We find the following three coset-relations which are such that each admissible character of the pair is one of the finite set of admissible characters (S-1 to S-15) :

\begin{table}[H] 
\centering
\begin{tabular}{|c  ||c |  c  c    c  c||c |  c  c  c   c||    c   ||c|}
    \rowcolor{lightgray}\hline
    \# &   Soln. & $c$ & $h_1$ & $h_2$ & $\text{m}_1$  & Soln. &  $\tilde{c}$ & $\tilde{h}_1$ & $\tilde{h}_2$ &  $\tilde{m}_1$ &  $(d_1, d_2)$ & ${\cal N}$\\ \hline
    C-4  &  S-9 &16 & $\frac{4}{5}$ & $\frac{6}{5}$ & 232 & S-6 & 8 & $\frac{1}{5}$ & $\frac{4}{5}$ & 134 & $(10,10)$ & $-288$\\\rowcolor{lightgray}
    C-5 & S-7 & $\frac{64}{7}$ & $\frac{2}{7}$ & $\frac{6}{7}$ & 136 & S-8 & $\frac{104}{7}$ & $\frac{5}{7}$ & $\frac{8}{7}$ & 188 & $(1,1)$ & $-288$\\
    C-6 & S-3 & $\frac{48}{7}$ & $\frac{1}{7}$ & $\frac{5}{7}$ & 78 & S-10 & $\frac{120}{7}$ & $\frac{6}{7}$ & $\frac{9}{7}$ & 156 & $(1, 1)$ & $-432$\\\hline
\end{tabular}
\caption{$\mathbf{[3,3]}$ admissible characters in a coset-bilinear relation with $c^{{\cal H}} = 24$ and  $( n_1, n_2 ) = ( 1, 2 )$. The last column indicates the meromorphic character $j + {\cal N}$.}\label{t7}
\end{table}
The last column indicates the meromorphic character $j + {\cal N}$. It is interesting to note that each of the meromorphic characters of table \ref{t7} corresponds to an actual CFT : $j - 288$ is the character of the CFTs in \cite{Schellekens:1992db} with Schellekens no. 64, 65 and $j - 432$ with Schellekens no. 58, 59. We expect this fact to play a role in the analysis that will decide if these coset-relations between characters are CFT coset-relations. 

We further find the following coset-relation (table \ref{t8}) where both the $\mathbf{[3,3]}$ admissible characters are drawn from the infinite family of solutions S-16. C-7 may occupy only one row in a table but it actually stands for a large number of bilinear coset-relations. There is one coset-relation for every pair of non-negative integers $(m, n)$ chosen so that $m+n \equiv 4\,\mod\,8$, $\quad0\leq m\leq 1004$ and $\quad0\leq n\leq \frac{130556 - 130m}{m+130}.$ For example when $m = 0$, $n = 4, 12, 20, \ldots 1004$; thus there are $126$ coset-relations with $m = 0$. Similarly for $m = 1$, $n = 3, 11, 19, \ldots 995$; thus there are $125$ coset-relations with $m = 1$. And so on. There are a total of $23759$ coset relations.  It will be interesting to see which of these large number of coset-relations between admissible characters actually survive to be  CFT coset-relations.

\begin{table}[H] 
\begin{center}
\resizebox{\textwidth}{!}{
\renewcommand{\arraystretch}{1.4}
\begin{threeparttable}
\begin{tabular}{|c  ||c |  c  c    c  c||c |  c  c  c   c||    c   ||c|}
\hline 
    \rowcolor{lightgray}\hline
    \# & Soln. &  c & $h_1$ & $h_2$ & $\text{m}_1$  & Soln. &  $\tilde{c}$ & $\tilde{h}_1$ & $\tilde{h}_2$ &  $\tilde{m}_1$ & $(d_1, d_2)$ &  $\mathcal{N}$ \\ \hline
    C-7 & S-16 & 12 & $\frac{1}{2}$ & 1 & $22+n$ & S-16 &  12 & $\frac{1}{2}$ & 1 & $22+m$ &  $\left(\frac{1028+m+n}{8}, 130556 - 130(m+n) - mn\right)$  & $\frac{9 m + 9 n - 4572}{8}$  \\ \hline
\end{tabular} 
   \caption{$\mathbf{[3,3]}$ admissible characters in a coset-bilinear relation with $c^{{\cal H}} = 24$ and  $( n_1, n_2 ) = ( 1, 2 )$. $m$ and $n$ are positive integers such that $m+n \equiv 4\,\mod\,8$, $\quad0\leq m\leq 1004$ and $\quad0\leq n\leq \frac{130556 - 130m}{m+130}$. These are   $23759$ coset-relations.}\label{t8}
\end{threeparttable}
}
\end{center}
\end{table}

\subsubsection{Coset Bilinear Relations with $c^{{\cal H}} = 32$ }

We now consider the case when the meromorphic CFT has the central charge $c^{{\cal H}} = 32$ or $K = 4$. In this case, there are an infinite number of characters possible viz. $j^{\frac13} (j  + {\cal N})$, where ${\cal N} \geq - 992$ is an integer. Amongst these, only a finite number of them should correspond to CFTs. But it is presently not known which of these infinite characters correspond to CFT. A  partial list can be found in \cite{King}. Hence, here at $K = 4$, similar to the $K = 3$ case, we have that there are many possibilities for the meromorphic character and we will specify which in our tables below. Also, note that for $K = 4$, we have two solutions to \eqref{39p} viz. (i) $\{n_1,  n_2\} = \{1, 3\}$, (ii) $\{n_1,  n_2\} = \{2, 2\}$. Thus there are two kinds of bilinear relations. But it turns out that amongst the $\mathbf{[3,3]}$ admissible characters, there are no examples of coset-relations with $\{n_1,  n_2\} = \{1, 3\}$. The following coset-relations in table \ref{t9} between $\mathbf{[3,3]}$ admissible characters have $\{n_1,  n_2\} = \{2, 2\}$. In each the admissible characters are from the finite set of admissible characters (S-1 to S-15).

\begin{table}[H] 
\centering
\begin{tabular}{|c  ||c |  c  c    c  c||c |  c  c  c   c||    c   ||c|}
    \rowcolor{lightgray}\hline
    \# &   Soln. & $c$ & $h_1$ & $h_2$ & $\text{m}_1$  & Soln. &  $\tilde{c}$ & $\tilde{h}_1$ & $\tilde{h}_2$ &  $\tilde{m}_1$ &  $(d_1, d_2)$ & ${\cal N}$\\ \hline
    C-8  &  S-9 &16 & $\frac{4}{5}$ & $\frac{6}{5}$ & 232 & S-9 & 16 & $\frac{6}{5}$ & $\frac{4}{5}$ & 232 & $d_1 + d_2 = 100$ & $-528$\\\rowcolor{lightgray}
    C-9 & S-12 & $24$ & $\frac{6}{5}$ & $\frac{9}{5}$ & 30 & S-6 & $8$ & $\frac{4}{5}$ & $\frac{1}{5}$ & 134 & $(50,50)$ & $-828$\\
    C-10 & S-3 & $\frac{48}{7}$ & $\frac{1}{7}$ & $\frac{5}{7}$ & 78 & S-13 & $\frac{176}{7}$ & $\frac{13}{7}$ & $\frac{9}{7}$ & 14& $(1, 1)$ & $-900$\\\rowcolor{lightgray}
    C-11 & S-7 & $\frac{64}{7}$ & $\frac{2}{7}$ & $\frac{6}{7}$ & 136 & S-11 & $\frac{160}{7}$ & $\frac{12}{7}$ & $\frac{8}{7}$ & 40& $(1, 1)$ & $-816$\\
     C-12 & S-8 & $\frac{104}{7}$ & $\frac{5}{7}$ & $\frac{8}{7}$ & 188 & S-10 & $\frac{120}{7}$ & $\frac{9}{7}$ & $\frac{6}{7}$ & 156& $(1, 1)$ & $-648$\\\hline
\end{tabular}
\caption{$\mathbf{[3,3]}$ admissible characters in a coset-bilinear relation with $c^{{\cal H}} = 32$ and  $( n_1, n_2 ) = ( 2, 2 )$. The last column indicates the meromorphic character $j^{\frac13}(j + {\cal N})$.} \label{t9}
\end{table}

We further find the following coset-relation (table \ref{t10}) where both the $\mathbf{[3,3]}$ admissible characters are drawn from the infinite families of solutions S-16 and S-18. C-13 may occupy only one row in a table but it actually stands for $127$ bilinear coset-relations. It will be interesting to see which of these large number of coset-relations between admissible characters actually survive to be a CFT coset-relation. 

\begin{table}[H] 
\begin{center}
\resizebox{\textwidth}{!}{
\renewcommand{\arraystretch}{1.4}
\begin{threeparttable}
\begin{tabular}{|c  ||c |  c  c    c  c||c |  c  c  c   c||    c   ||c|}
\hline 
    \rowcolor{lightgray}\hline
    \# & Soln. &  c & $h_1$ & $h_2$ & $\text{m}_1$  & Soln. &  $\tilde{c}$ & $\tilde{h}_1$ & $\tilde{h}_2$ &  $\tilde{m}_1$ & $(d_1, d_2)$ &  $\mathcal{N}$ \\ \hline
    C-13 & S-16 & 12 & $\frac{1}{2}$ & 1 & $22+2n$ & S-18 &  20 & $\frac{3}{2}$ & 1 & $13+n$ &  $\left(16448 + 64 n, 65278 -260 n- 2 n^2\right)$  &  $3 n - 957$ \\ \hline
\end{tabular} 
   \caption{$\mathbf{[3,3]}$ admissible characters in a coset-bilinear relation with $c^{{\cal H}} = 32$ and  $( n_1, n_2 ) = ( 2, 2 )$. With $0 \leq n \leq 126$, these are $127$ coset relations. }\label{t10}
\end{threeparttable}
}
\end{center}
\end{table}

\subsubsection{Coset Bilinear Relations with $c^{{\cal H}} = 40$ }

We now consider the case when the meromorphic CFT has the central charge $c^{{\cal H}} = 40$ or $K = 5$. In this case, there are an infinite number of characters possible viz. $j^{\frac23} (j  + {\cal N})$, where ${\cal N} \geq - 1240$ is an integer. Amongst these, only a finite number of them should correspond to CFTs. But it is presently not known which of these infinite characters correspond to CFT. Hence, here at $K = 5$, similar to the $K = 4$ and $K = 3$ cases, we have that there are many possibilities for the meromorphic character and we will specify which in our tables below. Also, note that for $K = 5$, we have two solutions to \eqref{39p} viz. (i) $\{n_1,  n_2\} = \{1, 4\}$, (ii) $\{n_1,  n_2\} = \{2, 3\}$. Thus there are two kinds of bilinear relations. But it turns out that amongst the $\mathbf{[3,3]}$ admissible characters, there are no examples of coset-relations with $\{n_1,  n_2\} = \{1, 4\}$. The following coset-relations in table \ref{t11a} between $\mathbf{[3,3]}$ admissible characters have $\{n_1,  n_2\} = \{2, 3\}$. In each the admissible characters are from the finite set of admissible characters (S-1 to S-15).

\begin{table}[H] 
\centering
\begin{tabular}{|c  ||c |  c  c    c  c||c |  c  c  c   c||    c   ||c|}
    \rowcolor{lightgray}\hline
    \# &   Soln. & $c$ & $h_1$ & $h_2$ & $\text{m}_1$  & Soln. &  $\tilde{c}$ & $\tilde{h}_1$ & $\tilde{h}_2$ &  $\tilde{m}_1$ &  $(d_1, d_2)$ & ${\cal N}$\\ \hline
    C-14  &  S-6 & 8 & $\frac{1}{5}$ & $\frac{4}{5}$ & 134 & S-15 & 32 & $\frac{9}{5}$ & $\frac{11}{5}$ & 3 & $(1250, 1250)$ & $-1103$\\\rowcolor{lightgray}
    C-15 & S-9 & $16$ & $\frac{4}{5}$ & $\frac{6}{5}$ & 232 & S-12 & $24$ & $\frac{6}{5}$ & $\frac{9}{5}$ & 30 & $(250,250)$ & $-978$\\
    C-16 & S-7 & $\frac{64}{7}$ & $\frac{2}{7}$ & $\frac{6}{7}$ & 136 & S-14 & $\frac{216}{7}$ & $\frac{12}{7}$ & $\frac{15}{7}$ & 3& $(1, 1)$ & $-1101$\\\rowcolor{lightgray}
    C-17 & S-8 & $\frac{104}{7}$ & $\frac{5}{7}$ & $\frac{8}{7}$ & 188 & S-13 & $\frac{176}{7}$ & $\frac{9}{7}$ & $\frac{13}{7}$ & 14& $(1, 1)$ & $-1038$\\
     C-18 & S-10 & $\frac{120}{7}$ & $\frac{9}{7}$ & $\frac{6}{7}$ & 156 & S-11 & $\frac{160}{7}$ & $\frac{8}{7}$ & $\frac{12}{7}$ & 40& $(1, 1)$ & $-1044$\\\hline
\end{tabular}
\caption{$\mathbf{[3,3]}$ admissible characters in a coset-bilinear relation with $c^{{\cal H}} = 40$ and  $( n_1, n_2 ) = ( 2, 3 )$. The last column indicates the meromorphic character $j^{\frac23}(j + {\cal N})$.} \label{t11a}
\end{table}

We further find the following coset-relation (table \ref{t12a}) where both the $\mathbf{[3,3]}$ admissible characters are drawn from the infinite families of solutions (S-16 to S-19). C-19 may occupy only one row in a table but it actually stands for a large number of bilinear coset-relations.  There is one coset-relation for every pair of non-negative integers $(m, n)$ chosen so that $\quad0\leq m\leq 502$ and $\quad0\leq n \leq \frac{32639 - 65m}{m + 65}.$ For example when $m = 0$, $n = 0,  \ldots 502$; thus there are $503$ coset-relations with $m = 0$. Similarly for $m = 1$, $n = 0, \ldots 493$; thus there are $494$ coset-relations with $m = 1$. And so on. There are a total of $47733$ coset relations in C-19. C-20 contains a further $254$ coset-relations.  It will be interesting to see which of these large number of coset-relations between admissible characters actually survive to be a CFT coset-relation.

\begin{table}[H] 
\begin{center}
\resizebox{\textwidth}{!}{
\renewcommand{\arraystretch}{1.4}
\begin{threeparttable}
\begin{tabular}{|c  ||c |  c  c    c  c c ||c |  c c  c  c   c||    c   ||c|}
\hline 
    \rowcolor{lightgray}\hline
    \# & Soln. &  c & $h_1$ & $h_2$ & $\text{m}_1$  & $\text{m}_2$  & Soln. &  $\tilde{c}$ & $\tilde{h}_1$ & $\tilde{h}_2$ &  $\tilde{m}_1$ & $\text{m}_2$  & $(d_1, d_2)$ &  $\mathcal{N}$ \\ \hline
    C-19 & S-18 & 20 & $1$ & $\frac32$ & $13+n$ & $354 + 548n$ &  S-18 &  20 & $1$ & $\frac{3}{2}$ & $13+m$ & $354 + 548m$ & $\left(32639 - 65(n+m) - nm, 4096(514+n+m)\right)$  & $ m + n -1214$  \\ \rowcolor{lightgray} 
    C-20 & S-16 & 12 & $\frac12$ & $1$ & $22+n$ & $26+44n$ &  S-19 &  28 & $\frac32$ & $2$ & $4$ & $506 + 384n$ & $\left( 32896 + 64 n,  128(130556-260n-n^2)\right)$  & $n-1214$\\\hline
\end{tabular} 
   \caption{$\mathbf{[3,3]}$ admissible characters in a coset-bilinear relation with $c^{{\cal H}} = 40$ and  $( n_1, n_2 ) = ( 2, 3 )$. The first row has $47733$ coset-relations. In the second row, there are $254$ coset relations ($n \leq 253$). }\label{t12a}
\end{threeparttable}
}
\end{center}
\end{table}

\subsubsection{Coset Bilinear Relations with $c^{{\cal H}} = 48$ }

We now consider the case when the meromorphic CFT has the central charge $c^{{\cal H}} = 48$ or $K = 6$. In this case, there are an infinite number of characters possible viz. $j^2  + {\cal N}_1\,j + {\cal N}_2$, where  $\mathcal{N}_1$. and $\mathcal{N}_2$ are integers with $\mathcal{N}_1\geq-1488,~~744 \,\mathcal{N}_1 + \mathcal{N}_2\geq-947304$.  Amongst these, only a finite number of them should correspond to CFTs. But it is presently not known which of these infinite characters correspond to CFTs.  Hence, here at $K = 6$, we have that there are many possibilities for the meromorphic character and we will specify which in our tables below. Also, note that for $K = 6$, we have three solutions to \eqref{39p} viz. (i) $\{n_1,  n_2\} = \{1, 5\}$, (ii) $\{n_1,  n_2\} = \{2, 4\}$, (iii) $\{n_1,  n_2\} = \{3, 3\}$ Thus there are three kinds of bilinear relations. But it turns out that amongst the $\mathbf{[3,3]}$ admissible characters, there are no examples of coset-relations with $\{n_1,  n_2\} = \{1, 5\}$ and $\{n_1,  n_2\} = \{2, 4\}$. The following coset-relations in table \ref{t13} between $\mathbf{[3,3]}$ admissible characters have $\{n_1,  n_2\} = \{3, 3\}$. In each the admissible characters are from the finite set of admissible characters (S-1 to S-15).

We further find the following coset-relation (table \ref{t14}) where both the $\mathbf{[3,3]}$ admissible characters are drawn from the infinite families of solutions (S-16 to S-19). C-25 contains 127 coset-relations. It will be interesting to see which of these coset-relations between admissible characters actually survive to be  CFT coset-relations.

\begin{table}[H] 
\centering
\begin{tabular}{|c  ||c |  c  c    c  c||c |  c  c  c   c||    c   ||c c|}
    \rowcolor{lightgray}\hline
    \# &   Soln. & $c$ & $h_1$ & $h_2$ & $\text{m}_1$  & Soln. &  $\tilde{c}$ & $\tilde{h}_1$ & $\tilde{h}_2$ &  $\tilde{m}_1$ &  $(d_1, d_2)$ & ${\cal N}_1$ & ${\cal N}_2$\\ \hline
    C-21  &  S-9 & $16$ & $\frac{4}{5}$ & $\frac{6}{5}$ & 232 & S-15 & 32 & $\frac{11}{5}$ & $\frac{9}{5}$ & 3 & $(6250, 6250)$ & $-1253$ & $79200$ \\\rowcolor{lightgray}
    C-22 & S-12 & $24$ & $\frac{6}{5}$ & $\frac{9}{5}$ & 30 & S-12 & $24$ & $\frac{9}{5}$ & $\frac{6}{5}$ & 30 & $d_1 + d_2 = 2500$ & $-1428$ & $291600$ \\
    C-23 & S-10 & $\frac{120}{7}$ & $\frac{6}{7}$ & $\frac{9}{7}$ & 156 & S-14 & $\frac{216}{7}$ & $\frac{15}{7}$ & $\frac{12}{7}$ & 3& $(1, 1)$ & $-1329$ & $123120$\\\rowcolor{lightgray}
     C-24 & S-11 & $\frac{160}{7}$ & $\frac{8}{7}$ & $\frac{12}{7}$ & 40 & S-13 & $\frac{176}{7}$ & $\frac{13}{7}$ & $\frac{9}{7}$ & 14& $(1, 1)$ & $-1434$ & $247104$ \\\hline
\end{tabular}
\caption{$\mathbf{[3,3]}$ admissible characters in a coset-bilinear relation with $c^{{\cal H}} = 48$ and  $( n_1, n_2 ) = ( 3, 3 )$. The last two columns indicates the meromorphic character $j^2  + {\cal N}_1\,j + {\cal N}_2$.} \label{t13}
\end{table}

\begin{table}[H] 
\begin{center}
\resizebox{\textwidth}{!}{
\renewcommand{\arraystretch}{1.4}
\begin{threeparttable}
\begin{tabular}{|c  ||c |  c  c    c  c c ||c |  c c  c  c   c||    c   ||c c|}
\hline 
    \rowcolor{lightgray}\hline
    \# & Soln. &  c & $h_1$ & $h_2$ & $\text{m}_1$  & $\text{m}_2$  & Soln. &  $\tilde{c}$ & $\tilde{h}_1$ & $\tilde{h}_2$ &  $\tilde{m}_1$ & $\text{m}_2$  & $(d_1, d_2)$ &  $\mathcal{N}_1$ & ${\cal N}_2$ \\ \hline
    C-25 & S-18 & 20 & $1$ & $\frac32$ & $13+n$ & $354 + 548n$ &  S-19 &  28 & $2$ & $\frac{3}{2}$ & $4$ & $506 + 768n$ & $\left(256(32639 - 130 n - n^2), 16384 (257 + n)\right)$  & $ n -1471$ & $148032 + 576n$  \\ \rowcolor{lightgray} 
\hline
\end{tabular} 
   \caption{$\mathbf{[3,3]}$ admissible characters in a coset-bilinear relation with $c^{{\cal H}} = 48$ and  $( n_1, n_2 ) = ( 3, 3 )$. There are $127$ coset-relations ($n \leq 126$). The last two columns indicates the meromorphic character $j^2  + {\cal N}_1\,j + {\cal N}_2$. }\label{t14}
\end{threeparttable}
}
\end{center}
\end{table}

\subsubsection{Coset Bilinear Relations with $c^{{\cal H}} = 56$ }
We now consider the case when the meromorphic CFT has the central charge $c^{{\cal H}} = 56$ or $K = 7$. In this case, there are an infinite number of characters possible viz. $j^{\frac13} (j^2  + {\cal N}_1\,j + {\cal N}_2)$, where  $\mathcal{N}_1$. and $\mathcal{N}_2$ are integers with $\mathcal{N}_1\geq-1736,~~992 \mathcal{N}_1 + \mathcal{N}_2\geq-1320452$.  Amongst these, only a finite number of them should correspond to CFTs. But it is presently not known which of these infinite characters correspond to CFTs. Hence, here at $K = 7$, we have that there are many possibilities for the meromorphic character and we will specify which in our tables below. Also, note that for $K = 7$, we have three solutions to \eqref{39p} viz. (i) $\{n_1,  n_2\} = \{1, 6\}$, (ii) $\{n_1,  n_2\} = \{2, 5\}$, (iii) $\{n_1,  n_2\} = \{3, 4\}$ Thus there are three kinds of bilinear relations. But it turns out that amongst the $\mathbf{[3,3]}$ admissible characters, there are no examples of coset-relations with $\{n_1,  n_2\} = \{1, 6\}$ and $\{n_1,  n_2\} = \{2, 5\}$. The following coset-relations in table \ref{t15a} between $\mathbf{[3,3]}$ admissible characters have $\{n_1,  n_2\} = \{3, 4\}$. In each the admissible characters are from the finite set of admissible characters (S-1 to S-15).

\begin{table}[H] 
\centering
\begin{tabular}{|c  ||c |  c  c    c  c||c |  c  c  c   c||    c   ||c c|}
    \rowcolor{lightgray}\hline
    \# &   Soln. & $c$ & $h_1$ & $h_2$ & $\text{m}_1$  & Soln. &  $\tilde{c}$ & $\tilde{h}_1$ & $\tilde{h}_2$ &  $\tilde{m}_1$ &  $(d_1, d_2)$ & ${\cal N}_1$ & ${\cal N}_2$\\ \hline
    C-26  &  S-12 & $24$ & $\frac{6}{5}$ & $\frac{9}{5}$ & 30 & S-15 & 32 & $\frac{9}{5}$ & $\frac{11}{5}$ & 3 & $(31250, 31250)$ & $-1703$ & $519300$ \\\rowcolor{lightgray}
    C-27 & S-13 & $\frac{176}{7}$ & $\frac{9}{7}$ & $\frac{13}{7}$ & 14 & S-14 & $\frac{216}{7}$ & $\frac{12}{7}$ & $\frac{15}{7}$ & 3 & $(1,1)$ & $-1719$ & $503604$ \\
   \hline
\end{tabular}
\caption{$\mathbf{[3,3]}$ admissible characters in a coset-bilinear relation with $c^{{\cal H}} = 56$ and  $( n_1, n_2 ) = ( 3, 4 )$. The last two columns indicate the meromorphic character $j^{\frac13}(j^2  + {\cal N}_1\,j + {\cal N}_2$).} \label{t15a}
\end{table}

We further find the following coset-relation (table \ref{t16a}) where both the $\mathbf{[3,3]}$ admissible characters are drawn from the infinite families of solutions (S-16 to S-19). C-28 may occupy only one row in a table but it actually stands for a large number of bilinear coset-relations.  There is one coset-relation for every pair of non-negative integers $(m, n)$ chosen so that $\quad0\leq m\leq 130196$ and $\quad0\leq n \leq \frac{2144592320 - 16472m}{m + 16472}.$ For example when $m = 0$, $n = 0,  \ldots 130196 $; thus there are $130197$ coset-relations with $m = 0$. Similarly for $m = 1$, $n = 0, \ldots 130187$; thus there are $130188$ coset-relations with $m = 1$. And so on. There are a total of $3137971936$ coset relations in C-28.   It will be interesting to see which of these large number of coset-relations between admissible characters actually survive to be CFT coset-relations. 

\begin{table}[H] 
\begin{center}
\resizebox{\textwidth}{!}{
\renewcommand{\arraystretch}{1.4}
\begin{threeparttable}
\begin{tabular}{|c  ||c |  c  c    c  c c ||c |  c c  c  c   c||    c   ||c c|}
\hline 
    \rowcolor{lightgray}\hline
    \# & Soln. &  c & $h_1$ & $h_2$ & $\text{m}_1$  & $\text{m}_2$  & Soln. &  $\tilde{c}$ & $\tilde{h}_1$ & $\tilde{h}_2$ &  $\tilde{m}_1$ & $\text{m}_2$  & $(d_1, d_2)$ &  $\mathcal{N}_1$ & ${\cal N}_2$ \\ \hline
    C-28 & S-19 & $28$ & $\frac32$ & $2$ & $4$ & $2 + 3 n$ &  S-19 &  28 & $\frac32$ & $2$ & $4$ & $2 + 3 m$ & $\left(64(131248+n+m), 2144592320 - 16472(n + m ) - mn\right)$  & $ -1728$ & $3(131248+n+m)$  \\ \rowcolor{lightgray} 
\hline
\end{tabular} 
   \caption{$\mathbf{[3,3]}$ admissible characters in a coset-bilinear relation with $c^{{\cal H}} = 56$ and  $( n_1, n_2 ) = ( 3, 4 )$. There are $3137971936$ coset-relations. The last two columns indicates the meromorphic character $j^2  + {\cal N}_1\,j + {\cal N}_2$. }\label{t16a}
\end{threeparttable}
}
\end{center}
\end{table}

\subsubsection{Coset Bilinear Relations with $c^{{\cal H}} = 64$ }
We now consider the case when the meromorphic CFT has the central charge $c^{{\cal H}} = 64$ or $K = 8$. In this case, there are an infinite number of characters possible viz. $j^{\frac23} (j^2  + {\cal N}_1\,j + {\cal N}_2)$, where  $\mathcal{N}_1$. and $\mathcal{N}_2$ are integers with $\mathcal{N}_1\geq-1984,~~1240 \mathcal{N}_1 + \mathcal{N}_2\geq-1755104$.  Amongst these, only a finite number of them should correspond to CFTs. But it is presently not known which of these infinite characters correspond to CFTs. Hence, here at $K = 8$, we have that there are many possibilities for the meromorphic character and we will specify which in our tables below. Also, note that for $K = 8$, we have four solutions to \eqref{39p} viz. (i) $\{n_1,  n_2\} = \{1, 7\}$, (ii) $\{n_1,  n_2\} = \{2, 6\}$, (iii) $\{n_1,  n_2\} = \{3, 5\}$, (iv) $\{ n_1, n_2 \} = \{4,4\}$. Thus there are four kinds of bilinear relations. But it turns out that amongst the $\mathbf{[3,3]}$ admissible characters, there are no examples of coset-relations with $\{n_1,  n_2\} = \{1, 7\}$,  $\{n_1,  n_2\} = \{2, 6\}$ and $\{n_1,  n_2\} = \{3, 5\}$. The following coset-relation in table \ref{t17a} is a self-coset relation with $\{n_1,  n_2\} = \{4, 4\}$. 
\begin{table}[H] 
\centering
\begin{tabular}{|c  ||c |  c  c    c  c||c |  c  c  c   c||    c   ||c c|}
    \rowcolor{lightgray}\hline
    \# &   Soln. & $c$ & $h_1$ & $h_2$ & $\text{m}_1$  & Soln. &  $\tilde{c}$ & $\tilde{h}_1$ & $\tilde{h}_2$ &  $\tilde{m}_1$ &  $(d_1, d_2)$ & ${\cal N}_1$ & ${\cal N}_2$\\ \hline
    C-29  &  S-15 & $32$ & $\frac{9}{5}$ & $\frac{11}{5}$ & $30$ & S-15 & 32 & $\frac{9}{5}$ & $\frac{11}{5}$ & 3 & $d_1 + d_2 = 1562500$ & $-1978$ & $822625$ \\\rowcolor{lightgray}
   \hline
\end{tabular}
\caption{$\mathbf{[3,3]}$ admissible characters in a coset-bilinear relation with $c^{{\cal H}} = 64$ and  $( n_1, n_2 ) = ( 4, 4 )$. The last two columns indicate the meromorphic character $j^{\frac23}(j^2  + {\cal N}_1\,j + {\cal N}_2$).} \label{t17a}
\end{table}

\section{ $\mathbf{[3,4]}$ Admissible Characters \label{4s}}
In this section, we study the $\mathbf{[3,4]}$ MLDE and it's admissible character solutions. In the first sub-section \ref{41ss}, we set up the $\mathbf{[3,4]}$ MLDE and outline a step-by-step procedure to obtain their admissible character solutions. In the subsequent subsection \ref{42ss} we implement the procedure and obtain $\mathbf{[3,4]}$ admissible characters.

\subsection{The $\mathbf{[3,4]}$ MLDE  and a Solution-generating Procedure \label{41ss}}

The characters of every $\mathbf{[3,4]}$ MLDE are solutions to the $\mathbf{[3,4]}$ CFT : 
\bea \label{51p}
\mathcal{D}^3\chi_i + \mu_{-1,1}\frac{E_6}{E_4}\mathcal{D}^2\chi_i +\mu_{1,0}E_4\mathcal{D}\chi_i + \mu\frac{\Delta}{E_4^2}\mathcal{D}\chi_i + \mu_{0,1}E_6 \chi_i= 0,
\eea
with $i = 0,1,2$. As it stands, there are four parameters in the MLDE  viz. $\mu_{-1,1}$, $\mu_{1,0}$, $\mu$, $\mu_{0,1}$. The characters have a $q$-expansion,
\be \label{52p}
\chi_i(q)= q^{\alpha_i}\sum\limits_{n=0}^{\infty} f^{(i)}_n q^n
\ee
where the leading behaviour are the indices/characteristic exponents of the CFT :
\be \label{53p}
\alpha_0 = - \frac{c}{24}, \quad \alpha_1 = h_1 - \frac{c}{24}, \quad \alpha_2 = h_2 - \frac{c}{24}.
\ee
Similar to [3,3] case, We will employ the Frobenius-like method to solve the linear differential equation in \eqref{51p}. In this method, one determines the ratios $\frac{f^{(i)}_n}{f^{(i)}_0}$. For the identity character one imposes $f^{(0)}_0 = 1$ and we will write $f^{(0)}_n = \text{m}_n$ for $n > 0$ and further impose that the $\text{m}_n$’s are all non-negative integers. For the non-identity characters, the $f^{(i)}_0$'s  can be any non-zero positive integer. In our search for character-like solutions we will require that the ratios $\frac{f^{(i)}_n}{f^{(i)}_0}$ for $i = 1, 2$ need not be non-negative integers but only non-negative rational numbers such that upon multiplying by a suitable positive integer all of them become non-negative integers. This step depends on the order $n$ up to which one solves the differential equation. We will denote this positive integer by $D_1$ for the $\chi_1$ character and by $D_2$ for the $\chi_2$ character.  $f^{(1)}_0$ and $f^{(2)}_0$ cannot be determined per se; they can be any multiple of $D_1$ and $D_2$ respectively. 

At the leading order in the $q$-expansion, the differential equation \eqref{51p} gives the following cubic equation
\begin{equation} \label{54p}
1728 \alpha ^3+144 \alpha ^2 (\mu _{-1,1}-6)+12 \alpha  (\mu _{1,0}-2\mu _{-1,1}+8)+\mu _{0,1}=0
\end{equation}
which is the indicial equation whose solutions are the indices \eqref{53p}. Now,  from the valence formula, we have the following constraint for the indices $ \sum_{r=0}^{n-1}\alpha_r=\frac{n(n-1)}{12}-\frac{l}{6}$ which for $l = 4$ gives $\sum_{r=0}^{2}\alpha_r=-\frac{1}{6}$. Imposing this constraint in \eqref{54p} sets  $\mu_{-1,1} =8$ and the indicial equation is 
\begin{equation} \label{55p}
   1728 \alpha ^3+288 \alpha ^2+12 \alpha  \left(\mu _{\text{1,0}}-8\right)+\mu _{\text{0,1}}=0
\end{equation}
More importantly, the number of parameters in the MLDE \eqref{51p} has reduced to three viz. $\mu_{1,0}$, $\mu_{0,1}$, $\mu$. Note that only two of them are present in the indicial equation \eqref{55p}. This means that these two parameters are completely determined by the indices. Such parameters are termed ``rigid'' parameters. The other kind of parameters, the ones which are completely independent of the indices are termed ``non-rigid'' parameters. Here, for us in \eqref{55p}, we have that $\mu_{1,0}$ and $\mu_{0,1}$ are rigid parameters and $\mu$ is a non-rigid parameter. Typically this kind of a clean separation of parameters into rigid and non-rigid ones may not be possible. Here, in \eqref{55p}, we could achieve this separation because of the way the co-efficient of the no-derivative term was written : the numerator of this coefficient which needs to be a weight-$12$ modular form was written as $\mu_{1,0}\, E_4^3 + \mu\, \Delta$ instead of say $a\,E_4^3 + b\,E_6^2$. In the latter representation, both $a$ and $b$ would appear in the indicial equation and neither can be called a non-rigid parameter. In any case, we are left to have to solve the following three-parameter MLDE to obtain $\mathbf{[3,4]}$ CFTs :
\bea \label{56p}
\mathcal{D}^3\chi_i + 8\frac{E_6}{E_4}\mathcal{D}^2\chi_i +\mu_{1,0}E_4\mathcal{D}\chi_i + \mu\frac{\Delta}{E_4^2}\mathcal{D}\chi_i + \mu_{0,1}E_6 \chi_i= 0,
\eea

At the next to leading order in the $q$-expansion, the differential equation \eqref{56p} gives the following equation :
\begin{equation}\label{57p}
    \frac{f_1}{f_0}=\frac{-829440 \alpha^3+698112 \alpha^2+\alpha \left(-12 \mu-8640 \mu _{\text{1,0}}-103680\right)+24 \mu _{\text{0,1}}}{1728 \alpha^3+5472 \alpha^2+\alpha \left(12 \mu _{\text{1,0}}+5664\right)+12 \mu _{\text{1,0}}+\mu _{\text{0,1}}+1920}
\end{equation}

At the next to the next to the leading order in the $q$-expansion, the differential equation \eqref{56p} gives the following equation :

\begin{footnotesize}
\begin{align} \label{58p}
    &\frac{f_2}{f_0}=\Big(503078584320 \alpha^6+665074188288 \alpha^5+\alpha^4 (20404224 \mu+9331200000 \mu _{\text{1,0}}+258508578816)+\alpha^3 (24634368 \mu\notag\\&+192969216 \mu _{\text{0,1}}+5939868672 \mu _{\text{1,0}}+295546724352)+\alpha^2 (210816 \mu \mu _{\text{1,0}}+144 \mu^2+48833280 \mu _{\text{1,0}}^2+8847360 \mu\notag\\&+1249261056 \mu _{\text{0,1}}-4978879488 \mu _{\text{1,0}}+162087174144)+\alpha (-288 \mu \mu _{\text{0,1}}+210816 \mu \mu _{\text{1,0}}+144 \mu^2-206496 \mu _{\text{0,1}} \mu _{\text{1,0}}\notag\\&+48833280 \mu _{\text{1,0}}^2+4617216 \mu+1050603264 \mu _{\text{0,1}}-1587548160 \mu _{\text{1,0}}-36946575360)-288 \mu \mu _{\text{0,1}}+2152224 \mu _{\text{0,1}} \mu _{\text{1,0}}\notag\\&+197208 \mu _{\text{0,1}}^2+371893248 \mu _{\text{0,1}}\Big)\times\Big(2985984 \alpha^6+27869184 \alpha^5+\alpha^4 (41472 \mu _{\text{1,0}}+105753600)+\alpha^3 (3456 \mu _{\text{0,1}}+255744 \mu _{\text{1,0}}\notag\\&+208465920)+\alpha^2 (144 \mu _{\text{1,0}}^2+16128 \mu _{\text{0,1}}+588672 \mu _{\text{1,0}}+224787456)+\alpha (24 \mu _{\text{0,1}} \mu _{\text{1,0}}+432 \mu _{\text{1,0}}^2+27456 \mu _{\text{0,1}}+597888 \mu _{\text{1,0}}\notag\\&+125577216)+36 \mu _{\text{0,1}} \mu _{\text{1,0}}+\mu _{\text{0,1}}^2+288 \mu _{\text{1,0}}^2+16704 \mu _{\text{0,1}}+223488 \mu _{\text{1,0}}+28385280\Big)^{-1}
\end{align}
\end{footnotesize}

We now evaluate the equations \eqref{55p}, \eqref{57p} and \eqref{58p} for the identity character. This means we replace $\alpha_i$ in these three equations by $\alpha_0$ and we replace the left hand sides of the latter two equations by $\text{m}_1$ and $\text{m}_2$ respectively. Then we interpret these three equations as the ones to be solved simultaneously for the three parameters in terms of $\alpha_0$, $\text{m}_1$ and $\text{m}_2$. The solution is as follows : 

\begin{footnotesize}
\begin{align} \label{59p}
    &\mu _{\text{1,0}} =\frac{1}{-393768 \alpha _0^2+\alpha _0 \text{m}_1^2-2 \alpha _0 \text{m}_2+\text{m}_1^2}\Big(38973312 \alpha _0^4-16 \alpha _0^3 (27 \text{m}_1^2-9072 \text{m}_1-54 \text{m}_2+830520)\notag\\&-16 \alpha _0^2 (57 \text{m}_1^2-10008 \text{m}_1-114 \text{m}_2-80676)-16 \alpha _0 (40 \text{m}_1^2-936 \text{m}_1-77 \text{m}_2)-160 \text{m}_1^2\Big)
\end{align}
\end{footnotesize}

\begin{footnotesize}
\begin{align} \label{60p}
    &\mu_{\text{0,1}} =\frac{1}{-393768 \alpha _0^2+\alpha _0 \text{m}_1^2-2 \alpha _0 \text{m}_2+\text{m}_1^2}\Big(212751360 \alpha _0^5+288 \alpha _0^4 \left(12 \text{m}_1^2-6048 \text{m}_1-24 \text{m}_2+947448\right)\notag\\&+288 \alpha _0^3 \left(31 \text{m}_1^2-6672 \text{m}_1-74 \text{m}_2-185040\right)+288 \alpha _0^2 \left(26 \text{m}_1^2-624 \text{m}_1-52 \text{m}_2\right)+2016 \alpha _0 \text{m}_1^2\Big)
\end{align}
\end{footnotesize}

\begin{footnotesize}
\begin{align} \label{61p}
    &\mu=\frac{1}{-393768 \alpha _0^2+\alpha _0 \text{m}_1^2-2 \alpha _0 \text{m}_2+\text{m}_1^2}\Big(-418037760 \alpha _0^4+48 \alpha _0^3 (5184 \text{m}_1^2+482112 \text{m}_1-10368 \text{m}_2-266552640)\notag\\&+48 \alpha _0^2 (10800 \text{m}_1^2+1732536 \text{m}_1-30672 \text{m}_2+49295520)+48 \alpha _0 (7608 \text{m}_1^2-18 \text{m}_2 \text{m}_1+1053540 \text{m}_1-18744 \text{m}_2)\notag\\&+48 (1992 \text{m}_1^2-19 \text{m}_1 \text{m}_2)\Big)
\end{align}
\end{footnotesize}

We have thus determined the three parameters of the MLDE in terms of the objects associated with the identity character viz. the index $\alpha_0$, the first Fourier coefficient $\text{m}_1$ and second Fourier coefficient $\text{m}_2$. We now consider the differential equation \eqref{56p} at the next order in the $q$-expansion, which when evaluated for the identity character, is an equation for $\text{m}_3$. After using \eqref{59p}, \eqref{60p} and \eqref{61p}, we obtain the following polynomial equation in the variables $\alpha_0$, $\text{m}_1$, $\text{m}_2$ and $\text{m}_3$. 

\begin{scriptsize}
\begin{align} \label{62p}
   &\alpha _0^5
   +\frac{\alpha _0^4}{931598640} \Big(1581 \text{m}_1^3-432564 \text{m}_1^2-4743 \text{m}_2 \text{m}_1-54072366 \text{m}_1+865128 \text{m}_2+4743 \text{m}_3+11037470720\Big)\notag\\&
   +\frac{\alpha _0^3}{22358367360}\Big(111869 \text{m}_1^3-42 \text{m}_2 \text{m}_1^2-11444352 \text{m}_1^2-380715 \text{m}_2 \text{m}_1-126 \text{m}_3 \text{m}_1+168 \text{m}_2^2-2704710332 \text{m}_1+52158336 \text{m}_2\notag\\&+470931 \text{m}_3-49466918400\Big)\notag\\&
   +\frac{\alpha _0^2 }{268300408320}\Big(1443966 \text{m}_1^3-2068 \text{m}_2 \text{m}_1^2-9 \text{m}_3 \text{m}_1^2+3 \text{m}_2^2 \text{m}_1+13604720 \text{m}_1^2-4570434 \text{m}_2 \text{m}_1-1668 \text{m}_3 \text{m}_1+5248 \text{m}_2^2\notag\\&+9 \text{m}_2 \text{m}_3-10423131000 \text{m}_1+201438208 \text{m}_2+5564934 \text{m}_3\Big)\notag\\&
   +\frac{\alpha _0}{1609802449920}\Big(4091760 \text{m}_1^3-16056 \text{m}_2 \text{m}_1^2-129 \text{m}_3 \text{m}_1^2+55 \text{m}_2^2 \text{m}_1+158151072 \text{m}_1^2-7813260 \text{m}_2 \text{m}_1-936 \text{m}_3 \text{m}_1+14592 \text{m}_2^2\notag\\&+93 \text{m}_2 \text{m}_3\Big)\notag\\&
   +\frac{1}{1609802449920}\Big(750564 \text{m}_1^3-6672 \text{m}_2 \text{m}_1^2-75 \text{m}_3 \text{m}_1^2+38 \text{m}_2^2 \text{m}_1\Big) = 0
\end{align}
\end{scriptsize}
We observe that if we define a new variable 
\be \label{63p}
N = - 931598640 \, \alpha_0   
\ee
the polynomial equation \eqref{62p} becomes the following polynomial equation with indeterminates $N$, m$_1$, m$_2$ and m$_3$ : 
\begin{equation} \label{64p}
    \text{N}^5+p_1(\text{m}_1,\text{m}_2,\text{m}_3)\text{N}^4+p_2(\text{m}_1,\text{m}_2,\text{m}_3)\text{N}^3+p_3(\text{m}_1,\text{m}_2,\text{m}_3)\text{N}^2+p_4(\text{m}_1,\text{m}_2,\text{m}_3)\text{N}+p_5(\text{m}_1,\text{m}_2,\text{m}_3)=0
\end{equation}
where the $p_i(\text{m}_1,\text{m}_2,\text{m}_3)$'s are the following polynomials of the first three Fourier coefficients :
\begin{footnotesize}
\begin{align}\label{65p}
    p_1(\text{m}_1,\text{m}_2,\text{m}_3)=&-1581 \text{m}_1^3+432564 \text{m}_1^2+4743 \text{m}_2 \text{m}_1+54072366 \text{m}_1-865128 \text{m}_2-4743 \text{m}_3-11037470720\notag\\
    p_2(\text{m}_1,\text{m}_2,\text{m}_3)=&\quad 38816610 \Big(111869 \text{m}_1^3-42 \text{m}_2 \text{m}_1^2-11444352 \text{m}_1^2-380715 \text{m}_2 \text{m}_1-126 \text{m}_3 \text{m}_1+168 \text{m}_2^2\notag\\&-2704710332 \text{m}_1+52158336 \text{m}_2+470931 \text{m}_3-49466918400\Big)\notag\\
    p_3(\text{m}_1,\text{m}_2,\text{m}_3)=&-3013458423784200 \Big(1443966 \text{m}_1^3-2068 \text{m}_2 \text{m}_1^2-9 \text{m}_3 \text{m}_1^2+3 \text{m}_2^2 \text{m}_1+13604720 \text{m}_1^2-4570434 \text{m}_2 \text{m}_1\notag\\&-1668 \text{m}_3 \text{m}_1+5248 \text{m}_2^2+9 \text{m}_2 \text{m}_3-10423131000 \text{m}_1+201438208 \text{m}_2+5564934 \text{m}_3\Big)\notag\\
    p_4(\text{m}_1,\text{m}_2,\text{m}_3)=&\quad 467888961548984062248000\Big(4091760 \text{m}_1^3-16056 \text{m}_2 \text{m}_1^2-129 \text{m}_3 \text{m}_1^2+55 \text{m}_2^2 \text{m}_1+158151072 \text{m}_1^2\notag\\&-7813260 \text{m}_2 \text{m}_1-936 \text{m}_3 \text{m}_1+14592 \text{m}_2^2+93 \text{m}_2 \text{m}_3\Big)\notag\\
    p_5(\text{m}_1,\text{m}_2,\text{m}_3)=&-435884720250045845771912142720000 \Big(750564 \text{m}_1^3-6672 \text{m}_2 \text{m}_1^2-75 \text{m}_3 \text{m}_1^2+38 \text{m}_2^2 \text{m}_1\Big)\notag\\
\end{align}
\end{footnotesize}

Now let us consider \eqref{64p} as a polynomial equation to determine $N$, which at this stage is a rational number. Given that the coefficient of $N^5$ is $1$ (which is the reason we made the definition \eqref{63p}) and that all other coefficients are integers, we can use the integer-root-theorem to conclude that $N$ has to be an integer. This is an important stage of our analysis. $N$ which was just rational is now restricted to be an integer. This means that the central charge, given in terms of $N$ by
\be \label{66p}
    c=\frac{\text{N}}{38816610},
\ee
is a rational number, whose denominator (not necessarily in the $\frac{p}{q}$ form with $p$, $q$ coprime) is always $38816610$. Our problem has become a ``finite'' problem in the following sense.  If one is searching for CFTs in a certain range of central charge, say $0 < c \leq 1$, one only needs to do $38816610$ computations for $ c = \frac{1}{38816610}, \frac{2}{38816610}, \ldots \frac{38816609}{38816610}, \frac{38816610}{38816610}$.

Every computation involves solving two Diophantine equations. One of the Diophantine equations is \eqref{64p}. The other Diophantine equation arises as follows. Via equations \eqref{59p}, \eqref{60p} and \eqref{61p}, we can think of the MLDE parameters as functions of $N$, $\text{m}_1$ and $\text{m}_2$ and subsequently the indicial equation \eqref{55p} as having coefficients which are polynomials of $N$, $\text{m}_1$ and $\text{m}_2$. The solutions are thus now functions of $N$, $\text{m}_1$ and $\text{m}_2$. In these terms, we know one of the solutions  viz. $-\frac{N}{931598640 }$, which is nothing but $\alpha_0$ (see \eqref{63p}). Factoring out this solution, the indicial equation \eqref{55p} reduces to a quadratic equation in $\alpha$ with  coefficients which are polynomials of $N$, $\text{m}_1$ and $\text{m}_2$. The roots of this quadratic equation determine the other two indices $\alpha_1$ and $\alpha_2$ and eventually $h_1$ and $h_2$. For rational roots the discriminant has to be the square of a rational number. But since the discriminant is an integer (on account of all coefficients being integers), it needs to be a perfect square to be able to result in rational roots. If we denote the discriminant by $k^2$ with $k$ chosen to be a positive integer, we have 
\begin{align} \label{67p}
    &-65073335754457462509504\text{N}^6+q_1(\text{m}_1,\text{m}_2)\text{N}^5+q_2(\text{m}_1,\text{m}_2)\text{N}^4+q_3(\text{m}_1,\text{m}_2)\text{N}^3\notag+q_4(\text{m}_1,\text{m}_2)\text{N}^2\\&+q_5(\text{m}_1,\text{m}_2)\text{N}+q_6(\text{m}_1,\text{m}_2)=k^2
\end{align}
where the $q_i(\text{m}_1, \text{m}_2)$ are the following functions of m$_1$ and m$_2$ :
\begin{footnotesize}
\begin{align}\label{68p}
    q_1(\text{m}_1,\text{m}_2)=&\quad 673760104630537049637120 \Big(7975 \text{m}_1^2-3675168 \text{m}_1-15950 \text{m}_2+456092724\Big)\notag\\
    q_2(\text{m}_1,\text{m}_2)=&\quad 4358847202500458457719121427200 \Big(3 \text{m}_1^4-1344 \text{m}_1^3-12 \text{m}_2 \text{m}_1^2-2687936 \text{m}_1^2+2688 \text{m}_2 \text{m}_1+12 \text{m}_2^2\notag\\&+583826688 \text{m}_1+6229216 \text{m}_2+17627418288\Big)\notag\\
    q_3(\text{m}_1,\text{m}_2)=&-1353565375272410566195877007858126336000 \Big(31 \text{m}_1^4-8480 \text{m}_1^3-112 \text{m}_2 \text{m}_1^2-6601428 \text{m}_1^2+8896 \text{m}_2 \text{m}_1\notag\\&+100 \text{m}_2^2+163807488 \text{m}_1+13941096 \text{m}_2\Big)\notag\\
    q_4(\text{m}_1,\text{m}_2)=&\quad 35027212854301869805269694281330550210160640000 \Big(1429 \text{m}_1^4-175104 \text{m}_1^3-4324 \text{m}_2 \text{m}_1^2-71381520 \text{m}_1^2\notag\\&+29952 \text{m}_2 \text{m}_1+2500 \text{m}_2^2\Big)\notag\\
    q_5(\text{m}_1,\text{m}_2)=&-65262607716116280120092624051406635932474732811059200000 \Big(403 \text{m}_1^4-7488 \text{m}_1^3-794 \text{m}_2 \text{m}_1^2\Big)\notag\\
    q_6(\text{m}_1,\text{m}_2)=&\quad 5137478031955338058226803982807184674280996101836847788916736000000 \text{m}_1^4\notag
\end{align}
\end{footnotesize}

The equation \eqref{67p} is a polynomial equation with indeterminates $N$, $\text{m}_1$, $\text{m}_2$ and $k$, all non-negative integers and hence a Diophantine equation. We thus have two Diophantine equations, \eqref{64p} and \eqref{67p}, involving five Diophantine variables viz. $N$, $\text{m}_1$, $\text{m}_2$, $\text{m}_3$ and $k$. $\text{m}_3$ and $k$ are each present in only one of the Diophantine equations while the other three are simultaneously present in both. 

The procedure to obtain character like solutions to the $\mathbf{[3,4]}$ MLDE can now be stated. 

(i) We solve the two Diophantine equations \eqref{64p} and \eqref{67p} simultaneously.  We fix a value of $N$ and and then obtain solutions to \eqref{64p}. Each such solution is a set of non-negative integers for $N$, $\text{m}_1$, $\text{m}_2$ and $\text{m}_3$. We take the $\text{m}_1$, $\text{m}_2$ and $N$ values of each such solution and see if we can find non-negative integer values for $k$ that satisfy \eqref{67p}. At the end of this procedure, we will have a set of five non-negative integers, for $N$, $\text{m}_1$, $\text{m}_2$, $\text{m}_3$ and $k$ which simultaneously solve \eqref{64p} and \eqref{67p}.

(ii) At this stage, we have the three indices: $\alpha_0$, $\alpha_1$ and $\alpha_2$ ($N$ gives $\alpha_0$ while $k$ the discriminant, together with $\text{m}_1$, $\text{m}_2$ and $\text{m}_3$ gives the other two indices), which further means that we have the $c$, $h_1$ and $h_2$ of the putative CFTs. But we have imposed positivity on the Fourier coefficients of the character only upto order $q^3$ and only for the identity character.  Hence, now in this next stage of the procedure, we impose positivity. That is we determine the Fourier coefficients of all the characters upto some high order (for us $q^{1000}$), which are after all functions of the index of the character and the parameters in the MLDE which in turn are functions of the known $\alpha_0$, $\text{m}_1$ and $\text{m}_2$ (\eqref{59p}, \eqref{60p}, \eqref{61p}) and check if they are positive. We reject solutions to the Diophantine equations which do not survive the positivity constraints. We keep all the other solutions and in the process also compute $D_1$ and $D_2$, the positive integers that when multiplied to $\chi_1$ and $\chi_2$, makes all their Fourier coefficients to be positive integers as well.

\subsection{Solutions to the $\mathbf{[3,4]}$ MLDE\label{42ss}}

We implement the just delineated procedure for  $c < 96$. To search for solutions within a central charge range of $1$, we will need to perform around $931$ million computations and for a central charge range of $96$ that is around $89.4$ billion computations. But we undertake a limited search. The factorisation of the denominator in \eqref{66p} is as follows : $38816610 = 2\times 3\times 5\times 7\times 17\times 83\times 131$. We look for solutions with central charges given by (a) $c=\frac{n}{2}$ with $2\nmid n$ (these are $96$ computations), (b) $c = \frac{n}{3}$ with $3\nmid n$ (these constitute  $192$ computations), (c) $c = \frac{n}{5}$ with $5\nmid n$ (these constitute  $384$ computations), (d) $c = \frac{n}{7}$ with $7\nmid n$ (these constitute $576$ computations) and (e) $c = n$ with $n \in \mathbf{Z}_{\geq 0}$ (these are $96$ computations), which gives us a total of $1344$ computations. \\

Using the valence formula \ref{windex} for [3,4] case, we have $\sum_i h_i =  \frac{c}{8} - \frac{1}{6}$, requiring the conformal dimensions to be positive gives us a lower bound on the central charge i.e $ c \geq \frac{4}{3}$ or $N \geq 51755480$, thus reducing our computation to 1327 .

Using the similar motivation as [3,3] we are doing a limited search (limiting to $1327$ cases as opposed to a $89.4$ billion cases) for solutions of $\mathbf{[3,4]}$ MLDE

\begin{table}[H] 
\begin{center}
\resizebox{\textwidth}{!}{
\renewcommand{\arraystretch}{1.4}
\begin{threeparttable}
   \caption{Discrete set of solutions to the Diophantine equations \eqref{64p} and \eqref{67p} with $\frac{4}{3} \leq c \leq 96$}\label{t11}
\begin{tabular}{c||ccccc||ccc||ccc||}
\hline \hline
\rowcolor{Mywhite}
\# & $N$ &  $\text{m}_1$ & $\text{m}_2$ & $\text{m}_3$ &  $k$ & $\mu$ & $\mu_{1,0}$ & $\mu_{0,1}$ & $c$ & $h_1$ & $h_2$   \\  
\hline \hline
\rowcolor{Mygrey}
    S'-1. & 636592404 & 410 & 64739 & 2089934 & 22787074700652379514040978033023622144 & -55296 & -$\frac{1071}{25}$ & 0 & $\frac{82}{5}$ & $\frac{41}{60}$ & $\frac{6}{5}$ \\ \rowcolor{Mywhite}
    S'-2. & 659882370 & 323 & 60860 & 2158575 & 45853329031423822791822069765573120000 & -55296 & -$\frac{189}{4}$ & 0 & 17 & $\frac{17}{24}$ & $\frac{5}{4}$ \\ \rowcolor{Mygrey}
    S'-3. & 698698980 & 234 & 59805 & 2482242 & 72067959168925880011559792323325184000 & -55296 & -55 & 0 & 18 & $\frac{3}{4}$ & $\frac{4}{3}$ \\\rowcolor{Mywhite}
    S'-4. & 729752268 & 188 & 62087 & 2923494 & 88181885687655607252799840662343242752 & -55296 & -$\frac{1539}{25}$ & 0 & $\frac{94}{5}$ & $\frac{47}{60}$ & $\frac{7}{5}$ \\\rowcolor{Mygrey}
    S'-5. & 776332200 & 140 & 69950 & 3983800 & 107123028848651267056905128194867200000 & -55296 & -72 & 0 & 20 & $\frac{5}{6}$ & $\frac{3}{2}$ \\\rowcolor{Mywhite}
    S'-6. & 822912132 & 106 & 84429 & 5825442 & 117541111174650568416870757866489919488 & -55296 & -$\frac{2079}{25}$ & 0 & $\frac{106}{5}$ & $\frac{53}{60}$ & $\frac{8}{5}$ \\\rowcolor{Mygrey}
    S'-7. & 853965420 & 88 & 99935 & 7846300 & 115346865865608882029083794591565440000 & -55296 & -91 & 0 & 22 & $\frac{11}{12}$ & $\frac{5}{3}$ \\\rowcolor{Mywhite}
    S'-8. & 892782030 & 69 & 131905 & 12195106 & 89958121037055261674858306367696384000 & -55296 & -$\frac{405}{4}$ & 0 & 23 & $\frac{23}{24}$ & $\frac{7}{4}$ \\\rowcolor{Mygrey}
    S'-9. & 916071996 & 59 & 164315 & 16778125 & 49159175864915842196370689437986998016 & -55296 & -$\frac{2691}{25}$ & 0 & $\frac{118}{5}$ & $\frac{59}{60}$ & $\frac{9}{5}$ \\\rowcolor{Mywhite}
 \hline \hline
\end{tabular} 
\end{threeparttable}
}
\end{center}
\end{table}

For each of  the $1327$ cases that we listed above, we implemented the solution-generating procedure that we have described in section \ref{41ss}. We first give an account of the (simultaneous) solutions to the Diophantine equations \eqref{64p} and \eqref{67p}. We collect the Diophantine solutions in table \ref{t11}. We found a discrete set of $9$ solutions which are given in table \ref{t11}. We will number the discrete set of solutions S'-1$-$S'-9. For each solution, we first give the five Diophantine variables $N$, $\text{m}_1$, $\text{m}_2$, $\text{m}_3$ and $k$.  We also give the values of the MLDE parameters or in other words the point in the space of MLDE parameters at which this solution exists. The values the MLDE parameters are obtained from the Diophantine variables from the formulae in \eqref{59p}, \eqref{60p}, \eqref{61p}. The last detail we include in the tables is the central charge $c$ and the conformal dimensions of the characters $h_1$ and $h_2$; the $c$ follows from $N$ and the conformal dimensions require further the value of $k$, $\text{m}_1$ and $\text{m}_2$. The $9$ solutions presented here are not all the solutions to the Diophantine equations \eqref{64p} and \eqref{67p}. They are those solutions which on computation of higher Fourier coefficients (up to $q^{1000}$) result in admissible characters.

\begin{table}[H]
\begin{center}
\resizebox{0.8\textwidth}{!}{
\renewcommand{\arraystretch}{1.2}
\begin{threeparttable}
\caption{Details of the individual characters of the discrete set of $\mathbf{[3,4]}$ admissible character solutions}\label{t12}
\begin{tabular}{c||ccc||ccc||cc||}
\hline \hline
\rowcolor{Mywhite}
\# & $c$ &  $h_1$ & $h_2$ & $\text{m}_1$ &  $\text{m}_2$ & $\text{m}_3$ & $D_1$ & $D_2$    \\
\hline \hline
\rowcolor{Mygrey}
    S'-1. & $\frac{82}{5}$ & $\frac{41}{60}$ & $\frac{6}{5}$ & 410 & 64739 & 2089934 & 1 & 902\\ \rowcolor{Mywhite}
    S'-2. & 17 & $\frac{17}{24}$ & $\frac{5}{4}$ & 323 & 60860 & 2158575 & 1 & 51\\\rowcolor{Mygrey}
    S'-3. & 18 & $\frac{3}{4}$ & $\frac{4}{3}$ & 234 & 59805 & 2482242 & 1 & 1\\ \rowcolor{Mywhite}
    S'-4. & $\frac{94}{5}$ & $\frac{47}{60}$ & $\frac{7}{5}$ & 188 & 62087 & 2923494 & 1 & 4794\\\rowcolor{Mygrey}
    S'-5. & 20 & $\frac{5}{6}$ & $\frac{3}{2}$ & 140 & 69950 & 3983800 & 1 & 5 \\ \rowcolor{Mywhite}
    S'-6. & $\frac{106}{5}$ & $\frac{53}{60}$ & $\frac{8}{5}$ & 106 & 84429 & 5825442 & 1 & 15847\\\rowcolor{Mygrey}
    S'-7. & 22 & $\frac{11}{12}$ & $\frac{5}{3}$ & 88 & 99935 & 7846300 & 1 & 22\\ \rowcolor{Mywhite}
    S'-8. & 23 & $\frac{23}{24}$ & $\frac{7}{4}$ & 69 & 131905 & 12195106 & 1 & 253\\\rowcolor{Mygrey}
    S'-9. & $\frac{118}{5}$ & $\frac{59}{60}$ & $\frac{9}{5}$ & 59 & 164315 & 16778125 & 1 & 32509\\ \rowcolor{Mywhite}
 \hline \hline
\end{tabular}
\end{threeparttable}
}
\end{center}
\end{table}

After having described the Diophantine aspects of the admissible character solutions, we give further details of the solutions in the table \ref{t12}. These tables collect details about the individual characters. We first give the triple of central charge and conformal dimensions of the two non-identity characters. Then we give the first four Fourier coefficients of the identity character viz. $\text{m}_1$, $\text{m}_2$, $\text{m}_3$ (with $\text{m}_0 =1$). After that we give the leading Fourier coefficient of the two non-identity characters $D_1$ and $D_2$. It is important to note that the leading Fourier coefficient of the non-identity characters could be any positive integral multiple of what we have given because even then the MLDE would be solved. We have given the minimal values for $D_1$ and $D_2$. We have maintained a consistent numbering of the solutions between tables \ref{t11} and tables \ref{t12}.

\subsection{\label{43s}Notes and Comments on $\mathbf{[3,4]}$ Admissible Characters}

We make two observations about all of the $\mathbf{[3,4]}$ admissible character solutions :

(i) From the tables, it is clear that for all the solutions S'-1 to S’-9 have the feature that the index of the first character vanishes i.e. $\alpha_1 = 0$. This means that the for each of the solutions, $h_1 = \frac{c}{24}$. Furthermore, from the explicit $q$-series expansions, available from appendix \ref{app3}, we can see that the character is the trivial one, the number $1$, the ``constant character.’’ Thus every $\mathbf{[3,4]}$ admissible character has one of the characters to be the constant character. 

(ii) Now we examine the non-constant characters. The list of central charges of the $\mathbf{[3,4]}$ admissible characters $\frac{82}{5}, 17, 18, \frac{94}{5}, 20, \frac{106}{5}, 22, 23, \frac{118}{5}$ coincides with the list of the central charges of $\mathbf{[2,2]}$ admissible characters. The list of conformal dimensions of the non-constant non-identity character $\frac{6}{5}, \frac{5}{4}, \frac{4}{3},  \frac{7}{5}, \frac{3}{2}, \frac{8}{5}, \frac{5}{3}, \frac{7}{4}, \frac{9}{5}$ also coincides with those of the  $\mathbf{[2,2]}$ non-identity characters. Furthermore, when we compare the explicit $q$-series of the two non-constant characters (available in appendix \ref{app3}), we find exact match with the two characters of the 
$\mathbf{[2,2]}$ admissible character solutions.

What we are seeing here is a phenomenon that was already observed in \cite{Das:2021uvd}. It was shown there that a $\mathbf{[n,\ell]}$ admissible character adjoined with a constant character is a $\mathbf{[n+1,n + \ell]}$ admissible character. What we are seeing here is for $\mathbf{n} = 2, \mathbf{\ell} = 2$ : a $\mathbf{[2,2]}$ admissible character solution adjoined with a constant character produces a $\mathbf{[3,4]}$ admissible character. This phenomenon was observed in \cite{Das:2021uvd} for $\mathbf{n} = 2, \mathbf{\ell} = 0$.

Note that, when one of the indices vanishes, the modular representation associated with the solution becomes a reducible representation. The $T$-transformation on the sub-space  spanned by the constant character is the identity transformation which implies that the $S$-transformation on that sub-space is also the identity. Thus the modular representation associated with a $\mathbf{[3,4]}$ admissible character is the direct sum of the trivial representation and the two dimensional representation associated with the $\mathbf{[2,2]}$ admissible character. 

The result we seem to have derived, purely from the MLDE and  by way of solving it explicitly, is that there are no $\mathbf{[3,4]}$ admissible characters whose associated modular representations are three-dimensional and irreducible. This kind of a statement is known to be true from studies of modular data for vector-valued-modular-forms  \cite{Kaidi:2021ent}.  It is known that one gets irreducible modular data only for $\mathbf{[3,\ell = 3m]}$, i.e.  when the Wronskian index is a multiple of $3$. In this paper, we are seeing the truth of this result for $\mathbf{\ell} = 4$; the paper \cite{Das:2021uvd} had seen the truth of this result for $\mathbf{\ell} = 2$.

\section{Conclusions and Future Directions \label{5s}}

\subsection*{\underline{$\mathbf{[3,3]}$} :}

In section \ref{3s}, we have obtained the admissible character solutions for the $\mathbf{[3,3]}$ MLDE.  We now note the following points with reference to where in the space of parameters  these solutions exist. 

\noindent \textbf{(i)} It is intriguing that all of them (except S-17) lie
 on the following plane (see table \ref{t2}) :
\be \label{58q}
10368 \, \mu_{1,0} - 1728 \, \mu_{3,-1} - \mu = 0.
\ee
The solution S-17 is, in any case, not a true three-character solution and hence we will ignore it. 

\noindent \textbf{(ii)} Our calculations in section \ref{3s} have established that all solutions live on this plane in the parameter space. We decided to study the $\mathbf{[3,3]}$ MLDE on this plane i.e. we substituted \eqref{58q} into \eqref{21p}. We now have a two-parameter MLDE and is an easier problem technically.  Skipping all the details, we get the analog of \eqref{31p} to be the following :
\be \label{59q}
c = \frac{\text{N}}{70},
\ee
which is remarkable for various reasons. The first is that this means that we need to do only $70$ computations for a central charge range of $1$ (compare to $17817590$ of section \ref{3s}). We do not have to make any restrictions and simply perform all the computations and we arrive at the $\mathbf{[3,3]}$ admissible characters quickly.   The second reason is that \eqref{59q} is the exact relation that was obtained for solutions of the $\mathbf{[3,0]}$ MLDE in \cite{Das:2021uvd}. 

\noindent \textbf{(iii)} Recall that $\mu$ does not appear in the indicial equation \eqref{19p} and hence is the non-rigid parameter while the other two parameters $\mu_{1,0}$ and $\mu_{3,-1}$ are rigid parameters. One way of interpreting equation \eqref{58q} is that the non-rigid parameter is fixed in terms of the rigid parameters. Hence, when we study the $\mathbf{[3,3]}$ MLDE on the plane and we get a two-parameter MLDE, we can think of these two parameters to be rigid parameters. Thus we have a situation similar to the $\mathbf{[3,0]}$ MLDE : a two-parameter MLDE with both parameters rigid. 

Combining the two facts (a) all solutions to the $\mathbf{[3,3]}$ MLDE lie on the plane \eqref{58q}, (b) the MLDE on the plane reduces to a two-parameter rigid MLDE, we can arrive at the following statement : the $\mathbf{[3,3]}$ MLDE is in some sense a two-parameter rigid MLDE in disguise. A better theory of MLDEs and their parameter spaces and solutions, incorporating considerations coming from series-expansions around other cusps (beyond the $\tau = i \infty, q = 0$ considered in this paper), has appeared in \cite{movpole}. Applying that theory to the $\mathbf{[3,3]}$ MLDE  would properly explain this statement\footnote{We thank Arpit Das and Sunil Mukhi for discussions on this point.}.

\subsection*{\underline{$\mathbf{[3,4]}$} :}

In section \ref{4s}, we have obtained the admissible character solutions for the $\mathbf{[3,4]}$ MLDE.  We now note the following points with reference to where in the space of parameters  these solutions exist.

\noindent \textbf{(i)} All of them lie on the following 
 line (see table \ref{t11}) :
\be \label{60q}
\mu_{0,1} = 0, \qquad \mu = -55296.
\ee

\noindent \textbf{(ii)} Our calculations in section \ref{4s} have established that all solutions live on this line in the parameter space. We can now study the $\mathbf{[3,4]}$ MLDE on this line i.e. we substitute \eqref{60q} into \eqref{56p}. We obtain a one-parameter MLDE (with $\mu_{1,0}$ as the parameter.)

\noindent \textbf{(iii)} Recall that $\mu_{1,0}$ is a rigid parameter (it appears in the indicial equation \eqref{55p}). Hence, on the line in parameter space \eqref{60q}, the $\mathbf{[3,4]}$ MLDE reduces to a one-parameter rigid MLDE. This is nothing but the $\mathbf{[2,2]}$ MLDE; one way of seeing this is to see that the indicial equation \eqref{55p}, on the line \eqref{60q}, reduces to the 
$\mathbf{[2,2]}$ indicial equation.

In this paper, we have studied the $\mathbf{[3,3]}$ and $\mathbf{[3,4]}$ MLDEs. We studied them as three-parameter non-rigid MLDEs and obtained their admissible character solutions for $c < 96$.  For the $\mathbf{[3,4]}$ case, the CFT understanding of the  admissible character solutions is clear viz. they are $\mathbf{[2,2]}$ CFTs adjoined with a constant character and as such do not correspond to three-character CFTs with irreducible modular matrices. For the $\mathbf{[3,3]}$ case, the CFT understanding of the admissible character solutions is not yet there. We hope the multiple coset-relations between admissible characters for a variety of meromorphic characters that we have obtained in section \ref{34ss} will contribute to this understanding. We will leave this for future work.

\begin{center}
\textbf{Acknowledgments}
\end{center}

CNG thanks Arpit Das and Sunil Mukhi for collaborations in \cite{Das:2022slz, Das:2022uoe, movpole}; he is especially indebted to the latter for providing understanding and insights into  contemporary RCFT theory.  CNG also thanks Iosif Bena and gratefully acknowledges the hospitality of CEA Saclay where some part of this work was done. CNG further thanks Bobby Acharya, Paolo Creminelli, Atish Dabholkar and gratefully acknowledges the hospitality of the High-Energy Section of the ICTP where some part of this work was done. JS would like to thank Arpit Das, Sunil Mukhi, and Suresh Govindarajan for valuable discussions. He gratefully acknowledges the hospitality of the School of Physical Sciences at NISER, Bhubaneswar, where most of this work was done. He would also like to acknowledge support from the Institute Postdoctoral Fund of IIT Madras.

\newpage

\appendix

\section{The $q$-series of $\mathbf{[3,3]}$ Admissible Characters\label{app1}}

We solved the $\mathbf{[3,3]}$ MLDE and obtained admissible character solutions for $0 < c \leq 96$ in section \ref{3s}. Here, we gather all relevant details of the solution in one place. The infinite families of solutions are in S-16 - S-19 while the discrete set of $15$ are in S-1 - S-15.

\vspace{7mm}\hrule\vspace{1mm}
\noindent \textbf{S-1 :} \\ 
$c = 5$ , $h_1 = \frac{1}{16}$, $h_2 = \frac{9}{16}$. \\
$\alpha_0 = - \frac{5}{24}$, $\alpha_1 = - \frac{7}{48}$, $\alpha_2 =  \frac{17}{48}$. \\
$N = 89087950,~ \text{m}_1 = 27,~\text{m}_2 = 106,~\text{m}_3 = 433,~k = 10342630691459386216552857600$. \\
$D_1 = 1, \quad D_2 = 25.$ \\
$\mu_{1,0} = -\frac{155}{16}, \quad  \mu_{3,-1} = -\frac{595}{32}, \quad \mu = -68310.$ 
\begin{align*}
    \chi_{0}&=q^{-\frac{5}{24}}(1+27q+106q^2+433q^3+1214q^4+3400q^5+8213q^6+O[q]^7)\\
    \chi_{\frac{1}{16}}&=q^{-\frac{7}{48}}(1+53q+404q^2+2082q^3+8259q^4+28319q^5+86679q^6+O[q]^7)\\
    \chi_{\frac{9}{16}}&=q^{\frac{17}{48}}(25+246q+1297q^2+5341q^3+18663q^4+58104q^5+165715q^6+O[q]^7)
\end{align*}
Kaidi-Lin-Parra-Martinez exponent triplet: $\{\frac{17}{48}, \frac{19}{24},\frac{41}{48}\}$\\
\hrule\vspace{2mm}
\noindent \textbf{S-2 :} \\ 
$c = 6$,  $h_1 = \frac18$,  $h_2 = \frac58$. \\
$\alpha_0 = - \frac{1}{4}$, $\alpha_1 = - \frac{1}{8}$, $\alpha_2 =  \frac{3}{8}$. \\
$N = 106905540,~  \text{m}_1 = 26,~ \text{m}_2 = 79, ~ \text{m}_3 = 326, ~k = 12678063428240537942871244800.$ \\
$D_1 = 1, \quad D_2 = 13.$ \\
$ \mu_{1,0} = -\frac{47}{4}, \quad \mu_{3,-1} = -\frac{81}{4}, \quad \mu = -86832.$
\begin{align*}
    \chi_{0}&=q^{-\frac{1}{4}}(1+26q+79q^2+326q^3+755q^4+2106q^5+4460q^6+O[q]^7)\\
    \chi_{\frac{1}{8}}&=q^{-\frac{1}{8}}(1+79q+755q^2+4460q^3+20165q^4+77352q^5+263019q^6+O[q]^7)\\
    \chi_{\frac{5}{8}}&=q^{\frac{3}{8}}(13+163q+1053q^2+5142q^3+20820q^4+73951q^5+237758q^6+O[q]^7)
\end{align*}
Kaidi-Lin-Parra-Martinez exponent triplet: $\{\frac{3}{8}, \frac{3}{4},\frac{7}{8}\}$\\
\hrule\vspace{2mm}
\noindent \textbf{S-3 :} \\ 
\noindent $c = \frac{48}{7}$ , $h_1 = \frac17$, $h_2 = \frac57$. \\
$\alpha_0 = - \frac{2}{7}$, $\alpha_1 = - \frac{1}{7}$, $\alpha_2 =  \frac{3}{8}$. \\
$N = 122177760,~ \text{m}_1 = 78,~ \text{m}_2 = 784,~\text{m}_3 = 5271, ~ k = 13670984061093056577243033600.$ \\
$D_1 = 1, \quad D_2 = 55.$ \\
$\mu_{1,0} = -\frac{116}{7}, \quad \mu_{3,-1} = -\frac{10368}{343}, \quad \mu =-\frac{41015808}{343}.$ 
\begin{align*}
    \chi_{0}&=q^{-\frac{2}{7}}(1+78q+784q^2+5271q^3+26558q^4+113756q^5+426720q^6+O[q]^7)\\
    \chi_{\frac{1}{7}}&=q^{-\frac{1}{7}}(1+133q+1618q^2+11024q^3+56532q^4+240968q^5+901796q^6+O[q]^7)\\
    \chi_{\frac{5}{7}}&=q^{\frac{3}{7}}(55+890q+6720q^2+37344q^3+168077q^4+657147q^5+2303406q^6+O[q]^7)
\end{align*}
Kaidi-Lin-Parra-Martinez exponent triplet: $\{\frac{3}{7}, \frac{5}{7},\frac{6}{7}\}$\\
\hrule\vspace{2mm}
\noindent \textbf{S-4 :} \\ 
\noindent $c = 7$, $h_1 = \frac{3}{16}$,  $h_2 = \frac{11}{16}$. \\
$\alpha_0 = - \frac{7}{24}$, $\alpha_1 = - \frac{5}{48}$, $\alpha_2 =  \frac{19}{48}$. \\
$N = 124723130$,~ $\text{m}_1 = 25$,~$\text{m}_2 = 53$,~$\text{m}_3 = 246$,~$k = 15013496165021689669189632000.$  \\
$D_1 = 1, \quad D_2 = 27$. \\
$\mu_{1,0} = -\frac{227}{16}, \quad \mu_{3,-1} = -\frac{665}{32}, \quad \mu = -111186$.
\begin{align*}
    \chi_{0}&=q^{-\frac{7}{24}}(1+25q+53q^2+246q^3+404q^4+1297q^5+2082q^6+O[q]^7)\\
    \chi_{\frac{3}{16}}&=q^{-\frac{5}{48}}(1+106q+1214q^2+8213q^3+42061q^4+180305q^5+679360q^6+O[q]^7)\\
    \chi_{\frac{11}{16}}&=q^{\frac{19}{48}}(27+433q+3400q^2+19358q^3+89349q^4+356231q^5+1271644q^6+O[q]^7)
\end{align*}
Kaidi-Lin-Parra-Martinez exponent triplet: $\{\frac{19}{48}, \frac{17}{24},\frac{43}{48}\}$\\
\hrule\vspace{2mm}
\noindent \textbf{S-5 :} \\ 
\noindent $c = 8$, $h_1 = \frac14$, $h_2 = \frac34$. \\
$\alpha_0 = - \frac{1}{3}$, $\alpha_1 = - \frac{1}{12}$, $\alpha_2 =  \frac{5}{12}$. \\
$N = 142540720,~ \text{m}_1 = 24,~ \text{m}_2 = 28, \text{m}_3 = 192, k = 17348928901802841395508019200.$  \\
$D_1 = 1,  \quad D_2 = 7.$ \\
$\mu_{1,0} = -17, \quad \mu_{3,-1} = -20 , \quad \mu = -141696$.
\begin{align*}
    \chi_{0}&=q^{-\frac{1}{3}}(1+24q+28q^2+192q^3+134q^4+864q^5+568q^6+O[q]^7)\\
    \chi_{\frac{1}{4}}&=q^{-\frac{1}{12}}(1+134q+1809q^2+13990q^3+80724q^4+384940q^5+1598789q^6+O[q]^7)\\
    \chi_{\frac{3}{4}}&=q^{\frac{5}{12}}(7+142q+1329q^2+8674q^3+45017q^4+199190q^5+781840q^6+O[q]^7)
\end{align*}
Kaidi-Lin-Parra-Martinez exponent triplet: $\{\frac{5}{12},\frac{2}{3},\frac{11}{12}\}$ \\
\hrule\vspace{2mm}
\noindent \textbf{S-6 :} \\ 
$c = 8$, $h_1 = \frac15$, $h_2 = \frac45$. \\
$\alpha_0 = - \frac{1}{3}$, $\alpha_1 = - \frac{2}{15}$, $\alpha_2 =  \frac{7}{15}$. \\
$N = 142540720,~ \text{m}_1 = 134, ~ \text{m}_2 = 1920, ~ \text{m}_3 = 15904, ~ k = 12999967382473207851576960000.$ \\
$D_1 = 1, \quad D_2 = 57.$ \\
$\mu_{1,0} = -\frac{524}{25} , \quad \mu_{3,-1} = -\frac{896}{25}, \quad \mu = -\frac{3884544}{25}.$ 
\begin{align*}
    \chi_{0}&=q^{-\frac{1}{3}}(1+134q+1920q^2+15904q^3+96084q^4+475534q^5+2031292q^6+O[q]^7)\\
    \chi_{\frac{1}{5}}&=q^{-\frac{2}{15}}(1+190q+2832q^2+22497q^3+132298q^4+637052q^5+2666272q^6+O[q]^7)\\
    \chi_{\frac{4}{5}}&=q^{\frac{7}{15}}(57+1159q+10526q^2+67888q^3+348404q^4+1531344q^5+5976508q^6+O[q]^7)
\end{align*}
Kaidi-Lin-Parra-Martinez exponent triplet: $\{\frac{7}{15}, \frac{2}{3},\frac{13}{15}\}$ \\
\hrule\vspace{2mm}
\noindent \textbf{S-7 :} \\ 
\noindent $c = \frac{64}{7}$, $h_1 = \frac27$, $h_2 = \frac67$ \\
$\alpha_0 = - \frac{8}{21}$, $\alpha_1 = - \frac{2}{21}$, $\alpha_2 =  \frac{10}{21}$. \\
$N = 162903680,~ \text{m}_1 = 136 ,~\text{m}_2 = 2417,~\text{m}_3 = 24520,~k = 15375588244086679187252889600.$ \\
$D_1 = 3, \quad D_2 = 117.$ \\
$\mu_{1,0} = -\frac{164}{7} , \quad \mu_{3,-1} = -\frac{10240}{343} , \quad \mu = -\frac{65622528}{343}.$ 
\begin{align*}
    \chi_{0}&=q^{-\frac{13}{21}}(1+136q+2417q^2+24520q^3+173412q^4+982128q^5+4721464q^6+O[q]^7)\\
    \chi_{\frac{2}{7}}&=q^{-\frac{2}{21}}(3+632q+10787q^2+98104q^3+650801q^4+3501336q^5+16220388q^6+O[q]^7)\\
    \chi_{\frac{6}{7}}&=q^{\frac{10}{21}}(117+2952q+32220q^2+239680q^3+1395019q^4+6852776q^5+29613973q^6+O[q]^7)
\end{align*}
Kaidi-Lin-Parra-Martinez exponent triplet:  $\{\frac{10}{21}, \frac{13}{21},\frac{19}{21}\}$\\
\hrule\vspace{2mm}
\noindent \textbf{S-8 :} \\ 
\noindent $c = \frac{104}{7}$, $h_1 = \frac57$,   $h_2 = \frac87$. \\
$\alpha_0 = - \frac{13}{21}$, $\alpha_1 =  \frac{2}{21}$, $\alpha_2 =  \frac{11}{21}$. \\
$N = 264718480,~ \text{m}_1 = 188,~\text{m}_2 = 17260,~\text{m}_3 = 442300,~k = 22829833773044248767796224000.$ \\
$D_1 = 44,\quad D_2 = 725$.\\
$\mu_{1,0} = -44 , \quad \mu_{3,-1} = \frac{18304}{343} , \quad \mu = -\frac{188103168}{343}.$ 
\begin{align*}
    \chi_{0}&=q^{-\frac{13}{21}}(1+188 q+17260 q^2+442300 q^3+6347860 q^4+64800584 q^5+522953984 q^6\\&+O[q]^7)\\
    \chi_{\frac{5}{7}}&=q^{\frac{2}{21}}(44+13002 q+424040 q^2+6945400 q^3+77057840 q^4+660071271 q^5\\&+4681946436 q^6+O[q]^7)\\
    \chi_{\frac{8}{7}}&=q^{\frac{11}{21}}(725+52316 q+1197468 q^2+16227560 q^3+159342950 q^4+1250627760 q^5\\&+8299119768 q^6+O[q]^7)
\end{align*}
Kaidi-Lin-Parra-Martinez exponent triplet:  $\{\frac{2}{21}, \frac{8}{21},\frac{11}{21}\}$ \\
\hrule\vspace{2mm}
\noindent \textbf{S-9 :} \\ 
\noindent $c = 16$, $h_1 = \frac45$,  $h_2 = \frac65$. \\
$\alpha_0 = - \frac{2}{3}$, $\alpha_1 =  \frac{2}{15}$, $\alpha_2 =  \frac{8}{15}$. \\
$N = 285081440,~ \text{m}_1 = 232,~ \text{m}_2 = 31076,~\text{m}_3 = 946432,~k = 22829833773044248767796224000$. \\
$D_1 = 9, \quad D_2 = 154$. \\
$\mu_{1,0} =  -\frac{1244}{25}, \quad \mu_{3,-1} =  \frac{2048}{25}, \quad \mu = -\frac{16436736}{25}.$ 
\begin{align*}
    \chi_{0}&=q^{-\frac{2}{3}}(1+232 q+31076 q^2+946432 q^3+15503250 q^4+176413440 q^5+1564851548 q^6\\&+O[q]^7)\\
    \chi_{\frac{4}{5}}&=q^{\frac{2}{15}}(9+2792 q+101678 q^2+1834576 q^3+22189962 q^4+205600000 q^5+1567892552 q^6\\&+O[q]^7)\\
    \chi_{\frac{6}{5}}&=q^{\frac{8}{15}}(154+13264 q+343781 q^2+5167688 q^3+55549156 q^4+472958432 q^5\\&+3381576787 q^6+O[q]^7)
\end{align*}
Kaidi-Lin-Parra-Martinez exponent triplet: $\{\frac{2}{15}, \frac{1}{3},\frac{8}{15}\}$ \\
\hrule\vspace{2mm}
\noindent \textbf{S-10 :} \\ 
\noindent $c = \frac{120}{7}$, $h_1 = \frac67$,   $h_2 = \frac97$. \\
$\alpha_0 = - \frac{5}{7}$, $\alpha_1 =  \frac{1}{7}$, $\alpha_2 =  \frac{4}{7}$. \\
$N = 305444400,~  \text{m}_1 = 156,~ \text{m}_2 = 28926, ~ \text{m}_3 = 1053508, ~ k = 33196534784141788026353433600.$\\
$D_1 = 78, \quad D_2 = 2108.$ \\
$\mu_{1,0} = -\frac{404}{7}, \quad \mu_{3,-1} = \frac{34560}{343}, \quad \mu = -\frac{264964608}{343}.$ 
\begin{align*}
    \chi_{0}&=q^{-\frac{5}{7}}(1+156 q+28926 q^2+1053508 q^3+19649202 q^4+248807808 q^5+2420830844 q^6\\&+O[q]^7)\\
    \chi_{\frac{6}{7}}&=q^{\frac{1}{7}}(78+28692 q+1194804 q^2+23968816 q^3+317691804 q^4+3193405164 q^5\\&+26228892733 q^6+O[q]^7)\\
    \chi_{\frac{9}{7}}&=q^{\frac{4}{7}}(2108+200787 q+5744052 q^2+94219478 q^3+1096040820 q^4+10033033950 q^5\\&+76726774624 q^6+O[q]^7)
\end{align*}
Kaidi-Lin-Parra-Martinez exponent triplet: $\{\frac{1}{7}, \frac{2}{7}, \frac{4}{7}\}$ \\
\hrule\vspace{2mm}
\noindent \textbf{S-11 :} \\ 
\noindent $c = \frac{160}{7}$ , $h_1 = \frac87$,  $h_2 = \frac{12}{7}$. \\
$\alpha_0 = - \frac{20}{21}$, $\alpha_1 =  \frac{4}{21}$, $\alpha_2 =  \frac{16}{21}$. \\
$N = 407259200,~ \text{m}_1 = 40,~\text{m}_2 = 60440,~\text{m}_3 = 5474720,~k = 80410293873630197603223552000.$ \\
$D_1 = 285, \quad D_2 = 27170.$ \\
$\mu_{1,0} = -\frac{740}{7}, \quad \mu_{3,-1} = \frac{81920}{343}, \quad \mu = -\frac{517501440}{343}.$ 
\begin{align*}
    \chi_{0}&=q^{-\frac{20}{21}}(1+40 q+60440 q^2+5474720 q^3+193667650 q^4+4102819328 q^5+62020336720 q^6\\&+O[q]^7)\\
    \chi_{\frac{8}{7}}&=q^{\frac{4}{21}}(285+227848 q+17128120 q^2+553982240 q^3+11094069220 q^4+161218786320 q^5\\&+1853338543872 q^6+O[q]^7)\\
    \chi_{\frac{12}{7}}&=q^{\frac{16}{21}}(27170+3857360 q+167410480 q^2+4010153280 q^3+65969584027 q^4\\&+831863960600 q^5+8578248681160 q^6+O[q]^7)
\end{align*}
Kaidi-Lin-Parra-Martinez exponent triplet: $\{\frac{1}{21},\frac{4}{21},\frac{16}{21}\}$ \\
\hrule\vspace{2mm}
\noindent \textbf{S-12 :} \\ 
\noindent $c = 24$, $h_1 = \frac65$,  $h_2 = \frac95$. \\
$\alpha_0 = - 1$, $\alpha_1 =  \frac{1}{5}$, $\alpha_2 =  \frac{4}{5}$. \\
$N = 427622160,~ \text{m}_1 = 30,~\text{m}_2 = 87786,~\text{m}_3 =  9614200,~k= 98877947399322534305678592000.$\\
$D_1 = 11, \quad D_2 = 1102.$ \\
$\mu_{1,0} = -\frac{2924}{25}, \quad \mu_{3,-1} = \frac{6912}{25}, \quad \mu = -\frac{42259968}{25}.$ 
\begin{align*}
    \chi_{0}&=q^{-1}(1+30 q+87786 q^2+9614200 q^3+386516850 q^4+9054264996 q^5+149012587830 q^6\\&+O[q]^7)\\
    \chi_{\frac{6}{5}}&=q^{\frac{1}{5}}(11+10212 q+853137 q^2+30093406 q^3+649724292 q^4+10099481832 q^5\\&+123468053842 q^6+O[q]^7)\\
    \chi_{\frac{9}{5}}&=q^{\frac{4}{5}}(1102+166953 q+7778844 q^2+199098630 q^3+3482557140 q^4+46497487770 q^5\\&+505898532064 q^6+O[q]^7)
\end{align*}
Kaidi-Lin-Parra-Martinez exponent triplet: $\{0,\frac{1}{5},\frac{4}{5}\}$\\
\hrule\vspace{2mm}
\noindent \textbf{S-13 :} \\ 
\noindent $c = \frac{176}{7}$,  $h_1 = \frac97$,  $h_2 = \frac{13}{7}$. \\
$\alpha_0 = - \frac{22}{21}$, $\alpha_1 =  \frac{5}{21}$, $\alpha_2 =  \frac{17}{21}$. \\
$N = 447985120,~ \text{m}_1 = 14,~ \text{m}_2 = 66512, ~\text{m}_3 =  8878186, ~k = 90093424404042464671113523200.$\\
$D_1 = 782, \quad D_2 = 50922.$ \\
$\mu_{1,0} =-\frac{884}{7}, \quad \mu_{3,-1} = \frac{119680}{343}, \quad \mu = -\frac{655907328}{343}.$ 
\begin{align*}
    \chi_{0}&=q^{-\frac{22}{21}}(1+14 q+66512 q^2+8878186 q^3+405729320 q^4+10505611404 q^5+188103111104 q^6\\&+O[q]^7)\\
    \chi_{\frac{9}{7}}&=q^{\frac{5}{21}}(782+718267 q+64206178 q^2+2419951472 q^3+55577638246 q^4+915050772000 q^5\\&+11805394882804 q^6+O[q]^7)\\
    \chi_{\frac{13}{7}}&=q^{\frac{17}{21}}(50922+8656740 q+441429616 q^2+12203476160 q^3+228549023426 q^4\\&+3246661592893 q^5+37402350252062 q^6+O[q]^7)
\end{align*}
Kaidi-Lin-Parra-Martinez exponent triplet: $\{\frac{5}{21},\frac{17}{21}, \frac{20}{21}\}$\\
\hrule\vspace{2mm}
\noindent \textbf{S-14 :} \\ 
\noindent $c = \frac{216}{7}$, $h_1 = \frac{12}{7}$,  $h_2 = \frac{15}{7}$. \\
$\alpha_0 = - \frac{9}{7}$, $\alpha_1 =  \frac{3}{7}$, $\alpha_2 =  \frac{6}{7}$. \\
$N = 549799920,~ \text{m}_1 = 3,~ \text{m}_2 = 52254,~\text{m}_3 = 20440112,~k = 75231152349388320481013664000.$ \\
$D_1 = 11495, \quad D_2 = 260623.$ \\
$\mu_{1,0} = -\frac{1268}{7}, \quad \mu_{3,-1} = \frac{279936}{343}, \quad \mu = -\frac{1127913984}{343}.$ 
\begin{align*}
    \chi_{0}&=q^{-\frac{9}{7}}(1+3 q+52254 q^2+20440112 q^3+1800753741 q^4+76812566202 q^5\\&+2081358133988 q^6+O[q]^7)\\
    \chi_{\frac{12}{7}}&=q^{\frac{3}{7}}(11495+10341870 q+1234169640 q^2+62369273760 q^3+1890682123182 q^4\\&+40398690731121 q^5+666181065657558 q^6+O[q]^7)\\
    \chi_{\frac{15}{7}}&=q^{\frac{6}{7}}(260623+74348634 q+5748163632 q^2+227080834606 q^3+5841200182488 q^4\\&+110731961943684 q^5+1665028734349888 q^6+O[q]^7)
\end{align*}
Kaidi-Lin-Parra-Martinez exponent triplet: $\{\frac{3}{7}, \frac{5}{7}, \frac{6}{7}\}$\\
\hrule\vspace{2mm}
\noindent \textbf{S-15 :} \\ 
\noindent $c = 32$, $h_1 = \frac95$,  $h_2 = \frac{11}{5}.$ \\
$\alpha_0 = -\frac{4}{3}$, $\alpha_1 =  \frac{7}{15}$, $\alpha_2 =  \frac{13}{15}$. \\
$N = 570162880,~ \text{m}_1 = 3,~ \text{m}_2 = 62500,~\text{m}_3 = 31015600,~k = 74986027395393703935850560000.$ \\
$D_1 = 19, \quad D_2 = 434.$ \\
$\mu_{1,0} = -\frac{4844}{25}, \quad \mu_{3,-1} = \frac{23296}{25}, \quad \mu = -\frac{18095616}{5}.$ 
\begin{align*}
    \chi_{0}&=q^{-\frac{4}{3}}(1+3 q+62500 q^2+31015600 q^3+3129750000 q^4+147591171852 q^5\\&+4341193437688 q^6+O[q]^7)\\
    \chi_{\frac{9}{5}}&=q^{\frac{7}{15}}(19+17100 q+2144482 q^2+114172642 q^3+3638878124 q^4+81532613800 q^5\\&+1406310849764 q^6+O[q]^7)\\
    \chi_{\frac{11}{5}}&=q^{\frac{13}{15}}(434+135997 q+11356819 q^2+479480100 q^3+13086623892 q^4+261842159600 q^5\\&+4138618150431 q^6+O[q]^7)
\end{align*}
Kaidi-Lin-Parra-Martinez exponent triplet: $\{\frac{7}{15}, \frac{2}{3},\frac{13}{15}\}$\\
\hrule\vspace{2mm}
\noindent \textbf{S-16 :} \\ 
\noindent $c = 12$, $h_1 = \frac12$,  $h_2$ = 1 \\
$\alpha_0 = -\frac{1}{2}$, $\alpha_1 =  0$, $\alpha_2 =  \frac{1}{2}$. \\
$N = 213811080,~ \text{m}_1 = 22+n,~ \text{m}_2 = 26 + 44 n, ~\text{m}_3 = 1652 + 718 n,$\\  $k = -162906859429488820195200 (-163452 + 196 n + n^2).$  \\
$D_1 = 1, \quad D_2 = 1.$ \\
$\mu_{1,0} = -32, \quad \mu_{3,-1} = 0, \quad \mu = -331776.$ 
\begin{align*}
    \chi_{0}&=q^{-\frac{1}{2}}(1+ (22 + n)q+ (26 + 44 n)q^2+ (1652 + 718 n)q^3+ (14063 + 7352 n)q^4+ (114734 + 56549 n)q^5\\&+ (708998 + 356388 n)q^6+O[q]^7)\\
    \chi_{\frac{1}{2}}&=q^{0}(1+256 q+6144 q^2+76800 q^3+671744 q^4+4640256 q^5+27009024 q^6+O[q]^7)\\
    \chi_{1}&=q^{\frac{1}{2}}(1+44 q+718 q^2+7352 q^3+56549 q^4+356388 q^5+1934534 q^6+O[q]^7)
\end{align*}
Kaidi-Lin-Parra-Martinez exponent triplet: not available\\
\hrule\vspace{2mm}
\noindent \textbf{S-17 :} \\ 
\noindent $c = 16$, $h_1 = 1$, $h_2$ = 1 \\
$\alpha_0 = -\frac{2}{3}$, $\alpha_1 = \frac13$, $\alpha_2 =  \frac{1}{3}$. \\
$N = 285081440,~\text{m}_1 = 497 + n,~\text{m}_2 = 69912 + 160 n,~\text{m}_3 =  2120356 + 5348 n,~k = 0.$ \\
$D_1 = 1, \quad D_2 = 1.$ \\
$\mu_{1,0} = -44, \quad \mu_{3,-1} = 128, \quad \mu = -718848.$ 
\begin{align*}
    \chi_{0}&=q^{-\frac{2}{3}}(1+ (497 + n)q+ (69912 + 160 n)q^2+ (2120356 + 5348 n)q^3+ (34812700 + 142080 n)q^4\\& + (405421546 + 10961546 n)q^5+ (5668339904 + 2169191680 n)q^6+O[q]^7)\\
    \chi_{1}&=q^{\frac{1}{3}}(1+160 q+5348 q^2+142080 q^3+10961546 q^4+2169191680 q^5+555552219992 q^6+O[q]^7)\\
    \chi_{1}&=q^{\frac{1}{3}}(1+160 q+5348 q^2+142080 q^3+10961546 q^4+2169191680 q^5+555552219992 q^6+O[q]^7)
\end{align*}
Kaidi-Lin-Parra-Martinez exponent triplet: not available\\
\hrule\vspace{2mm}
\noindent \textbf{S-18 :} \\ 
\noindent $c = 20$, $h_1 = 1$  $h_2 = \frac32.$ \\
$\alpha_0 = -\frac{5}{6}$, $\alpha_1 = \frac16$, $\alpha_2 =  \frac{2}{3}$. \\
$N = 356351800,~ \text{m}_1 = 13+n,~\text{m}_2 = 354 + 548n,~\text{m}_3 =  32322 + 31114 n,$\\ $k = -162906859429488820195200 (-279359 - 830 n + n^2).$ \\
$D_1 = 1, \quad D_2 = 5.$ \\ 
$\mu_{1,0} = -80, \quad \mu_{3,-1} = 160, \quad \mu = -1105920.$ 
\begin{align*}
    \chi_{0}&=q^{-\frac{5}{6}}(1+ (13 + n)q+ (354 + 548 n)q^2+ (32322 + 31114 n)q^3+ (795221 + 800472 n)q^4\\& + (13170199 + 13151027 n)q^5+ (160042698 + 160104836 n)q^6+O[q]^7)\\
    \chi_{1}&=q^{\frac{1}{6}}(1+548 q+31114 q^2+800472 q^3+13151027 q^4+160104836 q^5+1565937470 q^6+O[q]^7)\\
    \chi_{\frac{3}{2}}&=q^{\frac{2}{3}}(5+592 q+21160 q^2+424000 q^3+5918900 q^4+64093600 q^5+573188736 q^6+O[q]^7)
\end{align*}
Kaidi-Lin-Parra-Martinez exponent triplet: not available\\
\hrule\vspace{2mm}
\noindent \textbf{S-19 :} \\ 
\noindent $c = 28$,  $h_1 = \frac32$, $h_2 = 2$. \\
$\alpha_0 = -\frac{7}{6}$, $\alpha_1 = \frac13$, $\alpha_2 =  \frac{5}{6}$. \\
$N = 498892520,~ \text{m}_1 = 4,~\text{m}_2 = 2 + 3n,~\text{m}_3 =  61816 + 668 n,$ \\ $k = 977441156576932921171200 (65624 + n).$ \\
$D_1 = 1, \quad D_2 = 3.$ \\
$\mu_{1,0} = -152, \quad \mu_{3,-1} = 560, \quad \mu = -2543616.$ 
\begin{align*}
    \chi_{0}&=q^{-\frac{7}{6}}(1+ 4q+ (2 + 3 n)q^2+ (61816 + 668 n)q^3+ (3695259 + 42230 n)q^4+ (123596676 + 1403272 n)q^5\\&+ (2722658614 + 30944629 n)q^6+O[q]^7)\\
    \chi_{\frac{3}{2}}&=q^{\frac{1}{3}}(1+904 q+94244 q^2+4143232 q^3+109986186 q^4+2073681984 q^5+30385613528 q^6+O[q]^7)\\
    \chi_{2}&=q^{\frac{5}{6}}(3+668 q+42230 q^2+1403272 q^3+30944629 q^4+509921900 q^5+6736667386 q^6+O[q]^7)
\end{align*}
Kaidi-Lin-Parra-Martinez exponent triplet: not available\\
\hrule\vspace{2mm}

\section[{Kaidi-Lin-Parra-Martinez Exponent-triplets and their Index-sets \label{app2}}]{\resizebox{\textwidth}{!}{Kaidi-Lin-Parra-Martinez Exponent-triplets and their Index-sets \label{app2}}}

This appendix adjoins section \ref{332ss}. In the paper \cite{Kaidi:2021ent}, the authors give a list of exponent-triplets for $\mathbf{[3,3]}$ admissible characters. We have reproduced this list in table \ref{t5}. We have explained in section \ref{332ss} how each of these exponent-triplets results in an infinite number of index-sets. We only need those index-sets which have $0 < c \leq 96$. Here, in the following table, we list all such index-sets for every KLP exponent-triplet. In the left column we give the exponent-triplets and in the right column the corresponding index-sets. The left column also gives the number of index-sets, which is always either $66$ or $78$. 

\begin{table}[H]
   \caption{Kaidi-Lin-Parra-Martinez Exponent-triplets and their Index-sets} \label{app2t1}
\begin{center}
\begin{threeparttable}
\resizebox{\textwidth}{9.2cm}{
   \renewcommand{\arraystretch}{1.3}
   $
\begin{array}{c|l}
\rowcolor{Mygrey}&\left\{8,\frac{1}{3},\frac{2}{3}\right\} ,\left\{16,\frac{1}{3},\frac{5}{3}\right\},\left\{16,\frac{2}{3},\frac{4}{3}\right\} ,\left\{24,\frac{1}{3},\frac{8}{3}\right\},\left\{24,\frac{2}{3},\frac{7}{3}\right\},\left\{24,\frac{4}{3},\frac{5}{3}\right\},\left\{32,\frac{1}{3},\frac{11}{3}\right\},\left\{32,\frac{2}{3},\frac{10}{3}\right\},\left\{32,\frac{4}{3},\frac{8}{3}\right\},\left\{32,\frac{5}{3},\frac{7}{3}\right\}\\
\rowcolor{Mygrey}&\left\{40,\frac{1}{3},\frac{14}{3}\right\},\left\{40,\frac{2}{3},\frac{13}{3}\right\},\left\{40,\frac{4}{3},\frac{11}{3}\right\},\left\{40,\frac{5}{3},\frac{10}{3}\right\},\left\{40,\frac{7}{3},\frac{8}{3}\right\},\left\{48,\frac{1}{3},\frac{17}{3}\right\},\left\{48,\frac{2}{3},\frac{16}{3}\right\},\left\{48,\frac{4}{3},\frac{14}{3}\right\},\left\{48,\frac{5}{3},\frac{13}{3}\right\},\left\{48,\frac{7}{3},\frac{11}{3}\right\}\\
\rowcolor{Mygrey}&\left\{48,\frac{8}{3},\frac{10}{3}\right\},\left\{56,\frac{1}{3},\frac{20}{3}\right\},\left\{56,\frac{2}{3},\frac{19}{3}\right\},\left\{56,\frac{4}{3},\frac{17}{3}\right\},\left\{56,\frac{5}{3},\frac{16}{3}\right\},\left\{56,\frac{7}{3},\frac{14}{3}\right\},\left\{56,\frac{8}{3},\frac{13}{3}\right\},\left\{56,\frac{10}{3},\frac{11}{3}\right\},\left\{64,\frac{1}{3},\frac{23}{3}\right\},\left\{64,\frac{2}{3},\frac{22}{3}\right\}\\
\rowcolor{Mygrey}&\left\{64,\frac{4}{3},\frac{20}{3}\right\},\left\{64,\frac{5}{3},\frac{19}{3}\right\},\left\{64,\frac{7}{3},\frac{17}{3}\right\},\left\{64,\frac{8}{3},\frac{16}{3}\right\},\left\{64,\frac{10}{3},\frac{14}{3}\right\},\left\{64,\frac{11}{3},\frac{13}{3}\right\},\left\{72,\frac{1}{3},\frac{26}{3}\right\},\left\{72,\frac{2}{3},\frac{25}{3}\right\},\left\{72,\frac{4}{3},\frac{23}{3}\right\},\left\{72,\frac{5}{3},\frac{22}{3}\right\}\\
\end{array}$
}
\end{threeparttable}
\end{center}
\end{table}

\begin{table}[H]
\begin{center}
\begin{threeparttable}
\resizebox{\textwidth}{11cm}{
   \renewcommand{\arraystretch}{1.3}
   $\begin{array}{c|l}
\rowcolor{Mygrey}&\left\{\frac{72}{7},\frac{4}{7},\frac{5}{7}\right\},\left\{\frac{120}{7},\frac{2}{7},\frac{13}{7}\right\},\left\{\frac{120}{7},\frac{6}{7},\frac{9}{7}\right\},\left\{\frac{144}{7},\frac{1}{7},\frac{17}{7}\right\},\left\{\frac{144}{7},\frac{3}{7},\frac{15}{7}\right\},\left\{\frac{144}{7},\frac{8}{7},\frac{10}{7}\right\},\left\{\frac{240}{7},\frac{4}{7},\frac{26}{7}\right\},\left\{\frac{240}{7},\frac{5}{7},\frac{25}{7}\right\},\left\{\frac{240}{7},\frac{11}{7},\frac{19}{7}\right\}\\
\rowcolor{Mygrey}&\left\{\frac{240}{7},\frac{12}{7},\frac{18}{7}\right\},\left\{\frac{288}{7},\frac{2}{7},\frac{34}{7}\right\},\left\{\frac{288}{7},\frac{6}{7},\frac{30}{7}\right\},\left\{\frac{288}{7},\frac{9}{7},\frac{27}{7}\right\},\left\{\frac{288}{7},\frac{13}{7},\frac{23}{7}\right\},\left\{\frac{288}{7},\frac{16}{7},\frac{20}{7}\right\},\left\{\frac{312}{7},\frac{1}{7},\frac{38}{7}\right\},\left\{\frac{312}{7},\frac{3}{7},\frac{36}{7}\right\},\left\{\frac{312}{7},\frac{8}{7},\frac{31}{7}\right\}\\
\rowcolor{Mygrey}&\left\{\frac{312}{7},\frac{10}{7},\frac{29}{7}\right\},\left\{\frac{312}{7},\frac{15}{7},\frac{24}{7}\right\},\left\{\frac{312}{7},\frac{17}{7},\frac{22}{7}\right\},\left\{\frac{408}{7},\frac{4}{7},\frac{47}{7}\right\},\left\{\frac{408}{7},\frac{5}{7},\frac{46}{7}\right\},\left\{\frac{408}{7},\frac{11}{7},\frac{40}{7}\right\},\left\{\frac{408}{7},\frac{12}{7},\frac{39}{7}\right\},\left\{\frac{408}{7},\frac{18}{7},\frac{33}{7}\right\},\left\{\frac{408}{7},\frac{19}{7},\frac{32}{7}\right\}\\
\end{array}$
}
\end{threeparttable}
\end{center}
\end{table}

\begin{table}[H]
\begin{center}
\begin{threeparttable}
\resizebox{\textwidth}{11cm}{
   \renewcommand{\arraystretch}{1.3}
   $\begin{array}{c|l}
\rowcolor{Mygrey}&\left\{8,\frac{1}{4},\frac{3}{4}\right\},\left\{14,\frac{1}{4},\frac{3}{2}\right\},\left\{14,\frac{1}{2},\frac{5}{4}\right\},\left\{26,\frac{1}{2},\frac{11}{4}\right\},\left\{26,\frac{3}{4},\frac{5}{2}\right\},\left\{26,\frac{3}{2},\frac{7}{4}\right\},\left\{32,\frac{1}{4},\frac{15}{4}\right\},\left\{32,\frac{3}{4},\frac{13}{4}\right\},\left\{32,\frac{5}{4},\frac{11}{4}\right\},\left\{32,\frac{7}{4},\frac{9}{4}\right\},\left\{38,\frac{1}{4},\frac{9}{2}\right\}\\
\rowcolor{Mygrey}&\left\{38,\frac{1}{2},\frac{17}{4}\right\},\left\{38,\frac{5}{4},\frac{7}{2}\right\},\left\{38,\frac{3}{2},\frac{13}{4}\right\},\left\{38,\frac{9}{4},\frac{5}{2}\right\},\left\{50,\frac{1}{2},\frac{23}{4}\right\},\left\{50,\frac{3}{4},\frac{11}{2}\right\},\left\{50,\frac{3}{2},\frac{19}{4}\right\},\left\{50,\frac{7}{4},\frac{9}{2}\right\},\left\{50,\frac{5}{2},\frac{15}{4}\right\},\left\{50,\frac{11}{4},\frac{7}{2}\right\}\\
\rowcolor{Mygrey}&\left\{56,\frac{1}{4},\frac{27}{4}\right\},\left\{56,\frac{3}{4},\frac{25}{4}\right\},\left\{56,\frac{5}{4},\frac{23}{4}\right\},\left\{56,\frac{7}{4},\frac{21}{4}\right\},\left\{56,\frac{9}{4},\frac{19}{4}\right\},\left\{56,\frac{11}{4},\frac{17}{4}\right\},\left\{56,\frac{13}{4},\frac{15}{4}\right\},\left\{62,\frac{1}{4},\frac{15}{2}\right\},\left\{62,\frac{1}{2},\frac{29}{4}\right\},\left\{62,\frac{5}{4},\frac{13}{2}\right\}\\
\rowcolor{Mygrey}&\left\{62,\frac{3}{2},\frac{25}{4}\right\},\left\{62,\frac{9}{4},\frac{11}{2}\right\},\left\{62,\frac{5}{2},\frac{21}{4}\right\},\left\{62,\frac{13}{4},\frac{9}{2}\right\},\left\{62,\frac{7}{2},\frac{17}{4}\right\},\left\{74,\frac{1}{2},\frac{35}{4}\right\},\left\{74,\frac{3}{4},\frac{17}{2}\right\},\left\{74,\frac{3}{2},\frac{31}{4}\right\},\left\{74,\frac{7}{4},\frac{15}{2}\right\},\left\{74,\frac{5}{2},\frac{27}{4}\right\}\\
\end{array}$
}
\end{threeparttable}
\end{center}
\end{table}

\begin{table}[H]
\begin{center}
\begin{threeparttable}
\resizebox{\textwidth}{11cm}{
   \renewcommand{\arraystretch}{1.3}
   $\begin{array}{c|l}
\rowcolor{Mygrey}&\left\{\frac{21}{2},\frac{1}{2},\frac{13}{16}\right\},\left\{15,\frac{3}{16},\frac{27}{16}\right\},\left\{15,\frac{11}{16},\frac{19}{16}\right\},\left\{\frac{45}{2},\frac{5}{16},\frac{5}{2}\right\},\left\{\frac{45}{2},\frac{1}{2},\frac{37}{16}\right\},\left\{\frac{45}{2},\frac{21}{16},\frac{3}{2}\right\},\left\{\frac{69}{2},\frac{1}{2},\frac{61}{16}\right\},\left\{\frac{69}{2},\frac{13}{16},\frac{7}{2}\right\},\left\{\frac{69}{2},\frac{3}{2},\frac{45}{16}\right\},\left\{\frac{69}{2},\frac{29}{16},\frac{5}{2}\right\}\\
\rowcolor{Mygrey}&\left\{39,\frac{3}{16},\frac{75}{16}\right\},\left\{39,\frac{11}{16},\frac{67}{16}\right\},\left\{39,\frac{19}{16},\frac{59}{16}\right\},\left\{39,\frac{27}{16},\frac{51}{16}\right\},\left\{39,\frac{35}{16},\frac{43}{16}\right\},\left\{\frac{93}{2},\frac{5}{16},\frac{11}{2}\right\},\left\{\frac{93}{2},\frac{1}{2},\frac{85}{16}\right\},\left\{\frac{93}{2},\frac{21}{16},\frac{9}{2}\right\},\left\{\frac{93}{2},\frac{3}{2},\frac{69}{16}\right\},\left\{\frac{93}{2},\frac{37}{16},\frac{7}{2}\right\}\\
\rowcolor{Mygrey}&\left\{\frac{93}{2},\frac{5}{2},\frac{53}{16}\right\},\left\{\frac{117}{2},\frac{1}{2},\frac{109}{16}\right\},\left\{\frac{117}{2},\frac{13}{16},\frac{13}{2}\right\},\left\{\frac{117}{2},\frac{3}{2},\frac{93}{16}\right\},\left\{\frac{117}{2},\frac{29}{16},\frac{11}{2}\right\},\left\{\frac{117}{2},\frac{5}{2},\frac{77}{16}\right\},\left\{\frac{117}{2},\frac{45}{16},\frac{9}{2}\right\},\left\{\frac{117}{2},\frac{7}{2},\frac{61}{16}\right\},\left\{63,\frac{3}{16},\frac{123}{16}\right\}\\
\end{array}$
}
\end{threeparttable}
\end{center}
\end{table}

\begin{table}[H]
\begin{center}
\begin{threeparttable}
\resizebox{\textwidth}{11cm}{
   \renewcommand{\arraystretch}{1.3}
   $\begin{array}{c|l}
\rowcolor{Mygrey}&\left\{\frac{80}{7},\frac{4}{7},\frac{6}{7}\right\},\left\{\frac{104}{7},\frac{1}{7},\frac{12}{7}\right\},\left\{\frac{104}{7},\frac{5}{7},\frac{8}{7}\right\},\left\{\frac{152}{7},\frac{2}{7},\frac{17}{7}\right\},\left\{\frac{152}{7},\frac{3}{7},\frac{16}{7}\right\},\left\{\frac{152}{7},\frac{9}{7},\frac{10}{7}\right\},\left\{\frac{248}{7},\frac{4}{7},\frac{27}{7}\right\},\left\{\frac{248}{7},\frac{6}{7},\frac{25}{7}\right\},\left\{\frac{248}{7},\frac{11}{7},\frac{20}{7}\right\},\left\{\frac{248}{7},\frac{13}{7},\frac{18}{7}\right\}\\
\rowcolor{Mygrey}&\left\{\frac{272}{7},\frac{1}{7},\frac{33}{7}\right\},\left\{\frac{272}{7},\frac{5}{7},\frac{29}{7}\right\},\left\{\frac{272}{7},\frac{8}{7},\frac{26}{7}\right\},\left\{\frac{272}{7},\frac{12}{7},\frac{22}{7}\right\},\left\{\frac{272}{7},\frac{15}{7},\frac{19}{7}\right\},\left\{\frac{320}{7},\frac{2}{7},\frac{38}{7}\right\},\left\{\frac{320}{7},\frac{3}{7},\frac{37}{7}\right\},\left\{\frac{320}{7},\frac{9}{7},\frac{31}{7}\right\},\left\{\frac{320}{7},\frac{10}{7},\frac{30}{7}\right\}\\
\rowcolor{Mygrey}&\left\{\frac{320}{7},\frac{16}{7},\frac{24}{7}\right\},\left\{\frac{320}{7},\frac{17}{7},\frac{23}{7}\right\},\left\{\frac{416}{7},\frac{4}{7},\frac{48}{7}\right\},\left\{\frac{416}{7},\frac{6}{7},\frac{46}{7}\right\},\left\{\frac{416}{7},\frac{11}{7},\frac{41}{7}\right\},\left\{\frac{416}{7},\frac{13}{7},\frac{39}{7}\right\},\left\{\frac{416}{7},\frac{18}{7},\frac{34}{7}\right\},\left\{\frac{416}{7},\frac{20}{7},\frac{32}{7}\right\},\left\{\frac{416}{7},\frac{25}{7},\frac{27}{7}\right\}\\
\end{array}$
}
\end{threeparttable}
\end{center}
\end{table}

\begin{table}[H]
\begin{center}
\begin{threeparttable}
\resizebox{\textwidth}{11cm}{
   \renewcommand{\arraystretch}{1.3}
   $\begin{array}{c|l}
\rowcolor{Mygrey}&\left\{5,\frac{1}{8},\frac{1}{2}\right\},\left\{17,\frac{1}{2},\frac{13}{8}\right\},\left\{17,\frac{5}{8},\frac{3}{2}\right\},\left\{26,\frac{3}{8},\frac{23}{8}\right\},\left\{26,\frac{7}{8},\frac{19}{8}\right\},\left\{26,\frac{11}{8},\frac{15}{8}\right\},\left\{29,\frac{1}{8},\frac{7}{2}\right\},\left\{29,\frac{1}{2},\frac{25}{8}\right\},\left\{29,\frac{9}{8},\frac{5}{2}\right\},\left\{29,\frac{3}{2},\frac{17}{8}\right\}\\
\rowcolor{Mygrey}&\left\{41,\frac{1}{2},\frac{37}{8}\right\},\left\{41,\frac{5}{8},\frac{9}{2}\right\},\left\{41,\frac{3}{2},\frac{29}{8}\right\},\left\{41,\frac{13}{8},\frac{7}{2}\right\},\left\{41,\frac{5}{2},\frac{21}{8}\right\},\left\{50,\frac{3}{8},\frac{47}{8}\right\},\left\{50,\frac{7}{8},\frac{43}{8}\right\},\left\{50,\frac{11}{8},\frac{39}{8}\right\},\left\{50,\frac{15}{8},\frac{35}{8}\right\},\left\{50,\frac{19}{8},\frac{31}{8}\right\}\\
\rowcolor{Mygrey}&\left\{50,\frac{23}{8},\frac{27}{8}\right\},\left\{53,\frac{1}{8},\frac{13}{2}\right\},\left\{53,\frac{1}{2},\frac{49}{8}\right\},\left\{53,\frac{9}{8},\frac{11}{2}\right\},\left\{53,\frac{3}{2},\frac{41}{8}\right\},\left\{53,\frac{17}{8},\frac{9}{2}\right\},\left\{53,\frac{5}{2},\frac{33}{8}\right\},\left\{53,\frac{25}{8},\frac{7}{2}\right\},\left\{65,\frac{1}{2},\frac{61}{8}\right\},\left\{65,\frac{5}{8},\frac{15}{2}\right\}\\
\rowcolor{Mygrey}&\left\{65,\frac{3}{2},\frac{53}{8}\right\},\left\{65,\frac{13}{8},\frac{13}{2}\right\},\left\{65,\frac{5}{2},\frac{45}{8}\right\},\left\{65,\frac{21}{8},\frac{11}{2}\right\},\left\{65,\frac{7}{2},\frac{37}{8}\right\},\left\{65,\frac{29}{8},\frac{9}{2}\right\},\left\{74,\frac{3}{8},\frac{71}{8}\right\},\left\{74,\frac{7}{8},\frac{67}{8}\right\},\left\{74,\frac{11}{8},\frac{63}{8}\right\},\left\{74,\frac{15}{8},\frac{59}{8}\right\}\\
\end{array}$
}
\end{threeparttable}
\end{center}
\end{table}

\begin{table}[H]
\begin{center}
\begin{threeparttable}
\resizebox{\textwidth}{11cm}{
   \renewcommand{\arraystretch}{1.3}
   $\begin{array}{c|l}
\rowcolor{Mygrey}&\left\{11,\frac{7}{16},\frac{15}{16}\right\},\left\{\frac{25}{2},\frac{1}{16},\frac{3}{2}\right\},\left\{\frac{25}{2},\frac{1}{2},\frac{17}{16}\right\},\left\{\frac{49}{2},\frac{1}{2},\frac{41}{16}\right\},\left\{\frac{49}{2},\frac{9}{16},\frac{5}{2}\right\},\left\{\frac{49}{2},\frac{3}{2},\frac{25}{16}\right\},\left\{35,\frac{7}{16},\frac{63}{16}\right\},\left\{35,\frac{15}{16},\frac{55}{16}\right\},\left\{35,\frac{23}{16},\frac{47}{16}\right\},\left\{35,\frac{31}{16},\frac{39}{16}\right\}\\
\rowcolor{Mygrey}&\left\{\frac{73}{2},\frac{1}{16},\frac{9}{2}\right\},\left\{\frac{73}{2},\frac{1}{2},\frac{65}{16}\right\},\left\{\frac{73}{2},\frac{17}{16},\frac{7}{2}\right\},\left\{\frac{73}{2},\frac{3}{2},\frac{49}{16}\right\},\left\{\frac{73}{2},\frac{33}{16},\frac{5}{2}\right\},\left\{\frac{97}{2},\frac{1}{2},\frac{89}{16}\right\},\left\{\frac{97}{2},\frac{9}{16},\frac{11}{2}\right\},\left\{\frac{97}{2},\frac{3}{2},\frac{73}{16}\right\},\left\{\frac{97}{2},\frac{25}{16},\frac{9}{2}\right\},\left\{\frac{97}{2},\frac{5}{2},\frac{57}{16}\right\}\\
\rowcolor{Mygrey}&\left\{\frac{97}{2},\frac{41}{16},\frac{7}{2}\right\},\left\{59,\frac{7}{16},\frac{111}{16}\right\},\left\{59,\frac{15}{16},\frac{103}{16}\right\},\left\{59,\frac{23}{16},\frac{95}{16}\right\},\left\{59,\frac{31}{16},\frac{87}{16}\right\},\left\{59,\frac{39}{16},\frac{79}{16}\right\},\left\{59,\frac{47}{16},\frac{71}{16}\right\},\left\{59,\frac{55}{16},\frac{63}{16}\right\},\left\{\frac{121}{2},\frac{1}{16},\frac{15}{2}\right\}\\
\end{array}$
}
\end{threeparttable}
\end{center}
\end{table}

\section{The $q$-series of $\mathbf{[3,4]}$ Admissible Characters\label{app3}}

We solved the $\mathbf{[3,4]}$ MLDE and obtained admissible character solutions for $0 < c \leq 96$ in section \ref{4s}. Here, we gather all relevant details of the solution in one place.

\vspace{7mm}\hrule\vspace{1mm}
\noindent \textbf{S'-1 :} \\ 
$c = \frac{ 82}{5}$ , $h_1 = \frac{41}{60}$, $h_2 = \frac{6}{5}$. \\
$\alpha_0 = - \frac{41}{60}$, $\alpha_1 =  0$, $\alpha_2 =  \frac{6}{5}$. \\
$N = 636592404,~ \text{m}_1 = 410,~\text{m}_2 = 64739,~\text{m}_3 = 2089934,~k = 22787074700652379514040978033023622144$. \\
$D_1 = 1, \quad D_2 = 902.$ \\
$\mu = -55296, \quad  \mu_{1,0} = -\frac{1071}{25} , \quad \mu_{0,1} = 0.$ 
\begin{align*}
    \chi_{0}&=q^{-\frac{41}{60}}(1+410 q+64739 q^2+2089934 q^3+35855320 q^4+423577724 q^5+3881851095 q^6+O[q]^7)\\
    \chi_{\frac{41}{60}}&=q^{0}(1)\\
    \chi_{\frac{6}{5}}&=q^{\frac{31}{60}}(902+86428 q+2370415 q^2+37227590 q^3+414998023 q^4+3647565246 q^5+26834755594 q^6+O[q]^7)
\end{align*}
\hrule\vspace{2mm}
\noindent \textbf{S'-2 :} \\ 
$c =  17$ , $h_1 = \frac{17}{24}$, $h_2 = \frac{5}{4}$. \\
$\alpha_0 = - \frac{17}{24}$, $\alpha_1 =  0$, $\alpha_2 =  \frac{13}{24}$. \\
$N = 659882370,~ \text{m}_1 = 323,~\text{m}_2 = 60860,~\text{m}_3 = 2158575,~k = 45853329031423822791822069765573120000$. \\
$D_1 = 1, \quad D_2 = 51.$ \\
$\mu = -55296, \quad  \mu_{1,0} = -\frac{189}{4} , \quad \mu_{0,1} = 0.$ 
\begin{align*}
    \chi_{0}&=q^{-\frac{17}{24}}(1+323 q+60860 q^2+2158575 q^3+39638985 q^4+495253231 q^5+4763929125 q^6+O[q]^7)\\
    \chi_{\frac{17}{24}}&=q^{0}(1)\\
    \chi_{\frac{5}{4}}&=q^{\frac{13}{24}}(51+5083 q+146285 q^2+2399600 q^3+27840135 q^4+253892484 q^5+1933281769 q^6+O[q]^7)
\end{align*}
\hrule\vspace{1mm}
\noindent \textbf{S'-3 :} \\ 
$c =  18$ , $h_1 = \frac{3}{4}$, $h_2 = \frac{4}{3}$. \\
$\alpha_0 = - \frac{3}{4}$, $\alpha_1 =  0$, $\alpha_2 =  \frac{7}{12}$. \\
$N = 698698980,~ \text{m}_1 = 234,~\text{m}_2 = 59805,~\text{m}_3 = 2482242,~k = 72067959168925880011559792323325184000$. \\
$D_1 = 1, \quad D_2 = 1.$ \\
$\mu = -55296, \quad  \mu_{1,0} = -55 , \quad \mu_{0,1} = 0.$ 
\begin{align*}
    \chi_{0}&=q^{-\frac{3}{4}}(1+234 q+59805 q^2+2482242 q^3+51022782 q^4+699150744 q^5+7282887147 q^6+O[q]^7)\\
    \chi_{\frac{3}{4}}&=q^{0}(1)\\
    \chi_{\frac{4}{3}}&=q^{\frac{7}{12}}(1+106 q+3293 q^2+57922 q^3+716625 q^4+6936054 q^5+55835218 q^6+O[q]^7)
\end{align*}
\hrule\vspace{1mm}
\noindent \textbf{S'-4 :} \\ 
$c =  \frac{94}{5}$ , $h_1 = \frac{47}{60}$, $h_2 = \frac{7}{5}$. \\
$\alpha_0 = - \frac{47}{60}$, $\alpha_1 =  0$, $\alpha_2 =  \frac{37}{60}$. \\
$N = 729752268,~ \text{m}_1 = 188,~\text{m}_2 = 62087,~\text{m}_3 = 2923494,~k = 88181885687655607252799840662343242752$. \\
$D_1 = 1, \quad D_2 = 4794.$ \\
$\mu = -55296, \quad  \mu_{1,0} = -\frac{1539}{25} , \quad \mu_{0,1} = 0.$ 
\begin{align*}
    \chi_{0}=&q^{-\frac{47}{60}}(1+188 q+62087 q^2+2923494 q^3+65738853 q^4+969081250 q^5+10749673338 q^6+O[q]^7)\\
    \chi_{\frac{47}{60}}=&q^{0}(1)\\
    \chi_{\frac{7}{5}}=&q^{\frac{37}{60}}(4794+532134 q+17518686 q^2+325092138 q^3+4226142146 q^4+42826539073 q^5+359900395528 q^6\notag\\&+O[q]^7)
\end{align*}
\hrule\vspace{1mm}
\noindent \textbf{S'-5 :} \\ 
$c =  20$ , $h_1 = \frac{5}{6}$, $h_2 = \frac{3}{2}$. \\
$\alpha_0 = - \frac{5}{6}$, $\alpha_1 =  0$, $\alpha_2 =  \frac{2}{3}$. \\
$N = 776332200,~ \text{m}_1 = 140,~\text{m}_2 = 69950,~\text{m}_3 = 3983800,~k = 107123028848651267056905128194867200000$. \\
$D_1 = 1, \quad D_2 = 5.$ \\
$\mu = -55296, \quad  \mu_{1,0} = -72 , \quad \mu_{0,1} = 0.$ 
\begin{align*}
    \chi_{0}&=q^{-\frac{5}{6}}(1+140 q+69950 q^2+3983800 q^3+102455165 q^4+1683350628 q^5+20493356870 q^6+O[q]^7)\\
    \chi_{\frac{5}{6}}&=q^{0}(1)\\
    \chi_{\frac{3}{2}}&=q^{\frac{2}{3}}(5+592 q+21160 q^2+424000 q^3+5918900 q^4+64093600 q^5+573188736 q^6+O[q]^7)
\end{align*}
\hrule\vspace{1mm}
\noindent \textbf{S'-6 :} \\ 
$c =  \frac{106}{5}$ , $h_1 = \frac{53}{60}$, $h_2 = \frac{8}{5}$. \\
$\alpha_0 = - \frac{53}{60}$, $\alpha_1 =  0$, $\alpha_2 =  \frac{43}{60}$. \\
$N = 822912132,~ \text{m}_1 = 106,~\text{m}_2 = 84429,~\text{m}_3 = 5825442,~k = 117541111174650568416870757866489919488$. \\
$D_1 = 1, \quad D_2 = 15847.$ \\
$\mu = -55296, \quad  \mu_{1,0} = -\frac{2079}{25} , \quad \mu_{0,1} = 0.$ 
\begin{align*}
    \chi_{0}=&q^{-\frac{53}{60}}(1+106 q+84429 q^2+5825442 q^3+171303844 q^4+3133434836 q^5+41810345161 q^6+O[q]^7)\\
    \chi_{\frac{53}{60}}=&q^{0}(1)\\
    \chi_{\frac{8}{5}}=&q^{\frac{43}{60}}(15847+1991846 q+76895739 q^2+1657293358 q^3+24761617596 q^4+285705257932 q^5\notag\\&+2712124663327 q^6+O[q]^7)
\end{align*}
\hrule\vspace{1mm}
\noindent \textbf{S'-7 :} \\ 
$c =  22$ , $h_1 = \frac{11}{12}$, $h_2 = \frac{5}{3}$. \\
$\alpha_0 = - \frac{11}{12}$, $\alpha_1 =  0$, $\alpha_2 =  \frac{3}{4}$. \\
$N = 853965420,~ \text{m}_1 = 88,~\text{m}_2 = 99935,~\text{m}_3 = 7846300,~k = 115346865865608882029083794591565440000$. \\
$D_1 = 1, \quad D_2 = 22.$ \\
$\mu = -55296, \quad  \mu_{1,0} = -91 , \quad \mu_{0,1} = 0.$ 
\begin{align*}
    \chi_{0}&=q^{-\frac{11}{12}}(1+88 q+99935 q^2+7846300 q^3+252284835 q^4+4954497636 q^5+70231360540 q^6+O[q]^7)\\
    \chi_{\frac{11}{12}}&=q^{0}(1)\\
    \chi_{\frac{5}{3}}&=q^{\frac{3}{4}}(22+2871 q+116370 q^2+2627625 q^3+41008770 q^4+492895557 q^5+4862385000 q^6+O[q]^7)
\end{align*}
\hrule\vspace{1mm}
\noindent \textbf{S'-8 :} \\ 
$c =  23$ , $h_1 = \frac{23}{24}$, $h_2 = \frac{7}{4}$. \\
$\alpha_0 = - \frac{23}{24}$, $\alpha_1 =  0$, $\alpha_2 =  \frac{19}{24}$. \\
$N = 892782030,~ \text{m}_1 = 69,~\text{m}_2 = 131905,~\text{m}_3 = 12195106,~k = 89958121037055261674858306367696384000$. \\
$D_1 = 1, \quad D_2 = 253.$ \\
$\mu = -55296, \quad  \mu_{1,0} = -\frac{405}{4} , \quad \mu_{0,1} = 0.$ 
\begin{align*}
    \chi_{0}&=q^{-\frac{23}{24}}(1+69 q+131905 q^2+12195106 q^3+438460776 q^4+9406082277 q^5+143713413394 q^6+O[q]^7)\\
    \chi_{\frac{23}{24}}&=q^{0}(1)\\
    \chi_{\frac{7}{4}}&=q^{\frac{19}{24}}(253+34523 q+1483178 q^2+35422737 q^3+582761520 q^4+7359901238 q^5+76075808205 q^6+O[q]^7)
\end{align*}
\hrule\vspace{1mm}
\noindent \textbf{S'-9 :} \\ 
$c =  \frac{118}{5}$ , $h_1 = \frac{59}{60}$, $h_2 = \frac{9}{5}$. \\
$\alpha_0 = - \frac{59}{60}$, $\alpha_1 =  0$, $\alpha_2 =  \frac{49}{60}$. \\
$N = 916071996,~ \text{m}_1 = 59,~\text{m}_2 = 164315,~\text{m}_3 = 16778125,~k = 49159175864915842196370689437986998016$. \\
$D_1 = 1, \quad D_2 = 32509.$ \\
$\mu = -55296, \quad  \mu_{1,0} = -\frac{2691}{25} , \quad \mu_{0,1} = 0.$ 
\begin{align*}
    \chi_{0}=&q^{-\frac{59}{60}}(1+59 q+164315 q^2+16778125 q^3+645117269 q^4+14589857952 q^5+233098944158 q^6+O[q]^7)\\
    \chi_{\frac{59}{60}}=&q^{0}(1)\\
    \chi_{\frac{9}{5}}=&q^{\frac{49}{60}}(32509+4551260 q+202207691 q^2+4989354706 q^3+84644822885 q^4+1100382873030 q^5\notag\\&+11689161859550 q^6+O[q]^7)
\end{align*}
\hrule

\end{document}